\documentclass[pra,twocolumn,preprintnumbers,amsmath,amssymb,showpacs,nofootinbib,floatfix]{revtex4}

\usepackage{graphicx,bm}

\makeatletter
\def\graphicscale{\twocolumn@sw{0.3}{0.4}}
\def\graphicthreescale{\twocolumn@sw{0.3}{0.4}}

\begin{document}
\title{Universal quantum behaviors of interacting fermions in 1D traps: \\

from few particles to the trap thermodynamic limit}

\author{Adriano Angelone, Massimo Campostrini and Ettore Vicari}

\affiliation{Dip. di Fisica dell'Universit\`a di Pisa and INFN, 
Largo Pontecorvo 3, I-56127 Pisa, Italy}

\date{\today}

\begin{abstract}

We investigate the ground-state properties of trapped fermion systems
described by the Hubbard model with an external confining potential.
We discuss the universal behaviors of systems in different regimes:
from few particles, i.e. in dilute regime, to the trap thermodynamic
limit.

The asymptotic trap-size (TS) dependence in the dilute regime
(increasing the trap size $\ell$ keeping the particle number $N$
fixed) is described by a universal TS scaling controlled by the dilute
fixed point associated with the metal-to-vacuum quantum transition.
This scaling behavior is numerically checked by DMRG simulations of
the one-dimensional (1D) Hubbard model.  In particular, the particle
density and its correlations show crossovers among different regimes:
for strongly repulsive interactions they approach those of a spinless
Fermi gas, for weak interactions those of a free Fermi gas, and for
strongly attractive interactions they match those of a gas of
hard-core bosonic molecules.

The large-$N$ limit keeping the ratio $N/\ell$ fixed corresponds to a
1D trap thermodynamic limit.  We address issues related to the
accuracy of the local density approximation (LDA).  We show that the
particle density approaches its LDA in the large-$\ell$ limit.  When
the trapped system is in the metallic phase, corrections at finite
$\ell$ are $O(\ell^{-1})$ and oscillating around the center of the
trap. They become significantly larger at the boundary of the fermion
cloud, where they get suppressed as $O(\ell^{-1/3})$ only. This
anomalous behavior arises from the nontrivial scaling at the
metal-to-vacuum transition occurring at the boundaries of the fermion
cloud.

\end{abstract}


\maketitle



\section{Introduction}
\label{intro}

The progress in the experimental activity in atomic physics, quantum
optics and nanoscience has provided a great opportunity to investigate
the nature of the quantum dynamics, and the interplay between quantum
and statistical behaviors in particle systems.  In particular,
experiments with cold atoms realize systems which are accurately
described by microscopic theoretical models such as dilute atomic
Fermi and Bose gases, or Hubbard and Bose-Hubbard models in optical
lattices.  See e.g. Refs.~\cite{BDZ-08,GPS-08,Esslinger-10,GBL-13}.  A
peculiar feature of these experiments is the confinement of the atoms,
i.e. the presence of a inhomogeneous space-dependent (usually
harmonic) trapping potential.  The tunability of the confining
potential allows the realization of quasi-two and quasi-one
dimensional systems, by tightly confining the particles along one or
two transverse dimensions.

The inhomogeneity induced by the trapping potential gives rise to
peculiar effects, such as the possibility of simultaneously observing
different phases, in particular Mott incompressible (insulator) and
superfluid (metallic) phases, depending on the distance from the
center of the trap.~\cite{BDZ-08,GPS-08,Esslinger-10} This is
essentially due to the fact that the effective local chemical
potential decreases with increasing the distance from the center of
the trap center.  At continuous quantum transitions the presence of
the trap does not allow the development of critical modes with
diverging length scales.  However, in the limit of large trap size
$\ell$ the system develops peculiar critical modes, which give rise to
a universal trap-size scaling (TSS), controlled by the universality
class of the transition of the homogeneous system~\cite{CV-09,CV-10}.
Therefore, in experiments of trapped particle systems, a thorough
understanding of the quantum many-body dynamics calls for a
quantitative analysis of the trap effects.  This issue has been much
discussed by theoretical and experimental investigations, see e.g.
Refs.~\onlinecite{SGN-95,DSMS-96,BR-97,GWSZ-00,DLO-01,MS-02,RMBS-03,
  RFZZ-03,WATB-04,ABGP-04,MB-04,FWMGB-06,
  XPAT-06,GPTCCR-06,HLD-07,Orso-07,CCQH-07,FH-07,
  CCM-08,XA-08,KB-09,GZHC-09,Taylor-09,CV-09,
  ZKKT-09,RBTS-09,Trotzky-etal-10,CV-10,ZH-10,HZ-10,
  PPS-10,PPS-10b,NNCS-10,ZKKT-10,CV-10-XX,CV-10-BH,TU-10,
  HOF-10,STS-11,FCMCW-11,ZHTGC-11,CV-11,HM-11,SBE-11,CTV-12,
  Pollet-12,CT-12,KLS-12,CR-12,CTV-13,CNPV-13}.

In this paper we consider spin-1/2 fermion systems described by the
lattice Hubbard model, at zero temperature, therefore in their ground
state.  We investigate the scaling behavior of unpolarized systems,
i.e. with equal number of spin-down and spin-up particles, when
varying the size of the trap confining the particles.  We mostly
consider one-dimensional (1D) systems with both attractive and
repulsive interactions, in the dilute regime when the trap size $\ell$
increases keeping the particle number $N$ fixed, and in the {\em trap
  thermodynamic} limit defined as the large-$N$ limit keeping the
ratio $N/\ell$ fixed.  Note that 1D fermion systems are also
experimentally interesting, indeed their realizations, and evidences
of pairing, have been recently reported in
Refs.~\cite{Moritz-etal-05,Liao-etal-10}.

The asymptotic trap-size dependence in the dilute regime for a system
of $N$ particles can be described in the framework of the TSS theory,
whose scaling behavior is controlled by the dilute fixed point
associated with the metal-to-vacuum quantum transition.  This is
controlled by the trap exponent $\theta$ related to the
renormalization-group (RG) perturbation arising from the trapping
potential, and requires a nontrivial rescaling of the on-site
interaction in 1D systems.  The TSS predictions are compared with
numerical results based on DMRG simulations.  In particular, we
address issues related to the expected crossover from strongly
repulsive interactions, where the system assumes the properties of a
free spinless Fermi gas, to strongly attractive interactions, where
the system is expected to be effectively constituted by hard-core
bosonic (spin zero) molecules of two fermions, passing through a free
Fermi gas in the absence of interactions.

We also investigate the trap thermodynamic limit of the Hubbard
model, i.e.  the large-$\ell$ and large-$N$ limit keeping the ratio
$N/\ell^d$ fixed.  In particular, we address issues related to the
accuracy of the local-density approximation (LDA), which approximates
the space-dependent particle density of a inhomogeneous trap by the
particle density of the homogenous system at the corresponding value
of the effective chemical potential.  LDA is usually used to determine
the particle density in inhomogeneous systems, providing an accurate
approximation when the inhomogeneity is sufficiently smooth.  Of
course, LDA is not exact, in particular at finite trap sizes.
Therefore, an analysis of the deviations from LDA, and therefore of
its accuracy, is required to get a robust confidence of its results.

LDA has been largely employed in studies of inhomogeneous interacting
fermion systems, see Refs.~\cite{BDZ-08,GPS-08,Esslinger-10,GBL-13}
and references therein, an in particular of 1D
systems~\cite{SGN-95,BR-97,GWSZ-00,RMBS-03,RFZZ-03,ABGP-04,MB-04,
  XPAT-06,GPTCCR-06,HLD-07,Orso-07,CCQH-07,FH-07,
  CCM-08,XA-08,KB-09,TU-10,HOF-10,STS-11,SBE-11}.  The comparison with
numerical results for the particle density shows that it provides a
good approximation in many cases.  A more quantitative analysis of the
deviations from LDA may be achieved by establishing whether LDA
becomes exact in the large trap-size limit, and how deviations get
suppressed at large, but finite, $\ell$.  We investigate this issue in
the trapped 1D Hubbard model, where LDA of the particle density can be
exactly computed by Bethe-Ansatz methods~\cite{LW-68,1DHM}, allowing
us to perform an accurate study of the deviations from LDA.  We show
that LDA of the particle density tends to become exact in the large
trap-size limit, with power-law suppressed corrections. In particular,
the corrections appear significantly larger at the boundary of the
fermion cloud, decreasing as $O(\ell^{-1/3})$, due to the critical
modes arising from the metal-to-vacuum transition at the edges of the
trap.

The paper is organized as follows.  In Sec.~\ref{model} we present the
Hubbard model in the presence of an external potential coupled to the
particle density; we provide the definitions of the observables and
correlations which are considered in the paper.  In
Sec.~\ref{dilutetss} we investigate the trap-size dependence at a
fixed particle number, i.e. in the dilute regime, in the framework of
the TSS theory.  In Sec.~\ref{thlim} we consider the thermodynamic
limit in a trap, i.e. the large trap-size limit keeping the ratio
$N/\ell$ fixed; in particular, we address the accuracy of the LDA of
the particle density and the peculiar scaling behavior at the boundary
of the cloud.  Finally, in Sec.~\ref{conclusions} we summarize our
main results and draw our conclusions.  Some appendices report
technical details of some results mentioned in the paper.

\section{The Hubbard model}
\label{model}

The Hamiltonian of the Hubbard model reads
\begin{eqnarray}
H =  - t \sum_{\sigma,\langle {\bf x}{\bf y}\rangle} 
(c_{\sigma {\bf x}}^\dagger c_{\sigma {\bf y}} + {\rm h.c.})
+ U \sum_{\bf x} n_{\uparrow{\bf x}} n_{\downarrow{\bf x}}
\label{hm}
\end{eqnarray}
where ${\bf x}$ are the sites of a cubic lattice, $\langle {\bf
  x}{\bf y}\rangle$ indicates nearest-neighbor sites,
$c_{\sigma{\bf
    x}}$ is a fermionic operator,
    $\sigma=\uparrow\downarrow$ labels the spin states, 
 and $n_{\sigma{\bf x}}\equiv
c_{\sigma{\bf x}}^\dagger c_{\sigma{\bf x}}$.
The particle number operators $\hat{N}_\sigma= \sum_{\bf x} n_{\sigma{\bf
    x}}$ are conserved, i.e. $[H,\hat{N}_\sigma]=0$. 
In the following we  consider balanced  Fermi systems, thus
\begin{equation}
N_\uparrow=N_\downarrow=N/2
\label{defN}
\end{equation}
where $N$ is the total number of particles.  In this symmetric case
$\langle n_{\uparrow{\bf x}}\rangle = \langle n_{\downarrow{\bf
    x}}\rangle$ and $\langle c_{\uparrow{\bf x}}^\dagger
c_{\uparrow{\bf y}} \rangle = \langle c_{\downarrow{\bf x}}^\dagger
c_{\downarrow{\bf y}} \rangle$.

The presence of a trapping potential can be taken into account by
adding a external-potential term in the Hamiltonian of the Hubbard
model,
\begin{eqnarray}
&&H_t =  H + H_e,\label{hmt}\\
&& H_e = \sum_{\bf x} V({\bf x}) \, n_{\bf x},\quad n_{\bf x} 
\equiv \sum_\sigma n_{\sigma{\bf x}}.
\label{exth}
\end{eqnarray}
For simplicity, we assume a rotational invariant potential
\begin{eqnarray}
V({\bf x})= {1\over p} v^p r^p,\qquad r\equiv |{\bf x}|,\label{potential}
\end{eqnarray}
where $r$ is the distance from the center of the trap, and $p$ is a
positive number.  We set the origin ${\bf x}=0$ at the center
of the trap.  The trap size $\ell$ is defined as
\begin{equation}
\ell \equiv {(2t)^{1/p}\over v},  
\qquad V({\bf x}) = 2t \,{r^p\over p \ell^p}
\label{trapsize}
\end{equation}
The trapping potential is effectively harmonic in most experiments,
i.e. $p=2$.~\cite{BDZ-08}  In the limit $p\to\infty$ we recover a homogenous
spherical system of size $L=2\ell$ with hard-wall boundary conditions.

The above definition of trap size $\ell$ naturally arises when we
consider a thermodynamic limit in a trap~\cite{BDZ-08,PGS-04}.
Indeed, in the presence of
trap, a {\em trap thermodynamic} limit can be consistently defined as the
large-$\ell$ limit keeping the ratio $N/\ell^d$ fixed. This is
equivalent to introducing a chemical potential $\mu$, by adding the
term
\begin{eqnarray}
H_\mu = - \mu \sum_{\bf x} n_{{\bf x}} \label{chempot}
\end{eqnarray}
to the Hubbard Hamiltonian (\ref{hm}).

In the rest of the paper we set the kinetic constant $t=1$ and
$\hslash = 1$; their dependence can be easily inferred by dimensional
analyses.

In our study we focus on the ground-state properties.  We consider
one-point observables such as the particle density
\begin{equation}
\rho({\bf x}) = \langle n_{\bf x} \rangle
\label{rhox}
\end{equation}
and the double occupancy
\begin{equation}
d_o({\bf x}) = \langle n_{\uparrow{\bf x}} n_{\downarrow{\bf x}} \rangle.
\label{dbocc}
\end{equation}
Moreover, we analyze the behavior of correlation functions such as the
one-particle correlation
\begin{eqnarray}
C({\bf x},{\bf y}) = \sum_\sigma \langle 
c_{\sigma{\bf x}}^\dagger c_{\sigma{\bf y}} 
+ {\rm h.c.} \rangle, 
\label{gdef}
\end{eqnarray}
the connected density-density  correlations
\begin{eqnarray}
&&G({\bf x},{\bf y}) = \langle n_{{\bf x}} n_{{\bf y}} \rangle - 
\langle n_{{\bf x}} \rangle \langle n_{{\bf y}} \rangle ,
\label{gndef}\\
&&M({\bf x},{\bf y}) = \langle n_{\uparrow{\bf x}} n_{\downarrow{\bf y}} \rangle-
\langle n_{\uparrow{\bf x}} \rangle \langle n_{\downarrow{\bf y}} \rangle ,
\label{mndef}
\end{eqnarray}
and the pair correlation
\begin{equation}
P({\bf x},{\bf y}) = \langle 
c_{\uparrow{\bf x}}^\dagger  c_{\downarrow{\bf x}}^\dagger 
c_{\downarrow{\bf y}} c_{\uparrow{\bf y}} + {\rm h.c.}\rangle.
\label{bdef}
\end{equation}

\begin{figure}[tbp]
\includegraphics*[scale=\graphicscale]{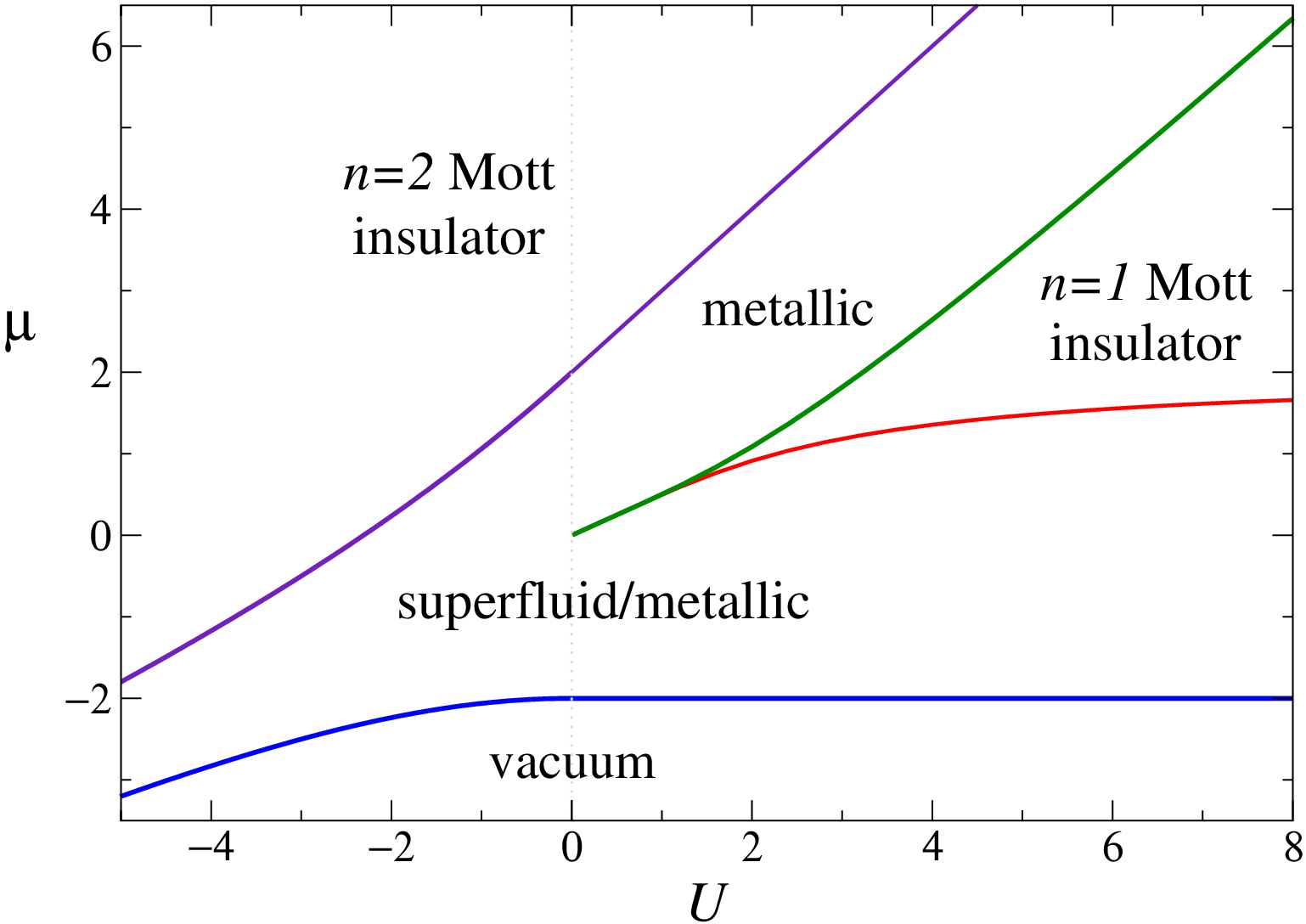}
\caption{(Color online) Phase diagram of the 1D unpolarized Hubbard
  model, which shows the different quantum phases related to the
  behavior of the particle density: vacuum, metal, $n=1$ and $n=2$
  Mott insulator phases.  
  }
\label{phdiad1}
\end{figure}

In the following we mostly consider 1D systems.  The homogeneous 1D
Hubbard model has been extensively studied, obtaining several exact
results, see e.g. Ref.~\cite{1DHM}. The phase diagram of homogenous 1D
unpolarized systems presents various quantum phases related to the
behavior of the particle density, such as vacuum, metallic and Mott
insulator (incompressible) phases, depending on the chemical potential
$\mu$ and the on-site interaction $U$. Their phase boundaries are
known exactly, by computations based on the Bethe Ansatz, see
e.g. Ref.~\cite{1DHM}.  In particular, the vacuum-to-metal transition
occurs at
\begin{eqnarray}
&&\mu_0 = -2 \qquad  {\rm for}\;\;U\ge 0,\label{mu0p}\\
&&\mu_0 = -2 \sqrt{1+{U^2\over 16}} \quad  {\rm for}\;\;U< 0.\label{mu0n}
\end{eqnarray}
Fig.~\ref{phdiad1} shows the phase diagram for the 1D unpolarized
Hubbard model, with the different phases characterized by different
behaviors of the particle density.  In the metallic phase region the
low-energy properties are characterized by algebraically decaying
correlations~\cite{1DHM,Giamarchi-book}.  At small attractive
interactions fermions form Cooper-pair-like bound states resembling
those of BCS superconductors, while in the limit of strong
interactions the pairs become tightly bounded within an extension of
the lattice spacing, giving effectively rise to a system of hard-core
bosons.

We finally mention that the 1D Hubbard model, even in the presence of
an external confining potential, can be exactly mapped into a model of
two interacting species (flavors) of hard-core bosons described by a
Bose-Hubbard model, whose Hamiltonian is formally analogous to that of
the Hubbard model with the fermionic operators $c_{\sigma x}$ replaced
by hard-core bosonic operators $b_{\sigma x}$. Some details are
reported in App.~\ref{equibh}.  The mapping between fermionic and
bosonic operators is nonlocal, but it directly maps the density
operator of fermions into that of bosons, i.e. $c_{\sigma x}^\dagger
c_{\sigma x}\to b_{\sigma x}^\dagger b_{\sigma x}$.  Thus the particle
density and its correlations for the fermionic Hubbard model are
identical to those of the two-flavor Bose-Hubbard model.

\section{Trap-size scaling in the dilute regime}
\label{dilutetss}

In this section we study the asymptotic trap-size dependence in the
dilute regime, when increasing the trap size while keeping the
particle number $N$ fixed. This issue is best addressed in the
framework of the TSS theory~\cite{CV-10}. Its universal features are
determined by the RG dimensions of the relevant perturbations at the
fixed point controlling the vacuum-to-metal transition of the Hubbard
model.

The scaling behavior of the Hubbard model in the dilute regime can be
inferred by a RG analysis of the corresponding quantum field theory,
see e.g. Ref.~\cite{Sachdev-book},
\begin{eqnarray}
Z_F &=& \int D\psi_\sigma^* D\psi_\sigma {\rm exp}\left(-\int_0^{1/T} 
d\tau \int d^dx 
\, {\cal L}_F \right),\label{zf}\\
{\cal L}_F &=& \sum_\sigma [\psi_\sigma^*{\partial \psi_\sigma\over d\tau} + 
{1\over 2m}|\nabla \psi_\sigma|^2 
- \mu |\psi_\sigma|^2]  + \nonumber \\
&+& u \, \psi_\uparrow^* \psi_\downarrow^*  
\psi_\downarrow \psi_\uparrow .
\nonumber
\end{eqnarray}
The dynamic critical exponent $z$ and the RG dimensions $y_\mu$ and
$y_u$ of the relevant parameter $\mu$ and $u$ at the {\em dilute}
fixed point ($\mu=0$ and $u=0$) encode most important information on
the scaling properties in the dilute regime.  Since the dilute fixed
point is essentially Gaussian, they can be obtained by simple
dimensional analyses,
\begin{equation}
z=2,\quad y_\mu=2, \quad y_u=2-d.
\label{rgdims}
\end{equation}
In order to obtain the scaling behavior of the observables, such as
the particle density and the correlations introduced in
Sec.~\ref{model}, we also need the RG dimensions of the fermionic
field $\psi$, density operator $n = \psi^\dagger \psi$, and pair
operator $p = \psi_\uparrow\psi_\downarrow$, which are respectively
$y_\psi=d/2$ and $y_n = y_p = d$.

The trap effects in the dilute regime can be inferred by analyzing the
RG perturbation arising from an external confining potential $V({\bf
  x})$ such as that of Eq.~(\ref{potential}), i.e.
\begin{equation}
P_V({\bf x}) = V({\bf x}) \sum_\sigma |\psi_\sigma({\bf x})|^2,
\qquad V({\bf x}) = v^p |{\bf x}|^p,
\label{Pv}
\end{equation}
at the dilute fixed point.  Proceeding analogously to the case of
spinless free-fermion systems~\cite{CV-10-XX,CV-10-BH}, the RG
dimension $y_v$ of the potential coupling $v$ can be obtained from the
RG relation
\begin{equation}
p y_v - p =  d+z-y_{n} = y_\mu.
\label{yU}
\end{equation}

This RG analysis tells us that the trap induces a length scale 
\begin{equation}
\xi \sim \ell^{\theta}
\label{xisca}
\end{equation}
with
\begin{equation}
\theta\equiv {1\over y_v} = {p\over p+y_\mu}={p\over p+2}.
\label{thetaexp}
\end{equation}
This implies that the  spatial coordinates ${\bf x}$ must be rescaled as
${\bf X}={\bf x}/\ell^\theta$ to get a nontrivial TSS limit.  In
particular, $\theta=1/2$ for the harmonic $p=2$ potential, in any
spatial dimension.

The knowledge of the above critical exponents and RG dimensions allows
us to write down the universal TSS ansatz for the observables
introduced in Sec.~\ref{model} in the dilute regime, which provides
the asymptotic behavior of their trap-size dependence. For example,
for the $n$-point correlation of a generic local operator ${\cal
  O}({\bf x})$, we expect~\cite{CV-10,CV-10-XX}
\begin{eqnarray}
F({\bf x}_1,...,{\bf x}_n; \ell,U, N ) &\equiv& 
\langle {\cal O}({\bf x}_1) ... {\cal O}({\bf x}_n) \rangle 
\label{genf}\\
&\approx&
\ell^{-\varepsilon} 
{\cal F}({\bf X}_1,...,{\bf X}_n; U\ell^{\theta y_u}, N )
\nonumber 
\end{eqnarray}
where
\begin{equation}
\varepsilon = n \theta y_o,\qquad {\bf X}_i={\bf x}_i/\ell^\theta,
\label{varepX}
\end{equation}
and $y_o$ is the RG dimension of the operator ${\cal O}({\bf x})$ at
the dilute fixed point.  Corrections to the above asymptotic behavior
are generally suppressed by further negative powers of $\ell$.  TSS
has some analogies with the standard FSS for homogeneous
systems~\cite{FBJ-73,PV-14} with two main differences: the
inhomogeneity due to the space dependence of the external field, and
the size $L$ replaced by $\ell^\theta$.

\subsection{TSS in a lattice gas of free spinless fermions}
\label{tts}

The trap-size dependence of the simplest noninteracting case $U=0$ can
be easily determined using the results obtained for trapped lattice
gases of spinless fermions, defined by
\begin{eqnarray}
H =  - \sum_{\langle {\bf x}{\bf y}\rangle} 
(c_{\bf x}^\dagger c_ {\bf y} + {\rm h.c.})
+ \sum_{\bf x} V({\bf x}) n_{\bf x},
\label{freef}
\end{eqnarray}
where $c_{\bf x}$ is a spinless fermionic operator, $n_{\bf x}=c_{\bf
  x}^\dagger c_ {\bf x}$, and the potential $V({\bf x})$ is given in
Eq.~(\ref{potential}).  In the following we report some of the results
of Refs.~\cite{V-12,CV-10-BH} which are useful for the rest of the
paper.

In arbitrary dimensions the asymptotic trap-size dependence of the
particle density of a free gas of $N$ spinless fermions ($N=\langle
\hat{N} \rangle$ and $\hat{N}=\sum_{\bf x} n_{\bf x}$) is given
by~\cite{CV-10-BH}
\begin{eqnarray}
\rho({\bf x},\ell,N) \approx \ell^{-d\theta} S_p({\bf X},N),\label{rhoxs0}
\end{eqnarray}
where $\theta$ is the same trap exponent (\ref{thetaexp}), and ${\bf
  X}\equiv {\bf x}/\ell^\theta$ As shown in
Refs.~\cite{CV-10,CV-10-BH}, the TSS limit corresponds to a continuum
limit in the presence of the trap.  Thus the TSS functions providing
the asymptotic trap-size dependence can be exactly derived from the
ground state of a trapped spinless Fermi gas defined in the continuum.
This is given by a Slater determinant of the lowest $N$ one-particle
eigenfunctions $\varphi_k({\bf x})$ of the Schr\"odinger problem
\begin{eqnarray}
H\varphi_k=\varepsilon_k \varphi_k,\qquad H = {1\over 2} 
{\bf p}^{\,2} 
+ {1\over p} |{\bf x}|^p,
\label{hpep}
\end{eqnarray}
with unit mass and trap size.  This allows us to write the TSS
function $S_p$ in Eq.~(\ref{rhoxs0}) as
\begin{eqnarray}
S_p({\bf X},N) = \sum_{k=1}^N \varphi_k({\bf X})^2. \label{spf}
\end{eqnarray}
In particular, in the case of 1D systems in a harmonic trap, it can be
written as~\cite{V-12,GWSZ-00}
\begin{eqnarray}
S_2(x,N) = {\sqrt{N}\over \sqrt{2}}\,
\left[ \varphi^\prime_{N+1}(x)\varphi_N(x) - 
\varphi^\prime_{N}(x)\varphi_{N+1}(x)\right]\quad
\label{ss2}
\end{eqnarray}
where 
\begin{eqnarray}
&&\varphi_k(x) = {H_{k-1}(x)\over
 \pi^{1/4} 2^{(k-1)/2} (k-1)!^{1/2}} \, e^{-x^2/2},
 \label{1deigf}
\nonumber
\end{eqnarray}
$H_k$ are the Hermite polynomials, and the corresponding
eigenvalues are $\varepsilon_k \propto k-1/2$.  In the
case of a 1D hard-wall trap, corresponding to $p\to\infty$,
\begin{eqnarray}
S_\infty(x,N) = {N\over 2} + {1\over 4} - {{\rm sin}[\pi(N+1/2)(1+x)]\over
4 \,{\rm sin}[\pi(1+x)/2]}\label{ssi}
\end{eqnarray}
with $|x|\le 1$.

The one particle correlation behaves as
\begin{eqnarray}
C({\bf x}_1,{\bf x}_2,\ell,N) \approx \ell^{-d\theta} E_p({\bf X}_1,{\bf X}_2,N),
\label{c0xs0}
\end{eqnarray}
where ${\bf X}_i= {\bf x}_i/\ell^\theta$ and
\begin{eqnarray}
E_p({\bf X}_1,{\bf X}_2,N)= \sum_{k=1}^N \varphi_k({\bf X}_1)\,\varphi_k({\bf X}_2).
\label{epf}
\end{eqnarray}
The connected density correlation scales asymptotically as
\begin{eqnarray}
&&G({\bf x}_1,{\bf x}_2,\ell,N) \approx 
\ell^{-2d\theta} Y_p({\bf X}_1,{\bf X}_2,N), \label{gn0xs0} \\
&&Y_p({\bf X}_1,{\bf X}_2,N) = - E_p({\bf X}_1,{\bf X}_2,N)^2 
\label{ypn}
\end{eqnarray}
for $|{\bf x}_1-{\bf x}_2|>0$, which can be derived from
the free fermion relation 
\begin{equation}
G({\bf x}_1,{\bf x}_2) 
=-|C({\bf x}_1,{\bf x}_2)|^2 + \delta({\bf x}_1-{\bf x}_2) C({\bf x}_1,{\bf x}_2).
\label{gcrel}
\end{equation}
Corrections to the above large-$\ell$ scaling behavior are
$O(\ell^{-2\theta})$ relatively to the leading term~\cite{CV-10}.

The above TSS behaviors are universal with respect to a large class of
short-range interactions.  For example, the same asymptotic behavior
is also expected in the presence of a density-density nearest-neighbour
interaction~\cite{CV-10-XX}
\begin{equation}
H_{nn} = w \sum_{\langle{\bf x}{\bf y}\rangle} n_{\bf x} n_{\bf y}.
\label{hdd}
\end{equation}
Indeed the RG dimension of the coupling $w$ is~\cite{Sachdev-book}
$y_w=-d$, thus the interaction $H_{nn}$ only induces
$O(\ell^{-d\theta})$ corrections to the asymptotic
behaviors.

The TSS functions of the particle density and the one-particle and
density correlations, cf. Eqs.~(\ref{rhoxs0}), (\ref{c0xs0}) and
(\ref{gn0xs0}), show also peculiar large-$N$ scaling behaviors.
Indeed,~\cite{V-12}
\begin{eqnarray}
S_p({\bf x},N) \approx  N^\theta {\cal S}_p(N^{(\theta-1)/d}{\bf X}),
\label{rhoxlnb}
\end{eqnarray}
and
\begin{eqnarray}
&&E_p({\bf X}_1,{\bf X}_2,N) \approx N^{\theta} {\cal E}_p(N^{\theta/d} {\bf X}_1,
N^{\theta/d} {\bf X}_2),\label{rxlab}\\
&&Y_p({\bf X}_1,{\bf X}_2,N) \approx N^{2\theta} {\cal Y}_p(N^{\theta/d} {\bf X}_1,
N^{\theta/d} {\bf X}_2).
\label{gnxlnb}
\end{eqnarray}
Note the different scaling of the space variable between the density
and the one-particle and connected density correlations.

In the case of 1D systems in a harmonic trap, the large-$N$ scaling
behavior can be derived from Eq.~(\ref{ss2}), obtaining
\begin{eqnarray}
&&S_2(X,N) \approx N^{1/2} 
{\cal S}_2(X/N^{1/2}) ,\label{sfgrho2}\\
&&{\cal S}_2(z) = {1\over\pi} \sqrt{2 - z^2}
\quad {\rm for} \;\; z\le z_b=\sqrt{2},\label{sfgrho3}
\end{eqnarray}
and ${\cal S}_2(z)=0$ for $z\ge z_b$.  The corrections to this
large-$N$ behavior are known~\cite{CV-10-BH}; they are $O(N^{-1})$
(relatively to the leading term) for $z<z_b$. In particular,
\begin{equation}
S_2(0,N) = {\sqrt{2}\over \pi} N^{1/2} \left[ 1 + {(-1)^{N+1}\over 4 N} + ...\right]
\label{rho0ln}
\end{equation}

We finally mention that around the boundary of the trap, i.e. at the
spatial point $z=\pm z_b=\sqrt{2}$ where the function ${\cal S}_2$
vanishes, a different large-$N$ scaling behavior sets
in:~\cite{CV-10-BH,Eisler-13}
\begin{eqnarray}
&&{\rm lim}_{N\to\infty} N^{-1/6}S_2(N^{1/2}z_b + N^{-1/6}z,N) = F(z),
\nonumber\\
&&F(z)= 2^{1/2} |{\rm Ai}^\prime(2^{1/2}z)|^2 -  2z|{\rm Ai}(2^{1/2}z)|^2 .
\label{fzsc}
\end{eqnarray}
This implies that at the boundary $z_b$ of the cloud the TSS function
of the particle density increases as $N^{1/6}$ only, instead of the
$O(N^{1/2})$ behavior for $|z|<z_b$. This is related to the fact that
the $z\to z_b$ limit of the $O(1/N)$ corrections is singular, see also
below.

In the following we extend these results to the Hubbard model, and in
general to lattice fermion gases with short-ranged interactions.

\subsection{TSS in the dilute regime of the Hubbard model}
\label{ttsdilp0}

\subsubsection{TSS for $d>2$}
\label{ttdgt2}

The RG analysis leading to Eqs.~(\ref{rgdims}) shows that the $U$ term
is irrelevant at the dilute fixed point for $d>2$, because 
its RG dimension $y_u$ is negative. 
Therefore, the
asymptotic trap-size dependence in the dilute regime turns out to be
the same as that of a free Fermi gases of $N$ particles with
$N_\uparrow=N_\downarrow=N/2$, independently of $U$, at least for
$U>U^*$ with $U^*<0$~\cite{Sachdev-book}.  

The asymptotic TSS can be easily determined from the results of
Sec.~\ref{tts}, obtaining
\begin{eqnarray}
&&\rho({\bf x},\ell,U,N) = \ell^{-d\theta} \left[2 S_p({\bf X},N/2)
 + O(\ell^{-\kappa})\right],\label{rhou0}\\ 
&&d_o({\bf x},\ell,U,N) = \ell^{-2d\theta} \left[S_p({\bf X},N/2)^2
 + O(\ell^{-\kappa})\right],\nonumber\\ 
&&C({\bf x}_1,{\bf x}_2,\ell,U,N)  =  
\ell^{-d\theta} 
\left[ 2 E_p({\bf X}_1,{\bf X}_2,N/2)  + O(\ell^{-\kappa})\right],
\nonumber\\
&&G({\bf x}_1,{\bf x}_2,\ell,U,N)  = \ell^{-2d\theta} 
\left[ 2 Y_p({\bf X}_1,{\bf X}_2,N/2)
  + O(\ell^{-\kappa})\right],\nonumber\\
&&M({\bf x}_1,{\bf x}_2,\ell,U,N)  =  O(\ell^{-2d\theta-\kappa}),\nonumber\\
&&P({\bf x}_1,{\bf x}_2,\ell,U,N)  =
\ell^{-2d\theta} 
\left[ E_p({\bf X}_1,{\bf X}_2,N/2)^2  + O(\ell^{-\kappa})\right],
\nonumber
\end{eqnarray}
where the exponent of the leading power law are determined by the RG
dimensions of the operators associated with the observables or the
correlations, see Eq.~(\ref{genf}).  The presence of the on-site
interaction induces scaling corrections with
\begin{equation}
\kappa = (d-2)\theta.
\label{kascco}
\end{equation}
For $d=3$ they dominate the scaling corrections expected within the
lattice model of free spinless fermion, i.e. the Hubbard model with
$U=0$, which are relatively suppressed as
$O(\ell^{-2\theta})$.~\cite{CV-10-BH}

The on-site interaction becomes marginal in 2D, thus a residual
weak dependence on $U$ is expected in the TSS limit, with 
at most logarithmic rescalings of the onsite interaction.

\subsubsection{TSS in 1D systems}
\label{ttdgt1d}

\begin{figure}[tbp]
\includegraphics*[scale=\graphicscale]{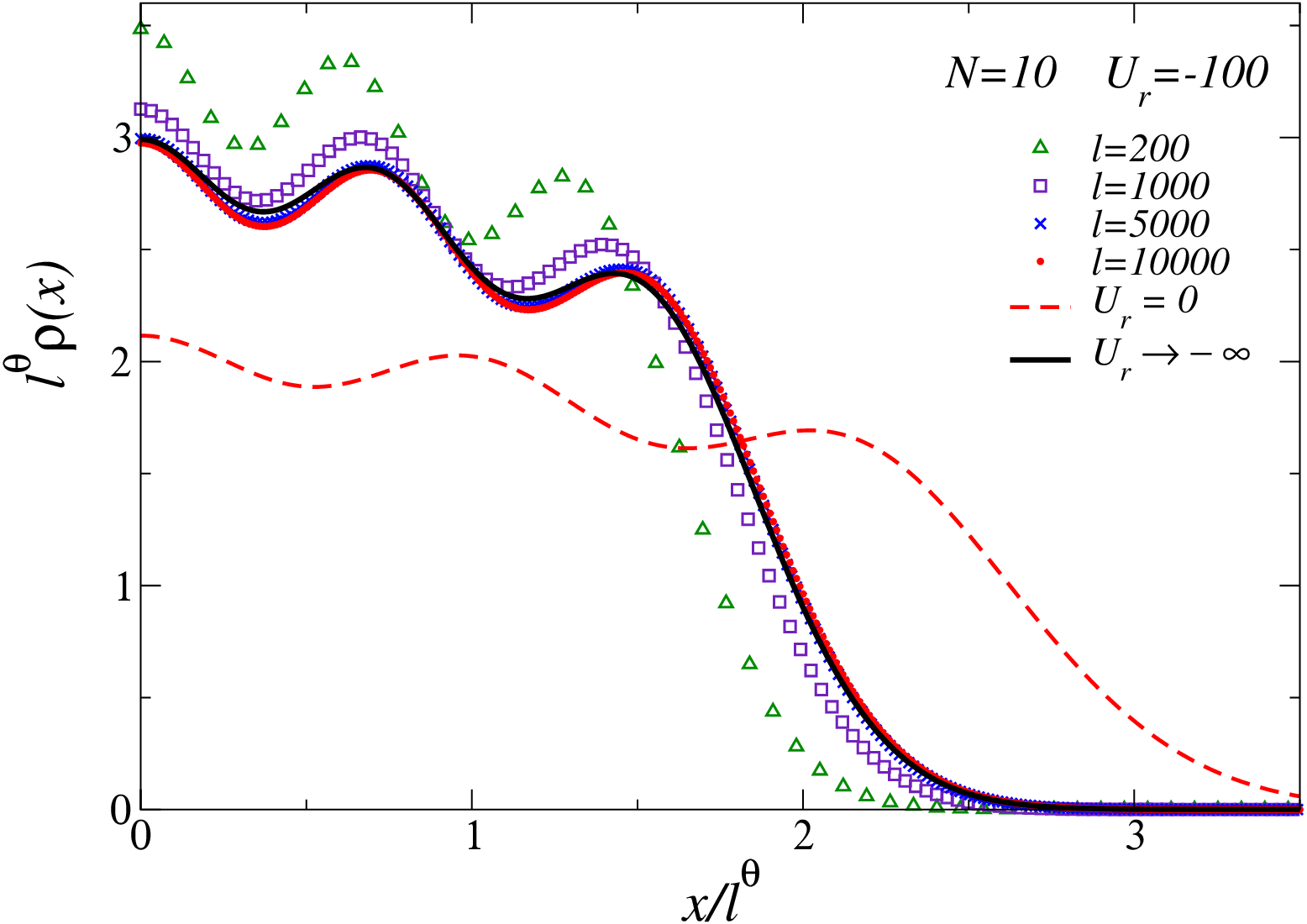}
\includegraphics*[scale=\graphicscale]{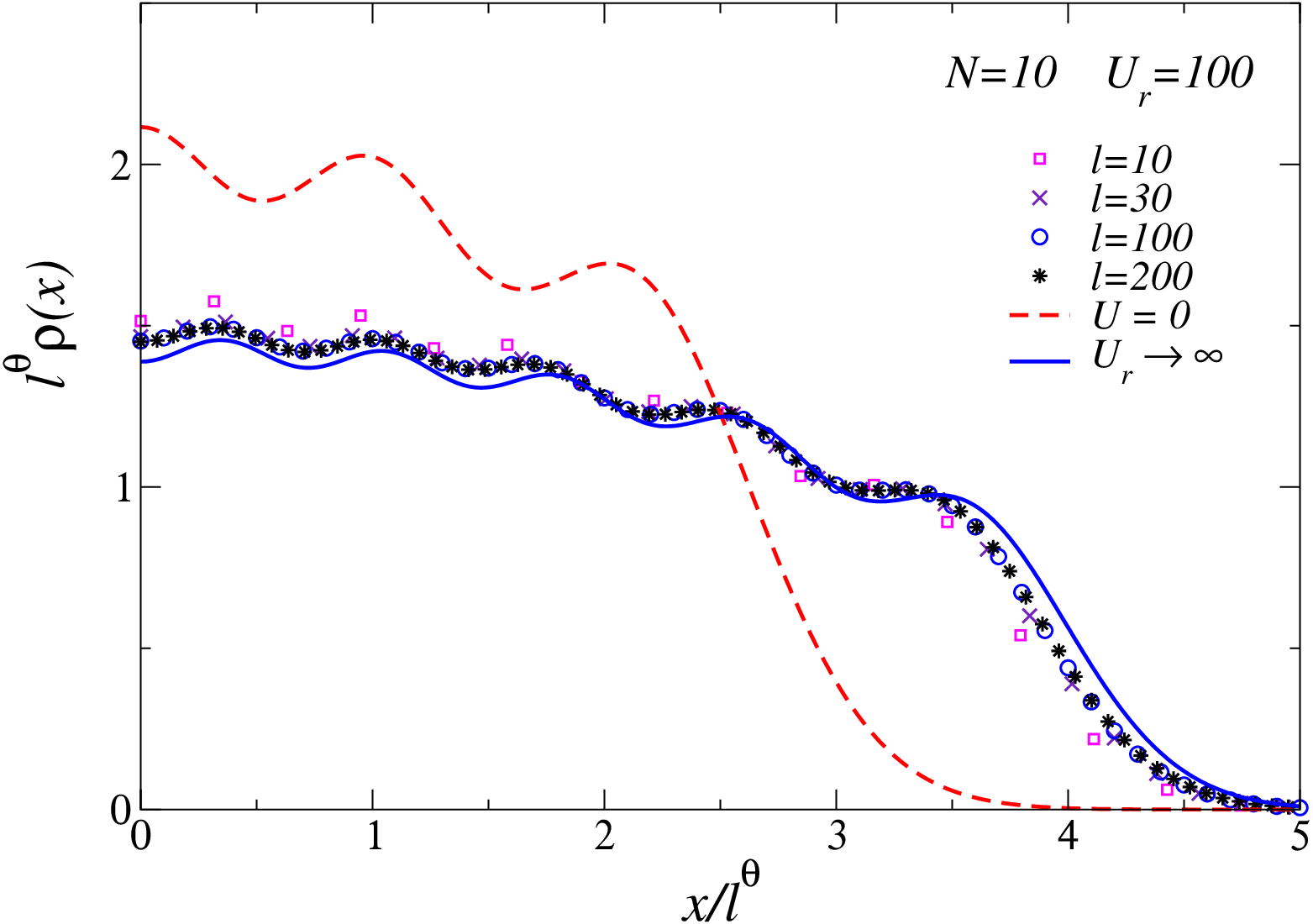}
\caption{(Color online) TSS of the particle density for $N=10$
  particles, at fixed $U_r\equiv U l^\theta=-100$ (top) and $U_r=100$
  (bottom).  In both cases the data appear to converge to a nontrivial
  curve with increasing $\ell$, which are quite close to the $U_r\to
  \infty$ and $U_r\to - \infty$ limits respectively,
  cf. Eqs.~(\ref{rhoui}) and (\ref{rhouim}), shown by the full lines.
  The dashed line is the $U_r=0$ curve for a free Fermi gas.  Due to
  the reflection symmetry with respect the center of the trap, we show
  only data for $x\ge 0$.  }
\label{rhotssN10}
\end{figure}

\begin{figure}[tbp]
\includegraphics*[scale=\graphicscale]{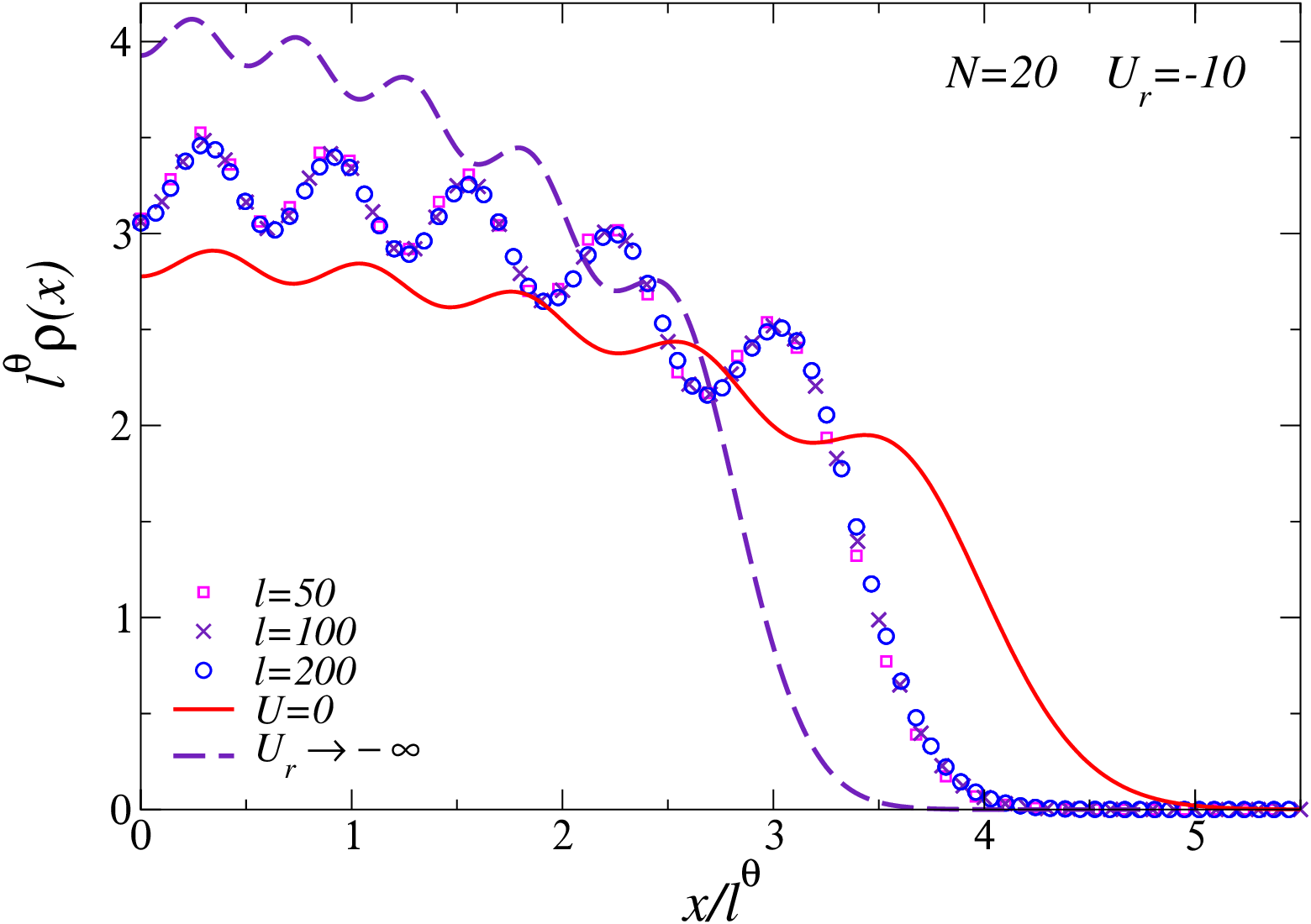}
\includegraphics*[scale=\graphicscale]{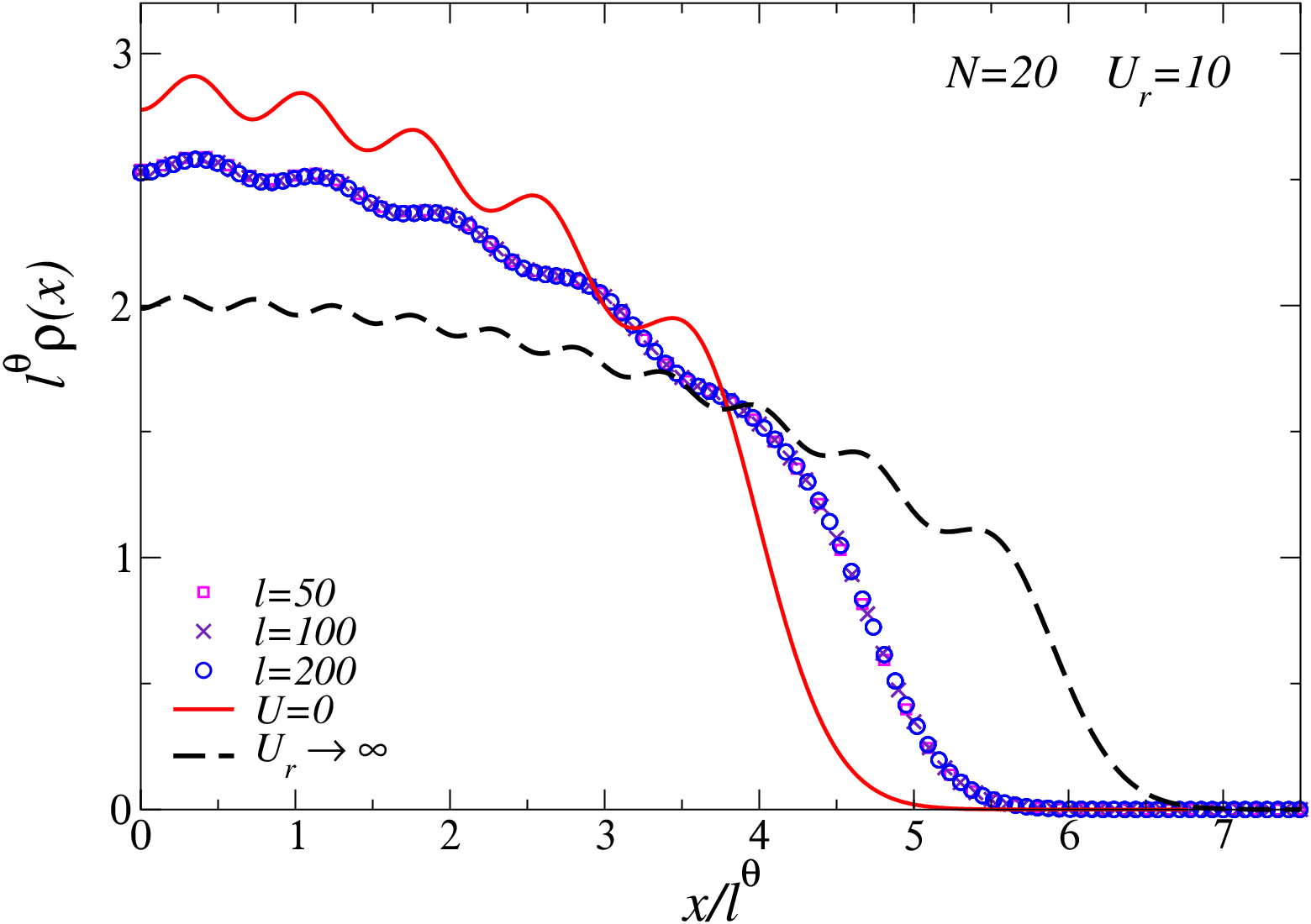}
\caption{(Color online) TSS of the particle density for $N=20$ particles,
  $U_r=-10$ (top) and $U_r=10$ (bottom). The lines show the 
  curves for $U_r=0$, $U_r\to\infty$ and $U_r\to -\infty$.  }
\label{rhotssN20}
\end{figure}

\begin{figure}[tbp]
\includegraphics*[scale=\graphicscale]{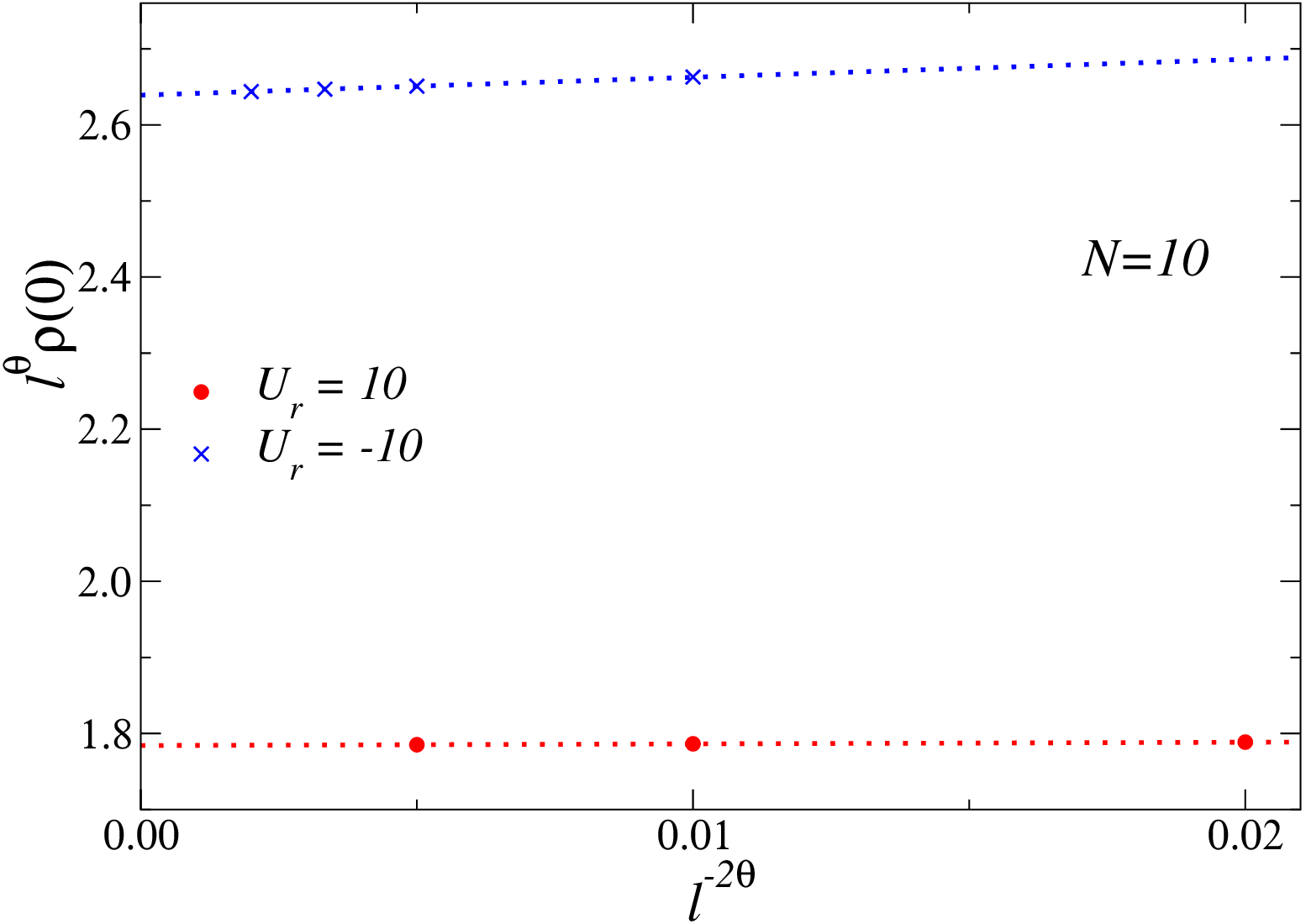}
\caption{(Color online) DMRG data of $\ell^{\theta} \rho(0)$ vs
  $\ell^{-2\theta}$ for $N=10$ and $U_r=-10,\,10$. The dotted lines
  show linear fits, which support the $O(\ell^{-2\theta})$ behavior of
  the corrections analogously to the free $U=0$ case.  }
\label{rhoN10sc}
\end{figure}

\begin{figure}[tbp]
\includegraphics*[scale=\graphicscale]{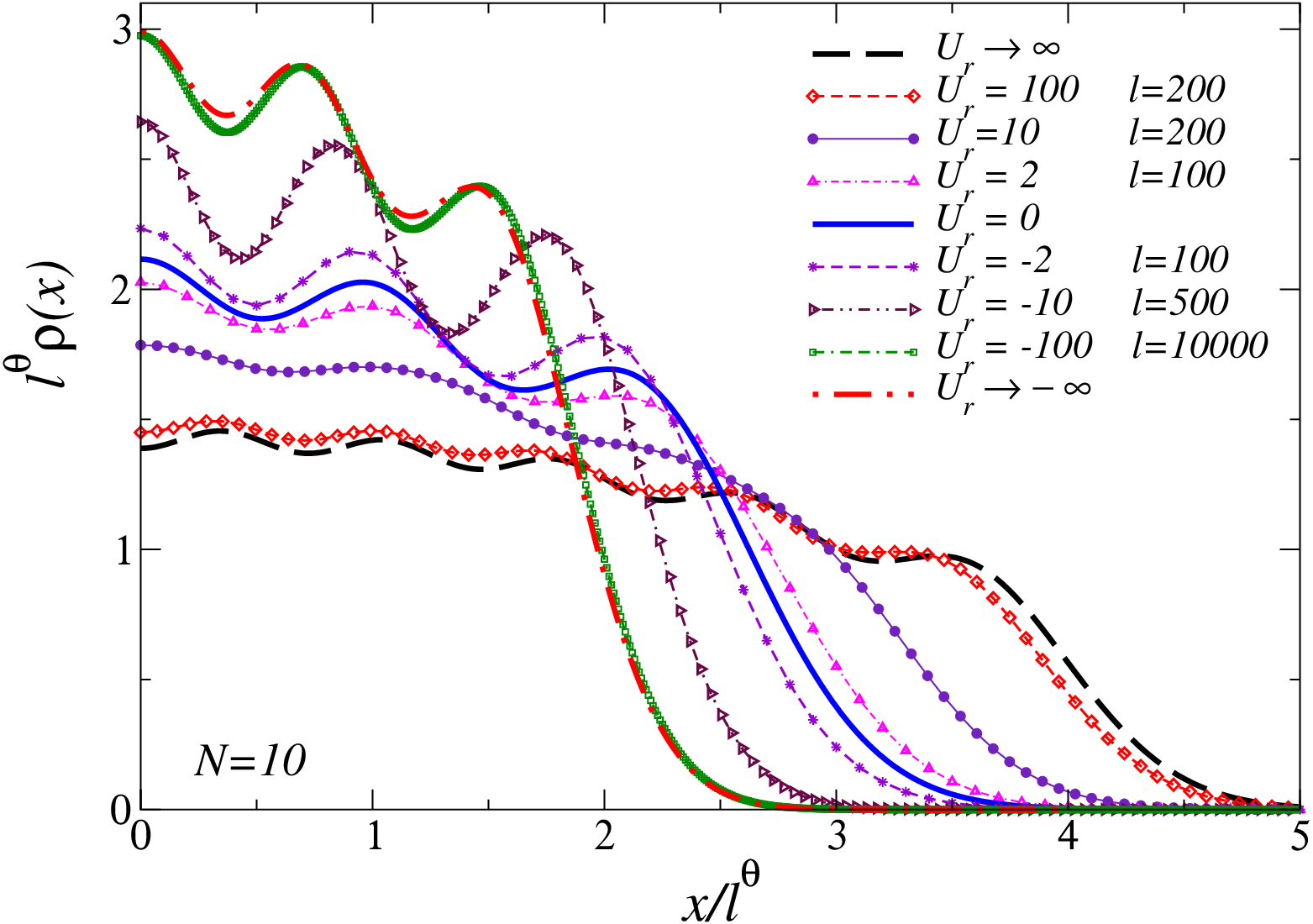}
\caption{(Color online) TSS functions of the particle density of
  $N=10$ particles, for various values of the scaling variable
  $U_r$. We report the exact curves for a free Fermi gas ($U_r=0$) and
  the limits $U_r\to \pm\infty$, cf. Eqs.~(\ref{rhoui}) and
  (\ref{rhouim}).  For the other values of $U_r$ we plot the DMRG
  results for a sufficiently large trap size, providing already the
  asymptotic curves with high precision, as shown in
  Figs.~\ref{rhotssN10} and \ref{rhotssN20} for some values of $U_r$.
}
\label{rhoN10}
\end{figure}

The relevance of the $U$ term in 1D gives rise to nontrivial TSS
limits, requiring an appropriate rescaling of the parameter
$U$. Indeed, we expect the large-$\ell$ scaling behavior
\begin{eqnarray}
\rho(x)  &\approx& \ell^{-\theta}  {\cal R}(X,U_r,N), 
\label{rhoxs}\\ 
d_o(x)  &\approx& \ell^{-2\theta}  {\cal D}(X,U_r,N), 
\label{doxs}\\ 
C(x_1,x_2) &\approx& \ell^{-\theta}  {\cal C}(X_1,X_2,U_r,N) ,
\label{gxs}\\
G(x_1,x_2) &\approx& \ell^{-2\theta} 
 {\cal G}(X_1,X_2,U_r,N),
\label{dexs}\\
M(x_1,x_2) &\approx& \ell^{-2\theta} 
 {\cal M}(X_1,X_2,U_r,N),
 \label{demxs}\\
P(x_1,x_2) &\approx& \ell^{-2\theta} 
 {\cal P}(X_1,X_2,U_r,N),
 \label{demfs}
\end{eqnarray}
where
\begin{eqnarray}
X_i=x_i/\ell^{\theta},  \qquad U_r = U \ell^{\theta}.
\label{XW}
\end{eqnarray}
$\theta$ is the same exponent of Eq.~(\ref{thetaexp}).  These TSS
behaviors are expected to be approached with power-law suppressed
corrections.  Of course, for $U_r=0$, i.e. for a strictly vanishing
$U$, we can derive the scaling functions from the results of
Sec.~\ref{tts}, taking into account that an unpolarized free Fermi
gases of $N$ particles is equivalent to two independent spinless Fermi
gases of $N/2$ particles.

\begin{figure}[tbp]
\includegraphics*[scale=\graphicscale]{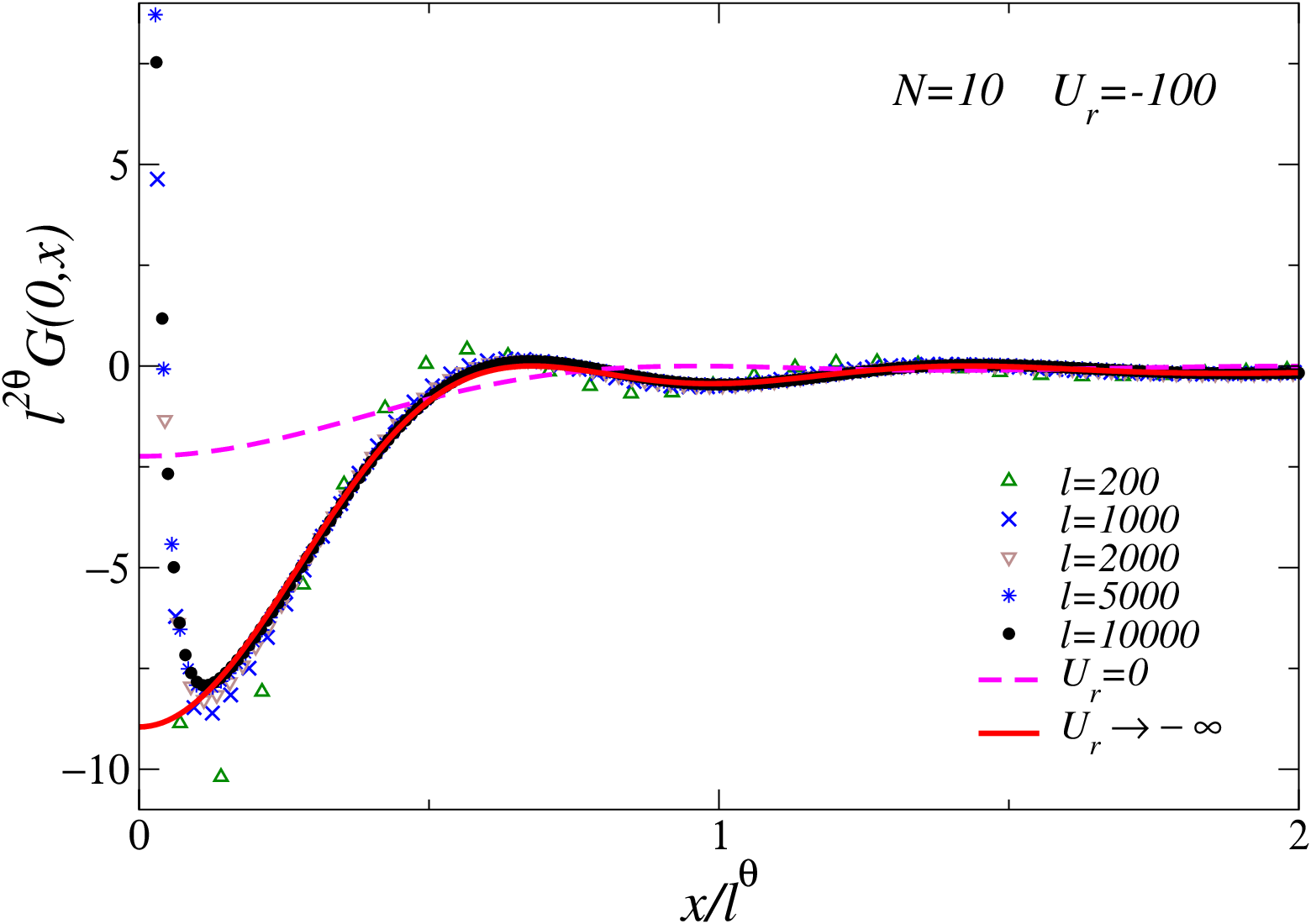}
\includegraphics*[scale=\graphicscale]{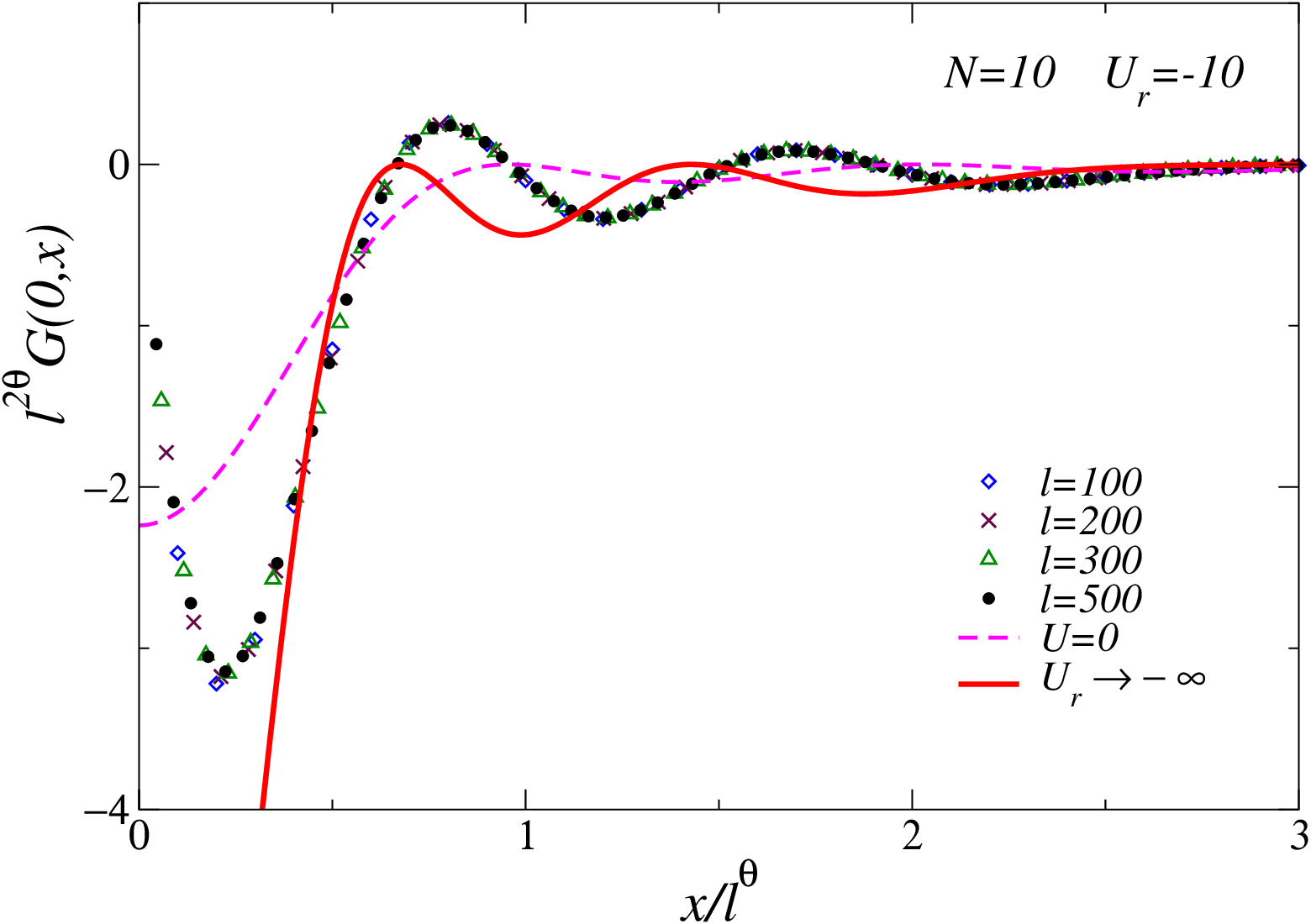}
\includegraphics*[scale=\graphicscale]{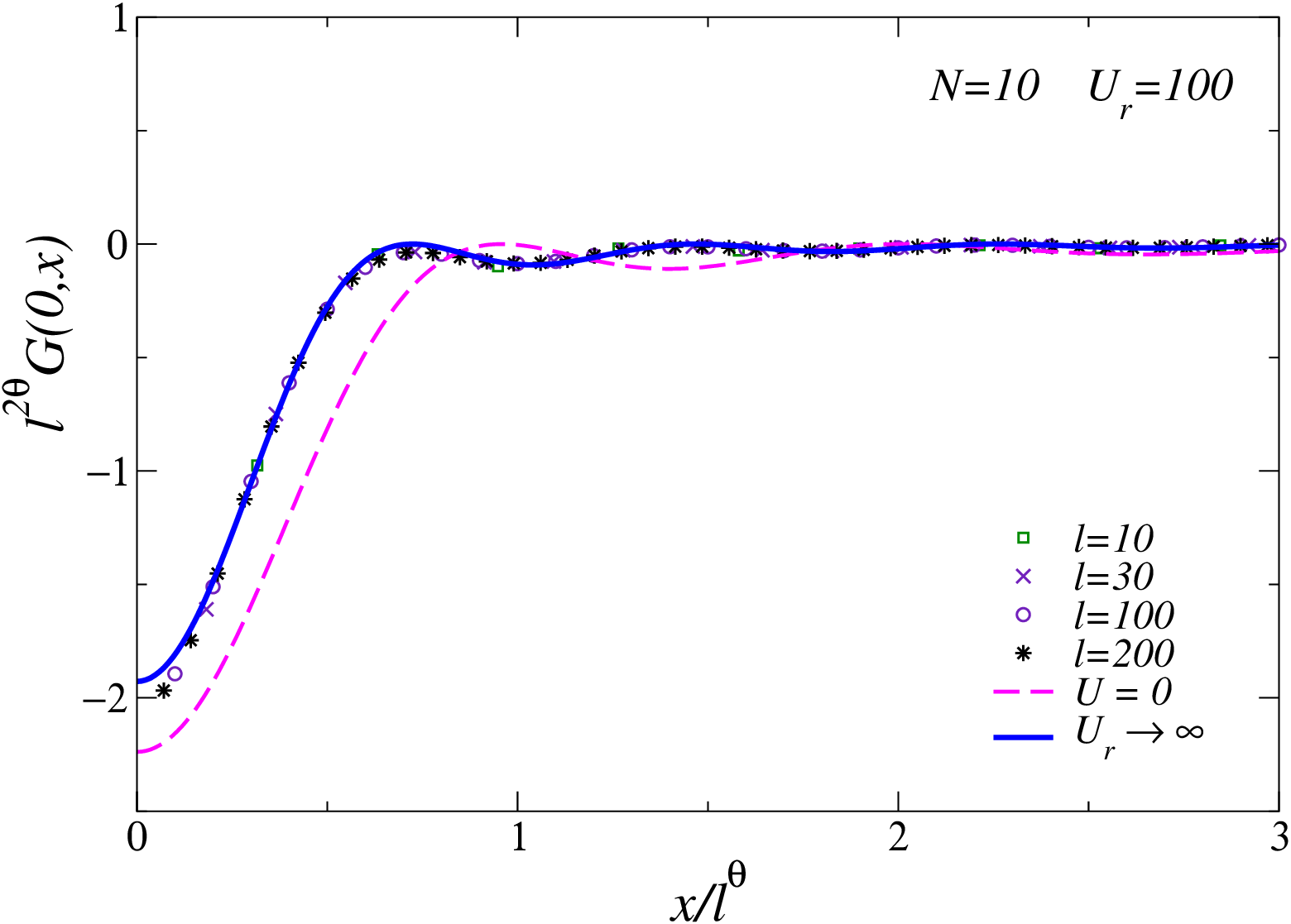}
\caption{(Color online) TSS of the density-density correlation
  $G(0,x)$, i.e. with a point fixed at the trap center, for $N=10$
  particles, and $U_r=-100, \,-10, \,100$.  We plot
  $\ell^{2\theta}G(0,x)$ vs $x/\ell^\theta$.  The lines show the $U=0$
  curve and the $U_r\to\infty$ and $U_r\to -\infty$ limits,
  cf. Eqs.~(\ref{dedeui}) and (\ref{dedeuim}).  }
\label{gn0N10}
\end{figure}

\begin{figure}[tbp]
\includegraphics*[scale=\graphicscale]{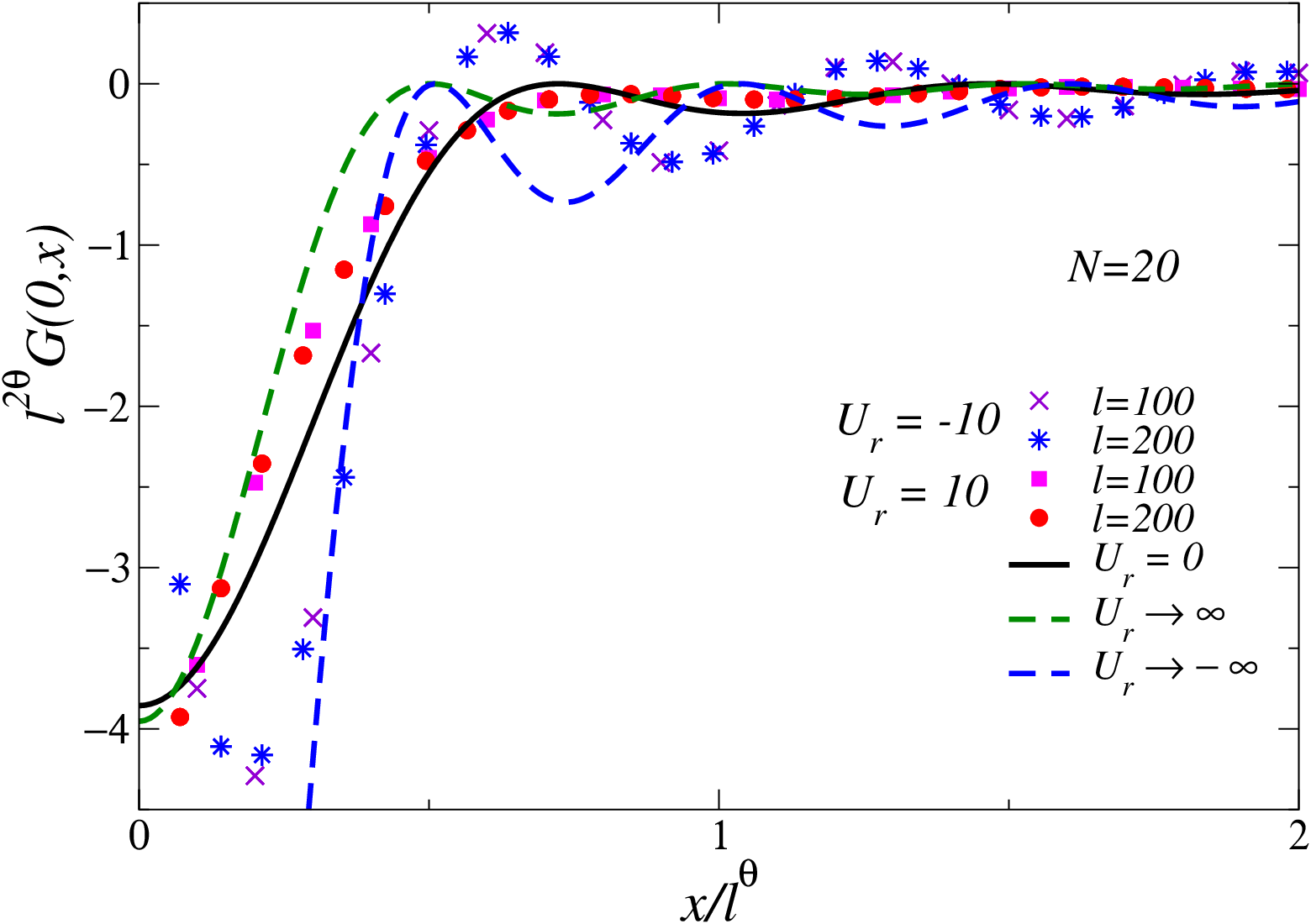}
\caption{(Color online) $\ell^{2\theta} G(0,x)$ versus $x/\ell^\theta$
  for $N=20$ and $U_r=-10,\,10$. The lines show the $U_r=0$ and
  $U_r\to\pm\infty$ limits.  }
\label{gn0N20}
\end{figure}

In order to check our predictions for the TSS behaviors, we present
numerical DMRG~\cite{Schollwock-05} results for the 1D
Hubbard model in the presence of a harmonic trap, for various 
values of $U$, $N$, and $\ell$. DMRG simulations are performed
for a chain of $L$ sites with open boundary conditions. The
size $L$ is chosen sufficiently large to make finite-size effects
negligible; in practice,
we set $L$ large enough to have $\rho < 10^{-15}$ at the edges
of the chain.  The number of states $M$ kept in the truncation is
$M\le1120$, which gives a maximum discarded weight below
$10^{-9}$.~\footnote{In a few computations at large $N$, the
maximum discarded weight is slightly larger than $10^{-9}$; however
the quality of the results is adequate for our needs: we estimate
(by varying $M$) that the truncation error is negligible in all
figures presented.}  
We use wavefunction prediction and exploit fully the conservation of
the total number of particles of each species.  The implicitly
restarted Arnoldi method is used to diagonalize the  Hamiltonian.

We note that, for large negative $U$, the
strong on-site attraction tends to bunch together the particles in the
middle of the trap: this effect reduces the chain length $L$ needed to
host all the particles, even down to $O(N)$ in the limit $U \to
-\infty$.  Moreover, it also reduces the number of states kept in the
DMRG truncation for a given accuracy. This fact allows us to get data
for larger and larger trap size as the attractive interaction
increases.

In Figs.~\ref{rhotssN10} and \ref{rhotssN20} we show results for the
particle density of systems with $N=10$ and $N=20$ particles, for some
positive and negative values of the rescaled on-site interaction
$U_r\equiv U\ell^\theta$.  They clearly confirm the TSS predicted by
Eq.~(\ref{rhoxs}), indeed the data for $\ell^\theta\rho(x)$ plotted
versus $x/\ell^\theta$ appear to approach a limiting curve
representing the TSS function ${\cal R}(X,U_r,N)$ at the given value
of $U_r$ and $N$.  The numerical results show that the asymptotic
behavior is generally approached with $O(\ell^{-2\theta})$ corrections
relative to the leading behavior, as in the case of noninteracting
fermion systems, see Sec.~\ref{tts}.  Some DMRG data for the approach
of the particle density at the origin to its asymptotic behavior are
reported in Fig.~\ref{rhoN10sc}.  The amplitude of these corrections
is significantly larger for attractive interactions, increasing with
increasing $|U_r|$, requiring larger and larger trap sizes to observe
the asymptotic behavior.  For example the ratio between the amplitudes
of the leading $O(\ell^{-1})$ corrections at $U_r=-10$ and $U_r=10$ is
approximately 10.

The asymptotic TSS curves of the particle density clearly depend on
the scaling variable $U_r$.  In Fig.~\ref{rhoN10} we show them for
$N=10$ and several values of $U_r$.  As expected, they extend to
larger regions when we pass from attractive to repulsive interactions.
We also note that the scaling density shows $N/2$ peaks for
$U_r\lesssim 10$, while they become $N$ for large $U_r$, see in
particular the data for $U_r=100$.  As we shall discuss below, when
varying $U_r$ from the strongly attractive to the strongly repulsive
regimes, the system experiences a crossover from a hard-core bosonic
gas of $N/2$ molecules to a free spinless fermion gas of $N$
particles.

\begin{figure}[tbp]
\includegraphics*[scale=\graphicscale]{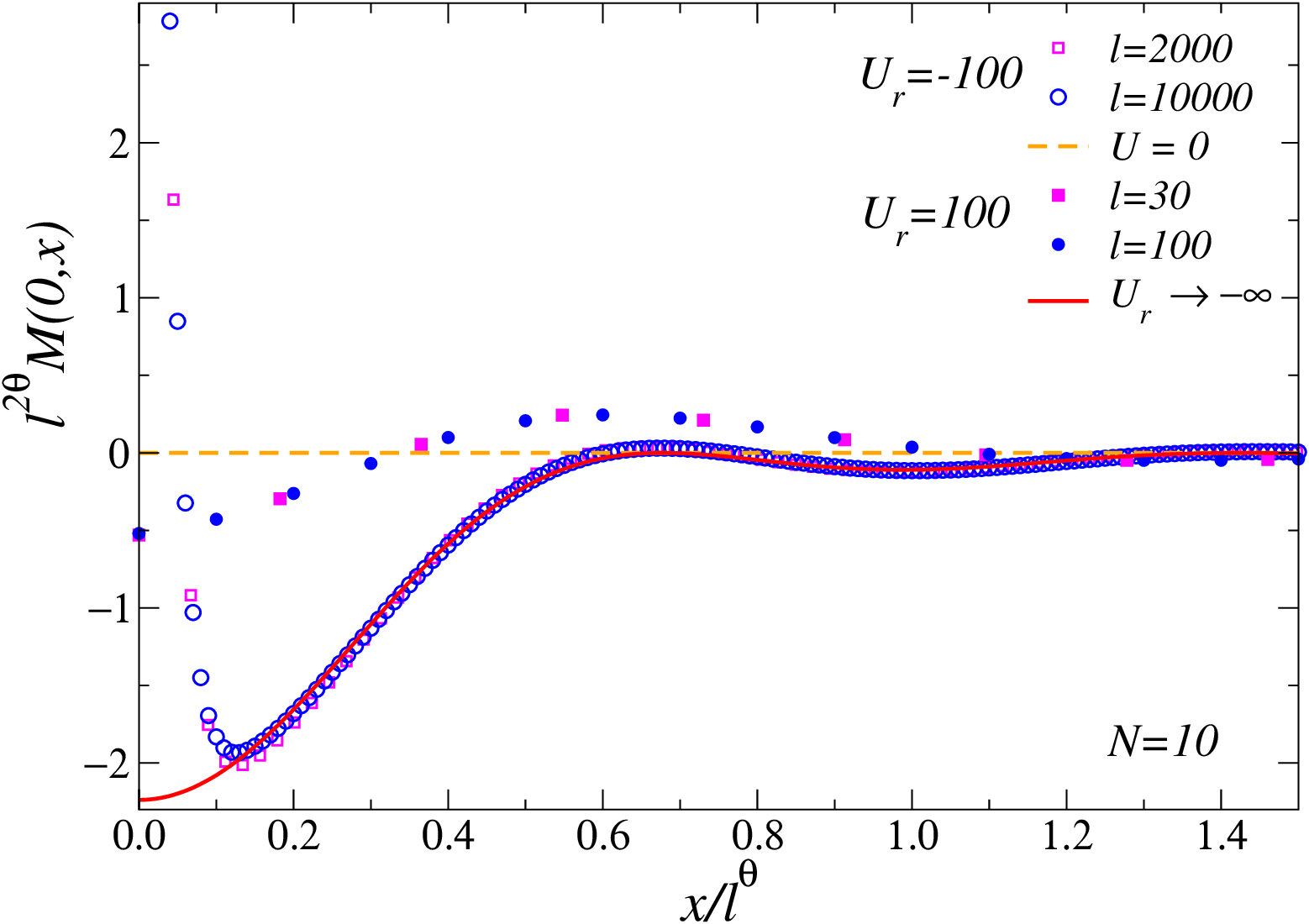}
\caption{(Color online) TSS of the correlation $M(0,x)\equiv \langle
  n_{0\uparrow} n_{x\downarrow}\rangle_c$ for $N=10$ and
  $U_r=-100,\,100$.  We plot $\ell^{2\theta} M(0,x)$ vs
  $x/\ell^\theta$.  The full line represents the $U_r\to -\infty$
  limit given by Eq.~(\ref{mixdeuim}).  }
\label{gnm0N10}
\end{figure}

\begin{figure}[tbp]
\includegraphics*[scale=\graphicscale]{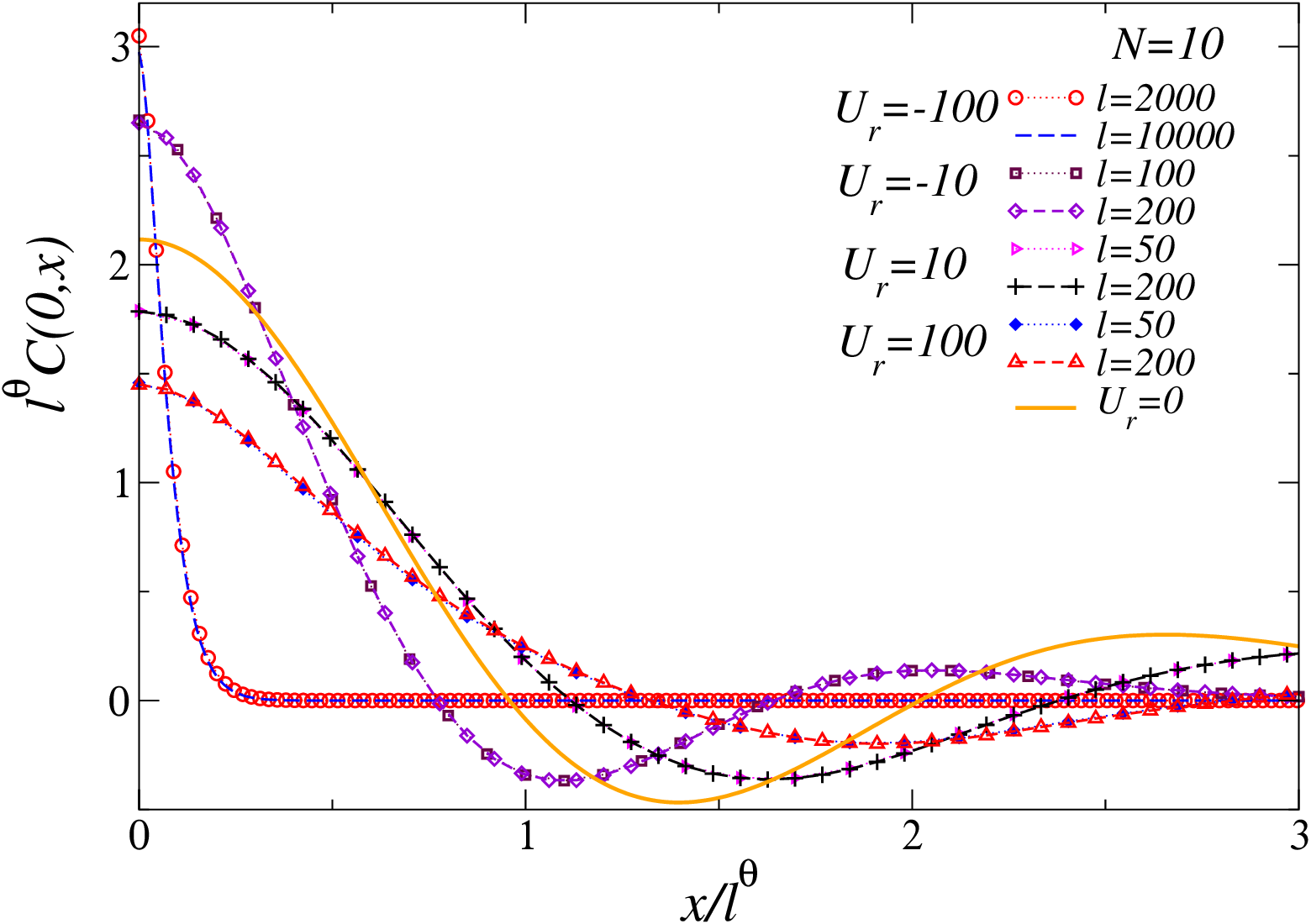}
\caption{(Color online) TSS of the one-particle correlation:
  $l^{\theta} C(0,x)$ versus $x/l^\theta$ for $N=10$ and various
  values of $U_r$. }
\label{c0N10}
\end{figure}

\begin{figure}[tbp]
\includegraphics*[scale=\graphicscale]{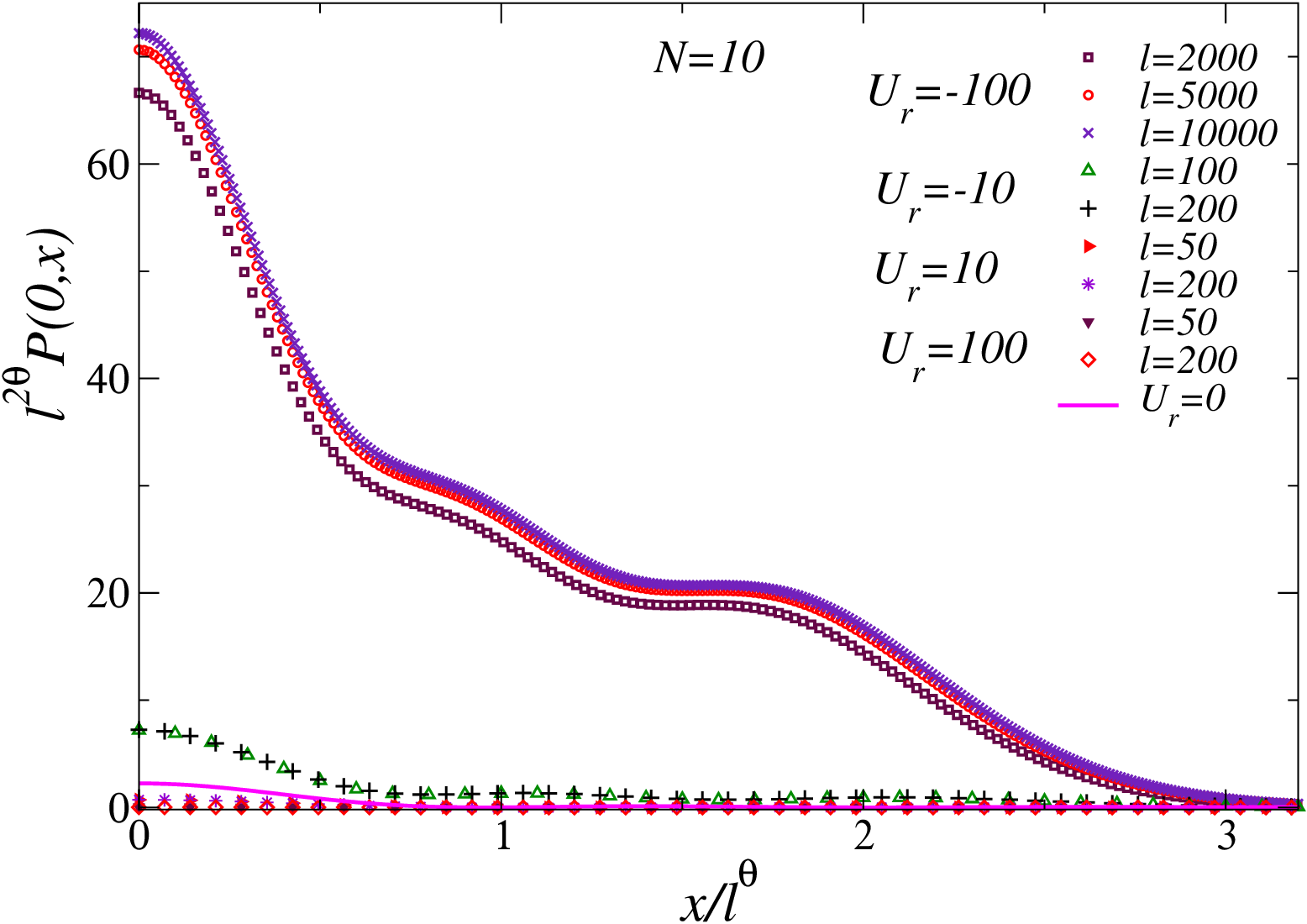}
\caption{(Color online) TSS of the pair correlation: $l^{2\theta}
  P(0,x)$ versus $x/l^\theta$ for $N=10$ and various
values of $U_r$. }
\label{f0N10}
\end{figure}

\begin{figure}[tbp]
\includegraphics*[scale=\graphicscale]{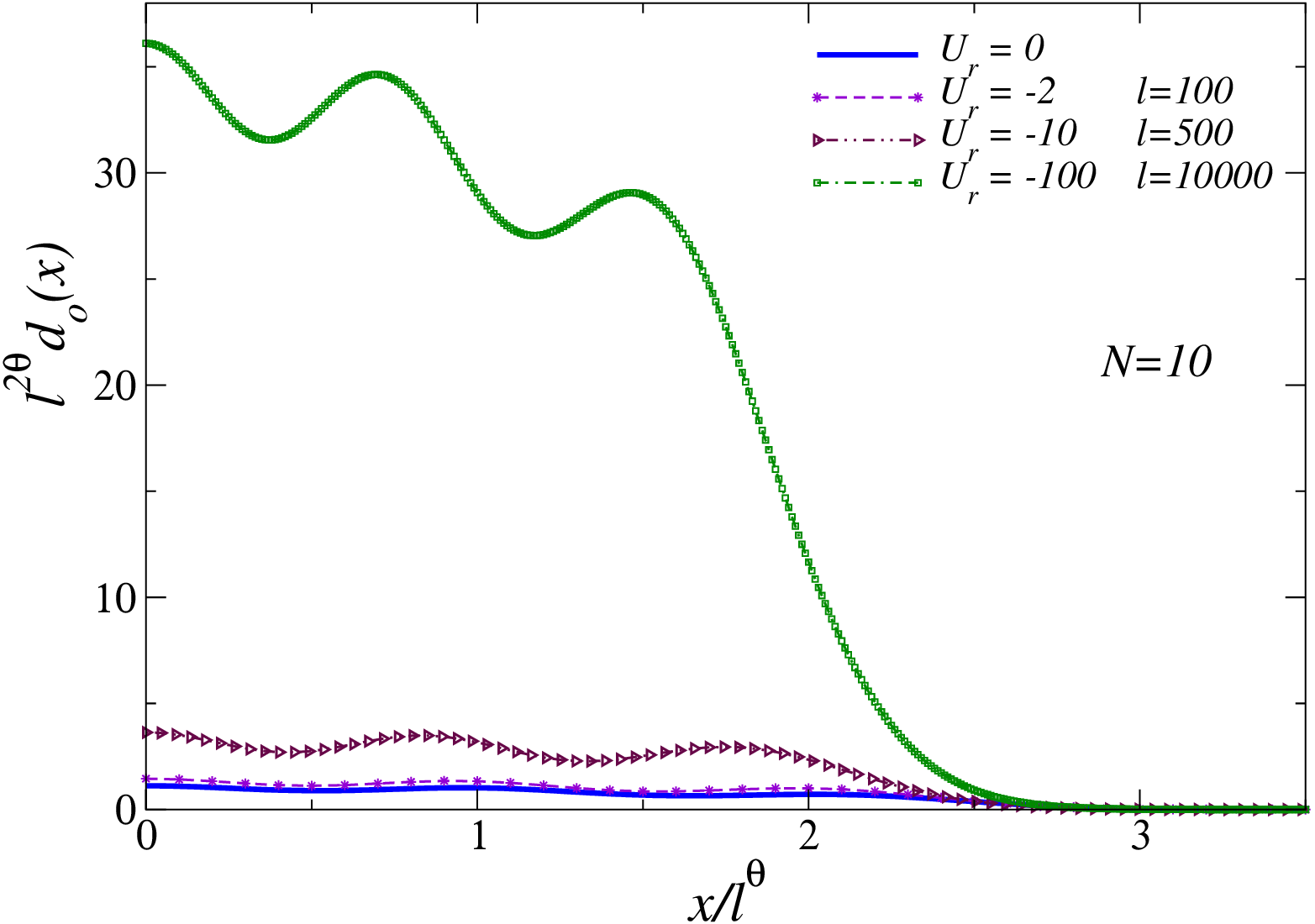}
\caption{(Color online) TSS functions of the double occupancy for
  $N=10$ and some values of $U_r$, obtained plotting $l^{2\theta}
  d_o(x)$ versus $x/l^\theta$.  Note that the double occupancy
  vanishes in the limit of strong repulsive interactions. }
\label{docN10}
\end{figure}

We also present results for the correlation functions introduced in
Sec.~\ref{model}, in particular we consider correlation functions with
a point fixed at the trap center $x=0$.  Figs.~\ref{gn0N10} and
\ref{gn0N20} show the scaling behavior of the connected density
correlation function $G(0,x)$, for $N=10$ and $N=20$ respectively, and
various values of $U_r$.  The data nicely support the TSS Ansatz
(\ref{dexs}).  In Fig.~\ref{gnm0N10} we show results for the connected
correlation $M(0,x)$ between up and down density.  TSS is also
confirmed by the data of the one-particle correlation, see
Fig.~\ref{c0N10}, and the pair correlation in Fig.~\ref{f0N10}. In
particular, the quantities which are more sensitive to the correlation
between up and down fermions, such as the up-down density correlation
$M(x,y)$ and the pair correlation $P(x,y)$, tend to become more and
more significant with decreasing $U_r$, as expected because up and
down fermions become more and more tightly correlated for large
negative values of the on-site interaction. This is also shown by the
behavior of the double occupancy shown in Fig.~\ref{docN10}.

An important issue concerns the universality of the TSS reported in
Eqs.~(\ref{rhoxs}-\ref{demfs}). They are expected to be universal
apart from a global multiplicative normalization, and normalizations
of the arguments of TSS functions. More precisely, we expect that they
are universal with respect to a large class of further short-ranged
interaction terms,  such as 
\begin{equation}
H_{nn} = \sum_{\sigma,\sigma'} w_{\sigma\sigma'} 
\sum_{\langle{\bf x}{\bf y}\rangle} n_{\sigma {\bf x}} n_{\sigma'{\bf y}}.
\label{hddhm}
\end{equation}
Indeed, $H_{nn}$ may only give rise to a change of the effective
quartic coupling $U$ (when adding $H_{nn}$ to the Hubbard Hamiltonian,
the effective relevant quartic coupling becomes
$U+2w_{\uparrow\downarrow}$), and to further $O(l^{-\theta})$
corrections, due to the fact that they introduce other irrelevant RG
perturbations of RG dimension $y_w=-d$ at the dilute fixed point.

\subsubsection{Crossover behavior as a function of the on-site interaction}
\label{ttdgt1dcross}

It is important to note that the TSS limit corresponds to a continuum
limit in the presence of the trap, i.e. it generally realizes a
continuum quantum field theory in the presence of an inhomogeneous
external field.  In the case of the Hubbard model in the dilute
regime, this continuum limit is given by the quantum field theory
(\ref{zf}), replacing the constant $\mu$ with a space-dependent
potential $\mu-V({\bf x})$.  This implies that the TSS of the
observables of the 1D trapped Hubbard model must approach the
solutions of the continuous problem of fermions with contact
interactions, which is also the so-called Gaudin-Yang (GY)
model~\cite{Gaudin-67,Yang-67}, with equal number of up and down
particles, $N_\uparrow=N_\downarrow=N/2$.  The GY Hamiltonian of a
trapped fermion gas can be written as
\begin{eqnarray}
H_{\rm GY} = \sum_{i=1}^N \left[ {p_i^2\over 2 m} + V(x_i)  \right]
+ g \sum_{i\ne j} \delta(x_i-x_j).
\label{GYmodels}
\end{eqnarray}
We expect that the TSS limit of the 1D Hubbard model at fixed $N$ is
related to the GY model with $g \sim U_r$.  More precisely, the TSS
functions entering formulas (\ref{rhoxs}-\ref{demfs}) are exactly
given by corresponding quantities of the GY problem with a trap of
unit size.

The equation of state of the homogenous GY model is exacty known for
both repulsive and attractive zero-range
interaction~\cite{Gaudin-67,Yang-67}.  It is characterized by
different asymptotic regimes with respect to the effective
dimensionless coupling $\gamma \equiv g/\rho$, where $\rho$ is the
particle density. At weak coupling $\gamma \ll 1 $ it behaves as a
perfect Fermi gas; in the strongly repulsive regime, $\gamma\gg 1$ the
equation of state approaches that of spinless Fermi gas; in the
strongly attractive regime $\gamma\to -\infty$ and for unpolarized
gases it matches that of a 1D gas of impenetrable
bosons~\cite{Girardeau-60}, more precisely hard-core bosonic molecules
of fermion pairs~\cite{ABGP-04,FRZ-04}.

The relation between the TSS of the trapped Hubbard model and the
continuum GY model can be exploited to determine the the TSS functions
of the particle density and its correlation, i.e.  ${\cal R}(X,U_r,N)$
and ${\cal G}(X_1,X_2,U_r,N)$ respectively, in the strongly repulsive
and attractive limits, i.e. $U_r\to\infty$ and $U_r\to -\infty$.

We know that in the $g\to\infty$ limit the particle density and its
correlations of the GY model become identical to those of a gas of $N$
spinless fermions~\cite{Schulz-90,GCWM-09,GBL-13}. This would imply
that the $U_r\to \infty$ limit of the TSS functions is
\begin{eqnarray}
 &&{\cal R}(X,U_r\to\infty,N) = S_{p}(X,N), \label{rhoui}\\
 &&{\cal G}(X_1,X_2,U_r\to\infty,N) = Y_{p}(X_1,X_2,N), \label{dedeui}
\end{eqnarray}
where $S_p$ and $Y_p$ are the same functions entering the spinless
free-fermion TSS, cf. Eqs.~(\ref{spf}) and (\ref{ypn}).  Moreover, the
TSS functions of the double occupancy and the pair correlation, i.e.
${\cal D}(X,U_r,N)$ and ${\cal P}(X_1,X_2,U_r,N)$, defined in
Eq.~(\ref{doxs}) and (\ref{demfs}) respectively, trivially vanishes in
the $U_r\to\infty$ limit.

In the $g\to -\infty$ limit the density properties of the GY model is
expected to match that of an ensemble of hard-core $N/2$ bosonic
molecules constituted by up and down fermions.  Indeed, with
increasing attraction, the pairing becomes increasingly localized in
space, and eventually the paired fermions form a tightly bound bosonic
molecule.  Actually, the results of Ref.~\cite{ABGP-04} for harmonic
traps, obtained by LDA, show that these bound states get trapped in a
smaller region, with an effective trap size $\ell_b=\ell/2$ in the
strongly attractive limit.  Thus, we expect that in the $g\to -\infty$
limit the particle density of the unpolarized GY model
(\ref{GYmodels}) with a harmonic trap matches that of $N/2$ hard-core
doubly-charged bosons with an effective trap size $\ell_b= \ell/2$,
which in turn can be mapped into a free gas of $N/2$ spinless
doubly-charged fermions in a harmonic trap of size $\ell_b$.  On the
basis of these arguments and using the results of Sec.~\ref{tts}, we
conjecture the following $U_r\to -\infty$ limit of the TSS functions
for harmonic traps:
\begin{eqnarray}
 &&{\cal R}(X,U_r\to -\infty,N) = 2^{3/2} S_{2}(\sqrt{2} X,N/2), 
\label{rhouim}\\
 &&{\cal G}(X_1,X_2,U_r\to -\infty,N) = 8 Y_{2}(\sqrt{2} X_1,\sqrt{2}X_2,N/2).
\qquad
  \label{dedeuim}
\end{eqnarray}
Moreover, since fermion pairs are tightly bounded in the strongly
attractive limit, we also predict
\begin{eqnarray}
{\cal M}(X_1,X_2,U_r\to -\infty,N) = 2 Y_{2}(\sqrt{2} X_1,\sqrt{2}X_2,N/2)
.\;\;  \label{mixdeuim}
\end{eqnarray}
Using analogous arguments we may also expect that in this $U_r\to
-\infty$ limit the TSS function of the double occupancy becomes
proportional to ${\cal R}(X,U_r\to -\infty,N)$. This is supported
by the DMRG data by comparing the data of the double occupancy and the
particle density for $N=10$ and $U_r=-100$, shown in Fig.~\ref{docN10}
and the top Fig.~\ref{rhotssN10} respectively. Moreover, in the same
limit the TSS function of the pair correlation should get proportional
to the one-particle correlation (density matrix) of a gas of $N/2$
hard-core bosons.  This can be checked by comparing the $U_r=-100$
data of Fig.~\ref{f0N10} with the results of Ref.~\cite{CV-10-BH} for
the one-particle correlation of hard-core Bose gases.

The curves corresponding to the $U_r\to\pm\infty$ limits are shown in
Figs.~\ref{rhotssN10}, \ref{rhotssN20}, \ref{rhoN10}, \ref{gn0N10},
and \ref{gn0N20} for both the particle density and its connected
correlations.  They are clearly approached by the data for large
$|U_r|$, see in particular the results for $N=10$ in
Figs.~\ref{rhotssN10} and \ref{gn0N10}.  Note that the limit $U_r\to
-\infty$ of the connected density-density correlations $G$ and $M$ is
apparently approached nonuniformly at small distance.  Indeed the data
for $U_r=-100$ shown in the top Fig.~\ref{gn0N10} appear to follow the
asymptotic curve (\ref{dedeuim}) for $x/\ell^{\theta}\gtrsim 0.1$,
while at smaller distances they show a sudden departure where the
correlation increases significantly. The comparison with the data for
$U_r=-10$, see the middle Fig.~\ref{gn0N10}, suggests that this occurs
at smaller and smaller distances with increasing $|U_r|$, being likely
related with the size of the molecules formed by the fermion pairs.

These data show that the dilute TSS of quantities related to the
particle density of the trapped Hubbard model experience a smooth
crossover from an effective hard-core bosonic gas of $N/2$ molecules
($U_r\to -\infty$) to an effective free gas of $N$ spinless fermions
($U_r\to\infty$), passing through two noninteracting fermion gases of
$N/2$ particles.  In these 1D systems the formation of a gas of
molecules of pair fermions has some analogies to the formation of a 3D
molecule BEC and the BCS-BEC crossover for 3D Fermi systems, which has
been recently observed in experiments with ultracold Fermi gases, see
e.g. Refs.~\cite{Esslinger-10,RGJ-04,Zwer-etal-04,Kin-etal-04,
  Chin-etal-04,Bou-etal-04,Par-etal-05,HDL-07}.

Let us note that the $U_r\to\infty$ limit of the TSS cannot be
obtained by first taking the $U\to\infty$ limit of the Hubbard model
and then the large trap-size limit. Indeed, on the one hand, the
$U_r\to\infty$ limit of the TSS reproduces the $g\to\infty$ limit of
the continuous GY model, essentially given by a gas of free spinless
fermions.  On the other hand, the ground state of the Hubbard model in
the $U\to\infty$ limit should be obtained by filling the central $N$
sites around the center of trap up to $|x|\lesssim N/2$ (actually we
expect some degeneration), without any particular scaling with respect
to the trap size.

Analogous considerations apply to the limit $U_r\to -\infty$, which
corresponds to a gas of $N/2$ bosonic molecules with hard-core
interactions.  Indeed, the ground state of the trapped Hubbard model
in the $U\to -\infty$ limit of the Hubbard model (at any finite trap
size $\ell$) is just obtained by completely filling the central $N/2$
sites around the center of trap (up to $|x|\le (N/2-1)/2$ if $N/2$ is
an odd number, in the case of even $N/2$ there are two degenerate
ground states filled for $-N/4\le x < N/4$ and $-N/4 < x \le N/4$),
without any particular scaling property with respect to the trap size.

\subsection{Large-$N$ limit of the dilute TSS functions}
\label{largen}

\begin{figure}[tbp]
\includegraphics*[scale=\graphicscale]{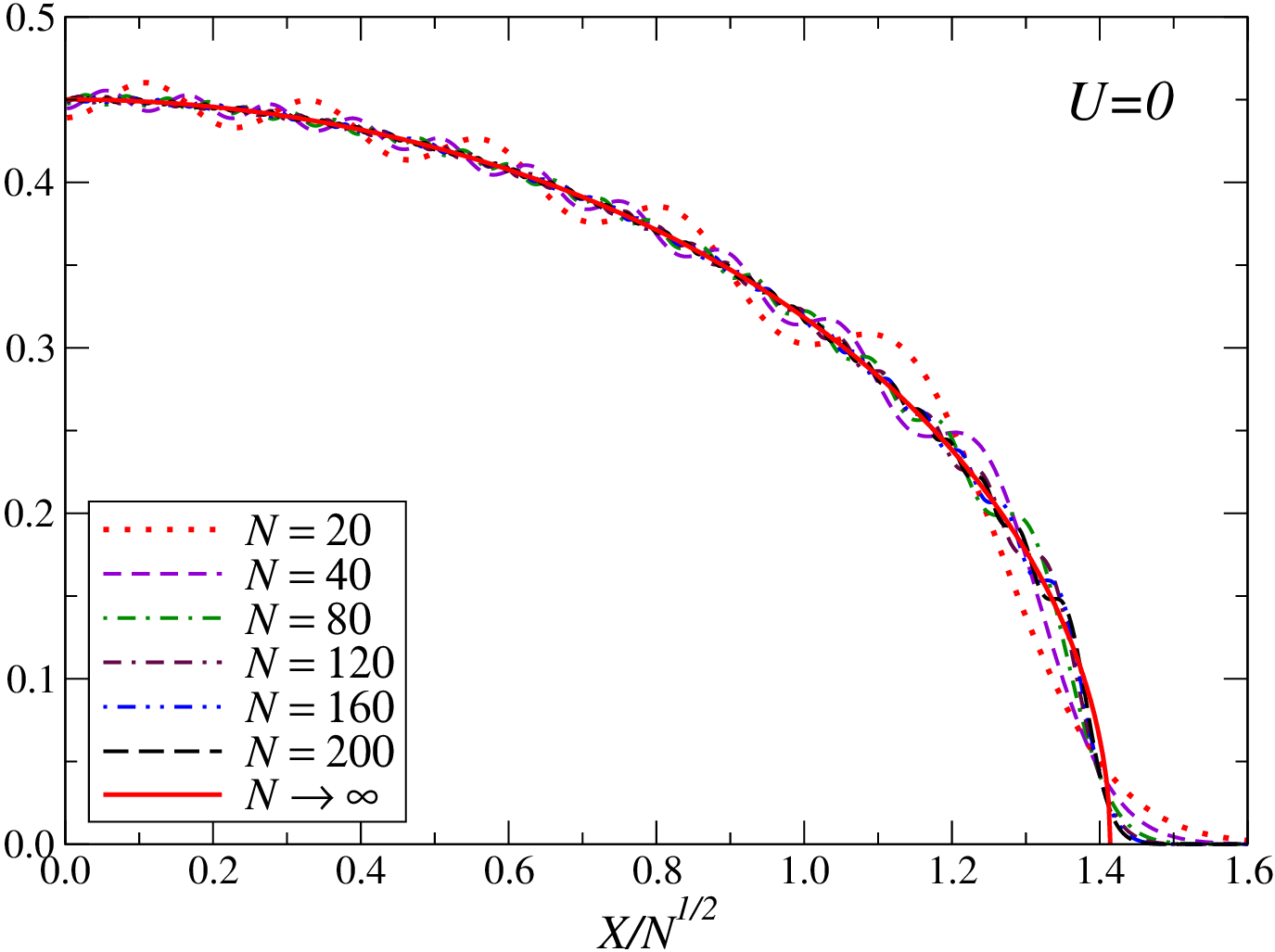}
\caption{(Color online) Large-$N$ scaling of the TSS function of the
  particle density of free Fermi gases: $N^{-1/2}{\cal R}(X,U_r=0,N)$
  versus $X/N^{1/2}$ for several values of $N$.  The finite-$N$
  curves, obtained using Eq.~(\ref{ss2}), clearly approach the
  large-$N$ limit ${\cal R}_\infty(z)$, cf. Eq.~(\ref{calru0}), with
  oscillations that get suppressed as $1/N$, in agreement with
  Eq.~(\ref{calru0}).  }
\label{rholnu0}
\end{figure}

\begin{figure}[tbp]
\includegraphics*[scale=\graphicscale]{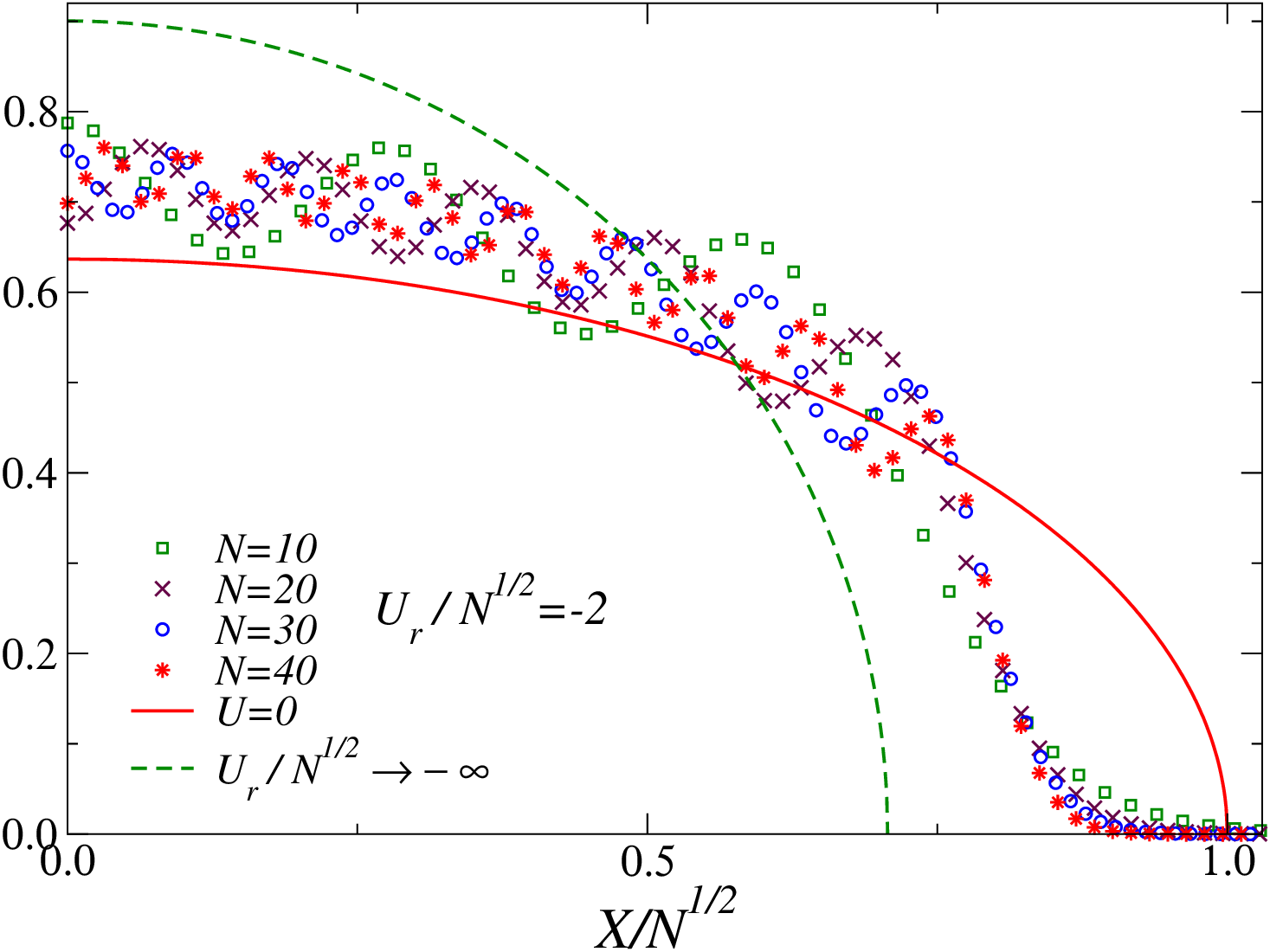}
\includegraphics*[scale=\graphicscale]{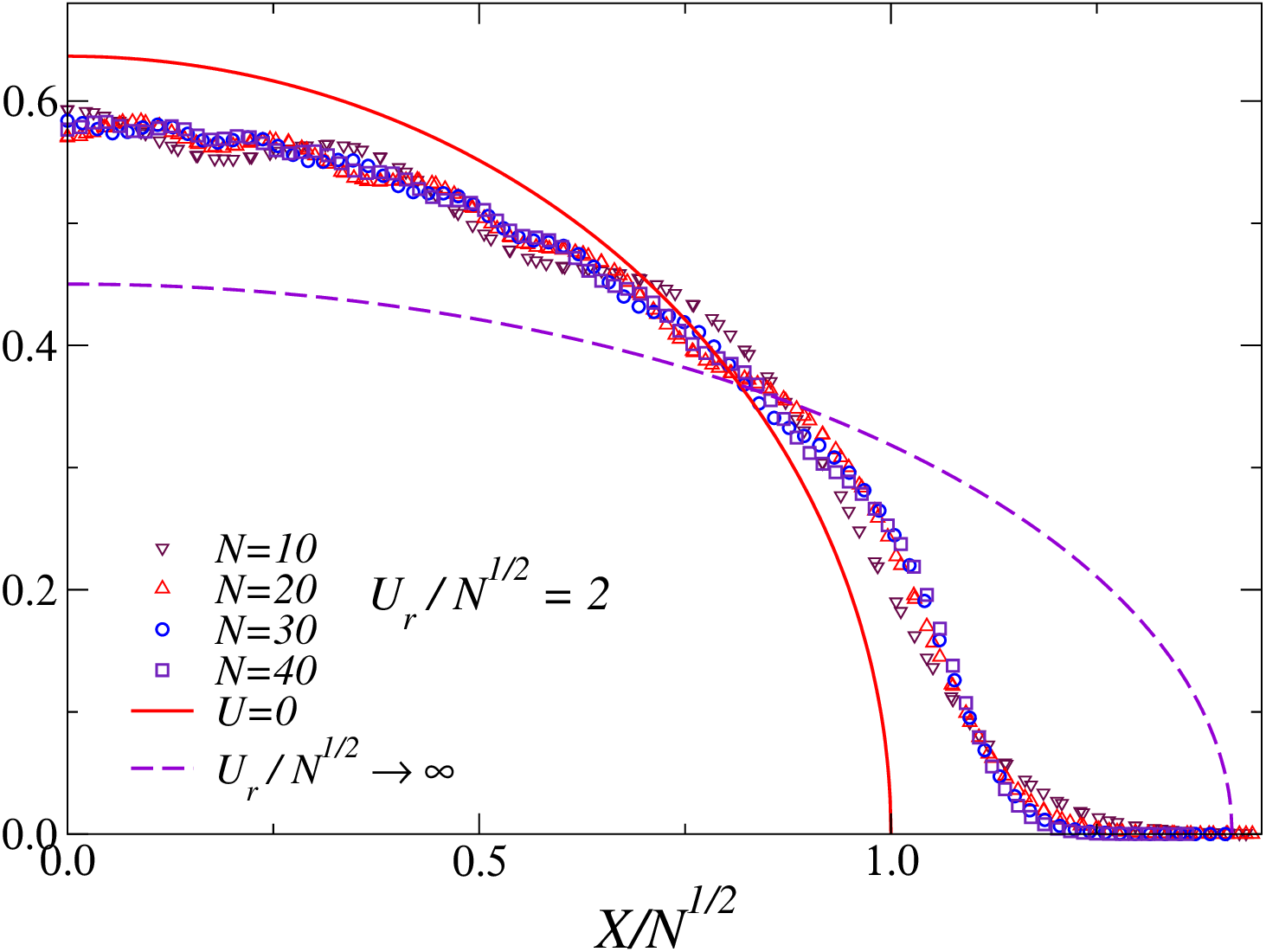}
\caption{(Color online) Large-$N$ scaling of the TSS function of the
  particle density: $N^{-1/2}{\cal R}(X,U_r,N)$ versus $X/N^{1/2}$ for
  $U_r/N^{1/2}=-2$ (top) and $U_r/N^{1/2}=2$ (bottom).  The TSS
  functions are derived from simulations with $\ell=100,\,200$, which
  are expected to be sufficiently large to effectively reproduce the
  $\ell\to\infty$ limit.  For comparison, we also show the asymptotic
  curve for $U=0$, cf. Eq.~(\ref{calru0}), and the expected asymptotic
  curves for $U_r/N^{1/2}\to \infty$ and $U_r/N^{1/2}\to -\infty$,
  given by Eq.~(\ref{calrui}) and (\ref{calruim}) respectively.  }
\label{rholn}
\end{figure}

We now discuss the large-$N$ scaling behavior of the TSS function
appearing on the r.h.s. of Eqs.~(\ref{rhoxs}-\ref{demfs}).  This also
corresponds to studying the large-$N$ behavior of the GY model
(\ref{GYmodels}), which is reproduced by the asymptotic TSS functions
of the Hubbard model in the large-$\ell$ limit.

On the basis of the results obtained at fixed $N$ and $U_r$, we extend
the Ansatz for the large-$N$ scaling of noninteracting Fermi gases,
see the end of Sec.~\ref{tts}, to the 1D Hubbard model, by allowing
for a large-$N$ rescaling of the on-site interaction.  We argue that
the TSS function of the particle density, cf. Eq.~(\ref{rhoxs}),
behaves asymptotically as
\begin{eqnarray}
{\cal R}(X,U_r,N) \approx N^{1/2} {\cal R}_\infty(X/N^{1/2},U_r/N^{1/2}) 
\label{lnrho1}
\end{eqnarray}
where ${\cal R}_\infty(z,u)$ is a nontrivial scaling function, and
power-law suppressed corrections are neglected. The large-$N$
rescaling of the on-site coupling $U_r$ may be inferred by noting that
in the continuum model (\ref{GYmodels}) the strength $g$ of the
interaction should be compared with the density of the gas around the
center of the trap, which is expected to increase as $N^{1/2}$ with
increasing $N$ (keeping $\ell$ fixed). This suggests that the
effective coupling strength should be $g/N^{1/2}\sim U_r/N^{1/2}$.  We
mention that this rescaling of $g$ was already considered in
Ref.~\cite{SBE-11} for the continuum GY model.  This large-$N$
rescaling may be also derived by requiring the consistency of this
scaling behavior with the dilute limit $N/\ell\to 0$ of the particle
density as obtained by the LDA, see Sec.~\ref{thlim}, which requires
$\rho=f_\rho(U,N/\ell,x/\ell)$ (taking into account that
asymptotically $N/\ell$ becomes a function of the chemical potential
$\mu$).

For $U_r=0$ we must recover the known results for spinless Fermi
gases,~\cite{KB-02,CV-10-BH} taking into account that it corresponds
to the large-$N$ limit of two identical Fermi gases of $N/2$
particles. Thus, we must have
\begin{eqnarray}
&&{\cal R}(X,0,N) = N^{1/2} \Bigl\{ 
{2\over \pi} \sqrt{1 - z^2} - {(-1)^{N/2}\over N}\times
\nonumber\\
&&\;\times
{{\rm cos}\bigl[N(z \sqrt{1-z^2} + {\rm arcsin}z)\bigr]\over 
2\pi  (1-z^2)}
+ O\left(N^{-2}\right)\Bigl\}
\label{calru0}
\end{eqnarray}
where $z=X/N^{1/2}$.  In Fig.~\ref{rholnu0} we report some results for
the noninteracting $U=0$ case, which show the above asymptotic
behaviors with increasing $N$, including the oscillations arising from
the $O(1/N)$ term. Note that the $O(1/N)$ corrections are singular at
$|z|=z_b=1$ where the leading term vanishes. As explained at the end
of Sec.~\ref{tts}, around $z_b$ a different large-$N$ scaling sets in,
so that at $z=z_b$ the particle density behaves as
\begin{equation}
\rho(\pm z_b)=O(N^{1/6}), 
\label{rhozbbe}
\end{equation}
instead of the $O(N^{1/2})$ scaling for $|z|<z_b$.  The difference
with the value of $z_b$ in Eq.~(\ref{sfgrho3}) is just due to the fact
that here fermions have two spin components, thus we have $N/2$
particles for each component.

The general interacting case shows similar behaviors.
Fig.~\ref{rholn} shows results for the asymptotic large-$\ell$ scaling
function ${\cal R}(X,U_r,N)$ versus $X/N^{1/2}$ at fixed
$U_r/N^{1/2}=-2,\,2$. They are obtained for a large trap size $\ell$,
sufficiently large to reproduce the asymptotic TSS at fixed $N$. These
results nicely support the expected large-$N$ scaling behavior
(\ref{lnrho1}).  Again we observe the phenomenon of the restriction of
the fermion cloud in the attractive regime.  Moreover, we note the
oscillations which should eventually get suppressed in the large-$N$
limit, likely as $O(1/N)$ around the center of the trap analogously to
the free case.  The comparison of such oscillations for different
values of $U_r/N^{1/2}$ shows that their amplitudes are larger for
attractive interactions.

We may also derive asymptotic behaviors in the limits $u\equiv
N^{-1/2}U_r \to\pm\infty$, corresponding to the pictures derived in
the limits $U_r\to\pm\infty$ in Sec.~\ref{ttdgt2}. The $u\to\infty$
limit corresponds to a free spinless fermi gas of $N$ particles,
thus
\begin{eqnarray}
&&{\cal R}_\infty(z,u\to\infty) = 
{2^{1/2}\over \pi}\sqrt{1 - {z^2\over 2}}.\label{calrui}
\end{eqnarray}
On the other hand, the $u\to -\infty$ is expected to reproduce the scaling
density of a hard-core bosonic gas in an effective half trap, leading
to
\begin{eqnarray}
{\cal R}_\infty(z,u\to -\infty) = 
{2^{3/2} \over \pi}\sqrt{1 - 2 z^2}.\label{calruim}
\end{eqnarray}
These curves are reported in Fig.~\ref{rholn}. They provide the
extreme behaviors when going from $U_r/N^{1/2}\to\infty$ to
$U_r/N^{1/2}\to -\infty$.  In both $U\to\pm\infty$ limits the
corrections are again expected to be $1/N$, since they the
corresponding particle densities map into those of free Fermi systems.

\begin{figure}[tbp]
\includegraphics*[scale=\graphicscale]{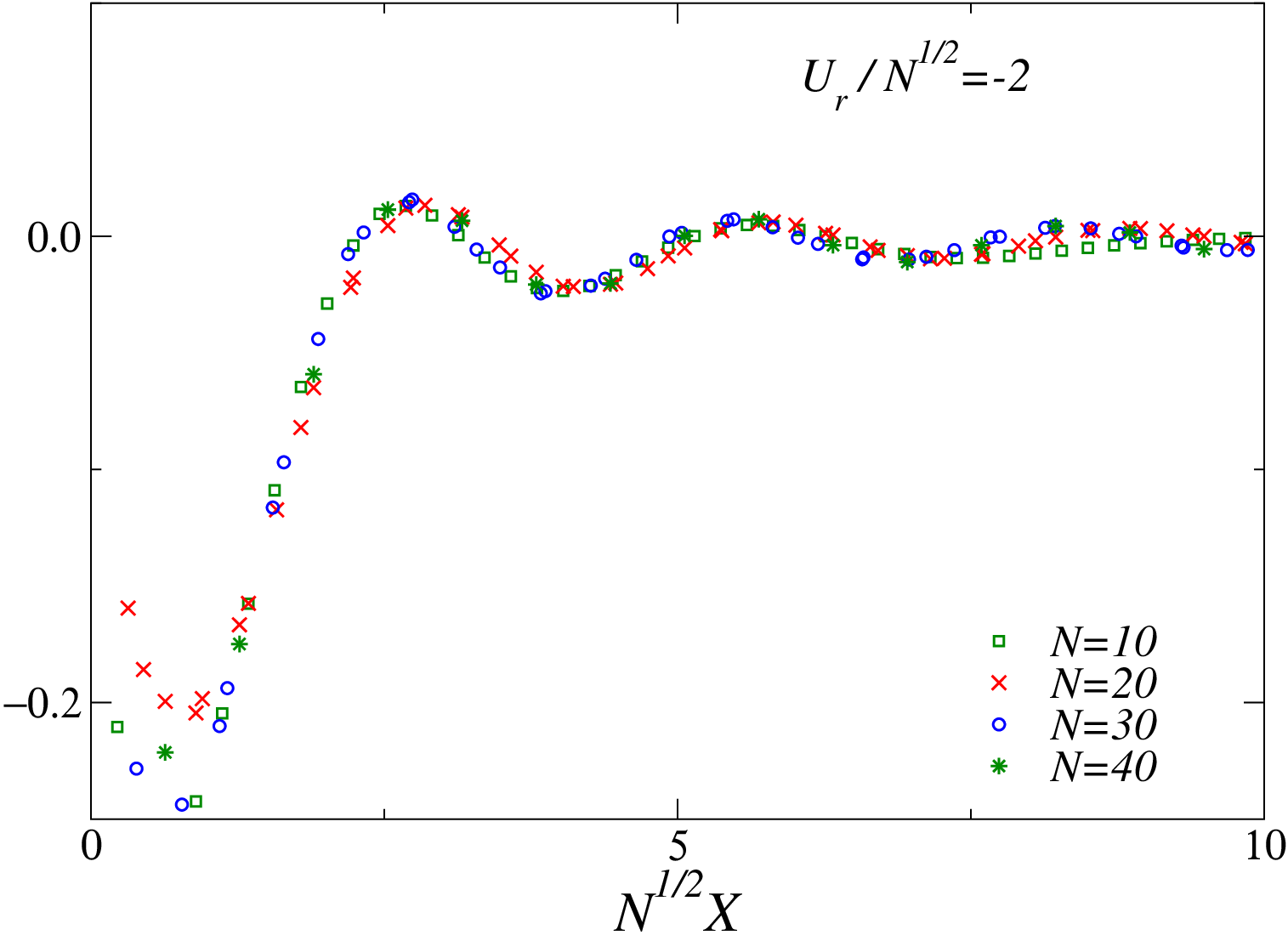}
\includegraphics*[scale=\graphicscale]{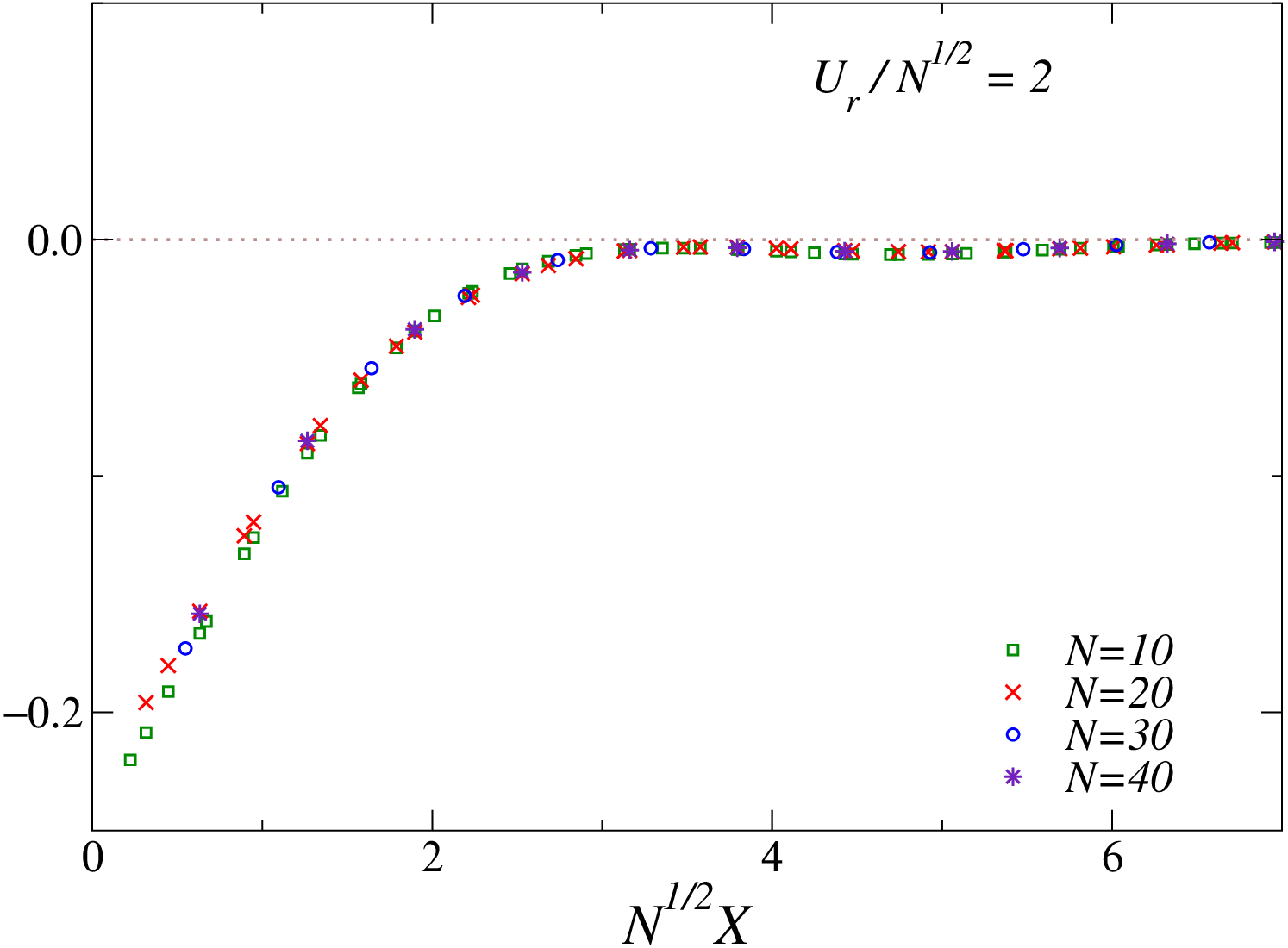}
\caption{(Color online) Large-$N$ scaling of the TSS function of the
  connected density correlation: $N^{-1}{\cal G}(0,X,U_r,N)$ versus
  $N^{1/2} X$ keeping $u\equiv U_r/N^{1/2}$ fixed, for $u=-2$ (top)
  and $u=2$ (bottom).  The data for negative $U$ appear to converge
  more slowly to their large-$N$ limit.  }
\label{gn0ln}
\end{figure}

Like the noninteracting $U=0$ case, see Sec.~\ref{tts}, the one-point
correlation and the connected density correlation scale differently
with respect to the distance from the trap center, as
\begin{eqnarray}
&&{\cal C}(X_1,X_2,U_r,N) \approx N^{1/2} 
{\cal C}_\infty(N^{1/2}X_i,U_r/N^{1/2}) ,\quad   
\label{g0s2}\\
&&{\cal G}(X_1,X_2,U_r,N) \approx N {\cal G}_\infty
(N^{1/2}X_i,U_r/N^{1/2}).  \label{gn0s2}
\end{eqnarray}
Clear evidence of this scaling behavior is shown by the DMRG data, see
e.g. Fig.~\ref{gn0ln} which shows data for $N^{-1}{\cal G}(0,X,U_r,N)$
versus $N^{1/2} X$ keeping $U_r/N^{1/2}$ fixed.

\section{The trap thermodynamic limit}
\label{thlim}

We now consider another large-$\ell$ limit, the so-called
thermodynamic limit in a trap, i.e. $\ell\to\infty,\;N\to\infty$
keeping the ratio
\begin{equation}
\upsilon \equiv N/\ell^d 
\label{rhot}
\end{equation}
fixed.
Note that $\upsilon$ becomes proportional to the filling $N/V$ in the
$p\to\infty$ limit of the trapping potential,
cf. Eq.~(\ref{potential}) (in 1D the filling would be $f=\upsilon/2$).
The above thermodynamic limit in a trap can be realized by introducing
a chemical-potential term, cf. Eq.~(\ref{chempot}).  Indeed,
$\upsilon$ and $\mu$ must be asymptotically related, i.e. $\upsilon =
\Upsilon(\mu)$.  We shall show that the function $\Upsilon(\mu)$ can
be exactly determined by LDA.

Notice that the asymptotic trap-size scaling keeping $\upsilon$ fixed
does not reproduce the continuum limit given by the GY model,
essentially because $\ell$ is considered in units of the lattice
spacing.

\subsection{Accuracy of the LDA of the particle density}
\label{acclda}

In the presence of a space-dependent confining potential, LDA
estimates the space-dependent particle density by the value of the
particle density $\rho_h(U,\mu)$ of the homogeneous system at the
effective chemical potential
\begin{equation}
\mu_{\rm eff}({\bf x}/\ell) \equiv \mu - V({\bf x}) = \mu - ({\bf x}/\ell)^2.
\label{mueff}
\end{equation}
This implies that the LDA particle density is a function of the ratio
${\bf x}/\ell$.  Indeed, we have
\begin{eqnarray}
\rho({\bf x},\ell,U,\mu) \approx 
\widetilde{\rho}_{{\rm lda}}({\bf x}/\ell, U,\mu) 
\equiv \rho_h[U,\mu_{\rm eff}({\bf x}/\ell)].
\label{lda}
\end{eqnarray} 
LDA usually provides a good approximation of the particle density when
the inhomogeneous external potential is sufficiently smooth.
Therefore, it is expected to provide better and better approximations
with increasing the trap size.  However, substantial deviations may be
present in the case the system develops long-range correlations.

LDA has been largely employed in studies of inhomogeneous fermion
systems, and in particular for 1D
systems~\cite{SGN-95,BR-97,GWSZ-00,RMBS-03,RFZZ-03,ABGP-04,MB-04,
  XPAT-06,GPTCCR-06,HLD-07,Orso-07,CCQH-07,FH-07,GPS-08,
  CCM-08,XA-08,KB-09,TU-10,HOF-10,STS-11,SBE-11,GBL-13}.  However, an
analysis of the deviations from LDA is called for, to get a robust
confidence of its results.  We investigate this issue in the trapped
1D Hubbard model (\ref{exth}), where the particle density
$\rho_h(U,\mu)$ of the homogenous system can be exactly computed using
the Bethe-Ansatz techniques, as a function of $U$ and
$\mu$.~\cite{1DHM} Some details are reported in App.~\ref{appBA}. This
allows us to check the accuracy of LDA with the Hubbard model, and
quantify the deviations of very accurate (practically exact) numerical
results obtained by DMRG simulations.

\begin{figure}[tbp]
\includegraphics*[scale=\graphicscale]{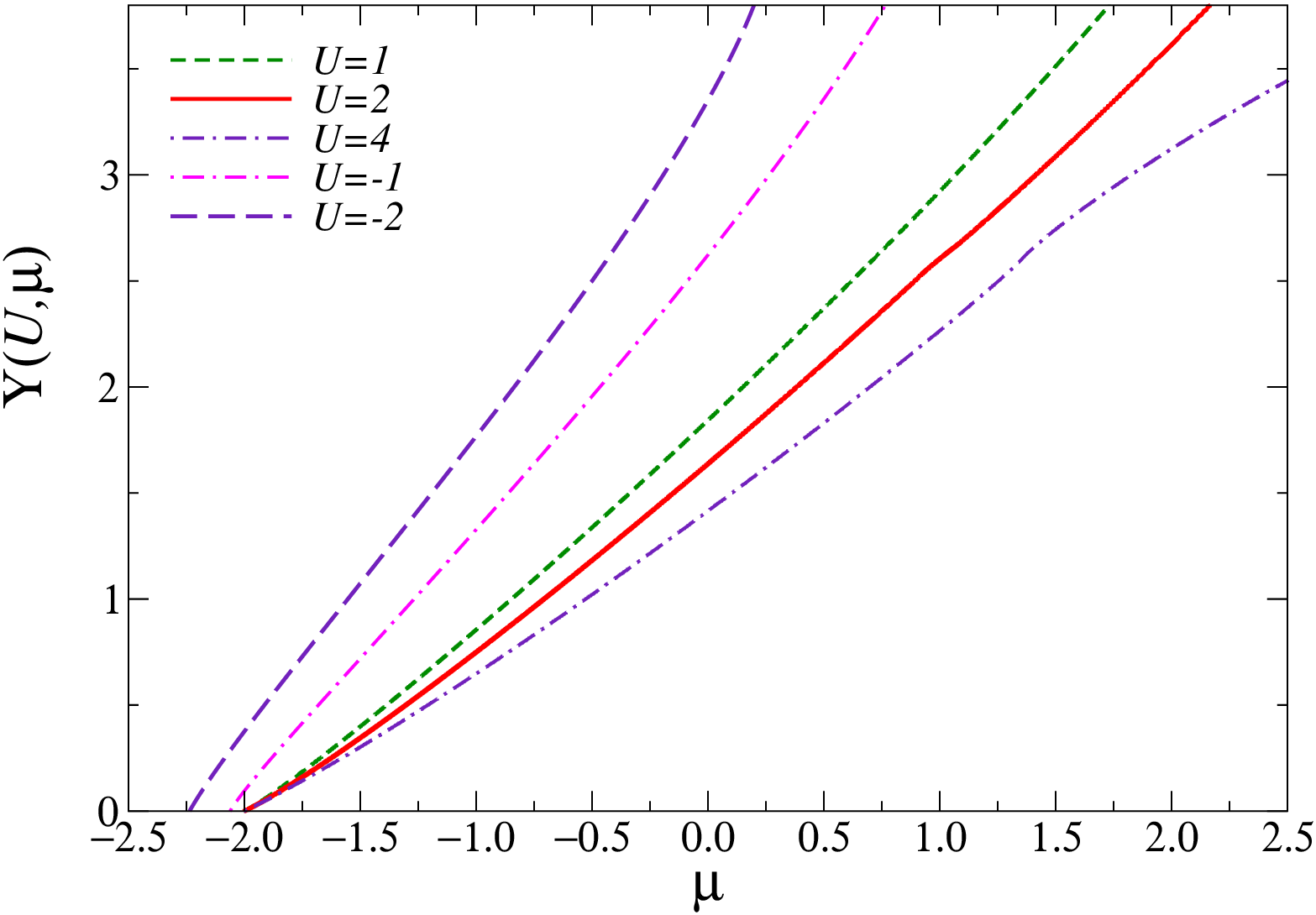}
\caption{(Color online) The function $\Upsilon(U,\mu)$ providing the
  asymptotic relation between $\upsilon\equiv N/\ell$ and $\mu$,
  cf. Eq.~(\ref{mufm1}), for various values of $U$.  }
\label{cmuplot}
\end{figure}

\begin{figure}[tbp]
\includegraphics*[scale=\graphicscale]{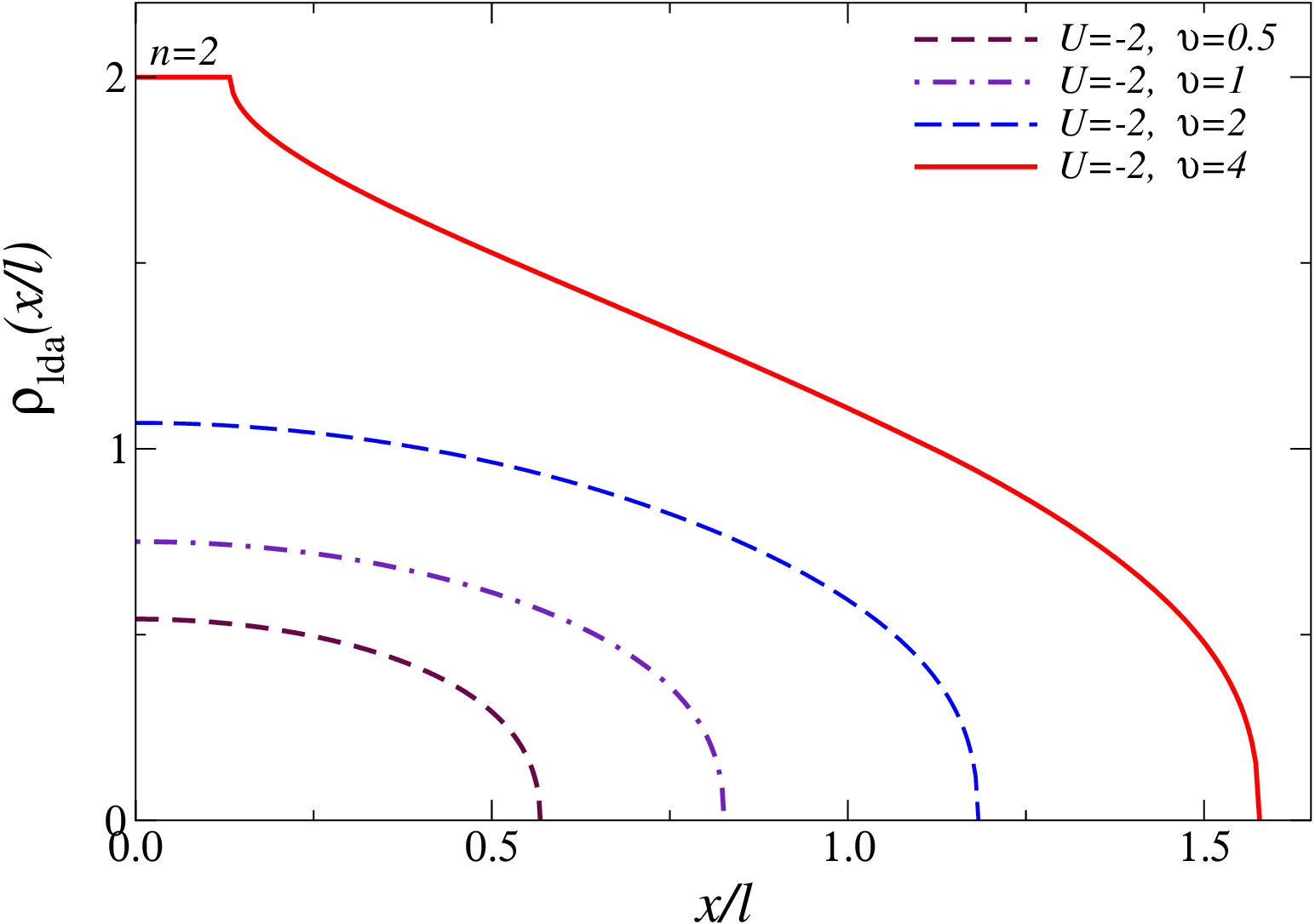}
\includegraphics*[scale=\graphicscale]{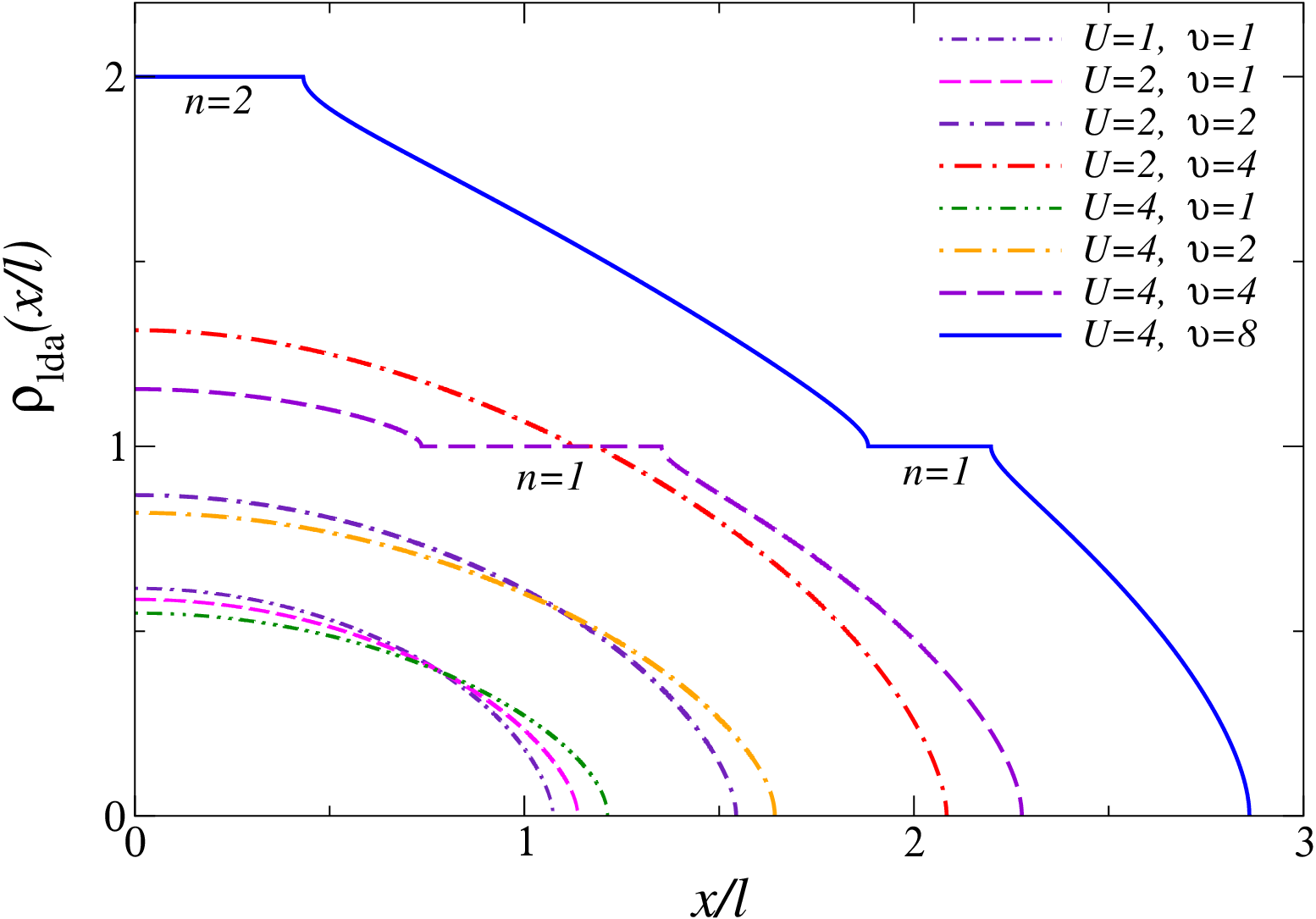}
\caption{(Color online) LDA of  the particle density for some
  negative (top) and positive (bottom) values of $U$.  Since the
  curves are symmetric for $x\to -x$, we show only them for $x\ge
  0$. Note that the $n=1$ and $n=2$ plateaus for some of the curves shown
  in the figures, which correspond to the Mott phases of Fig.~\ref{phdiad1}.  }
\label{ldafig}
\end{figure}

\begin{figure}[tbp]
\includegraphics*[scale=\graphicscale]{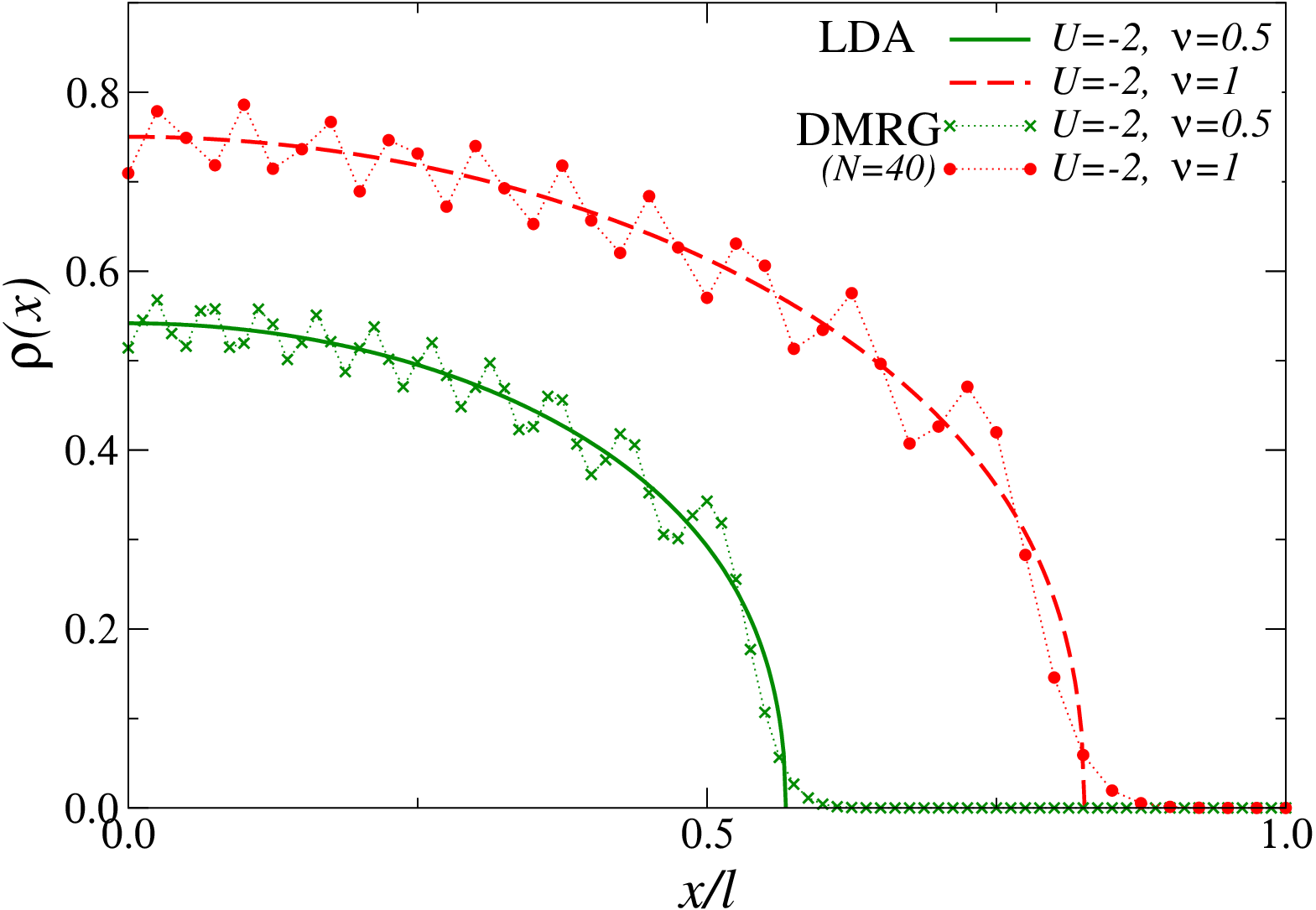}
\includegraphics*[scale=\graphicscale]{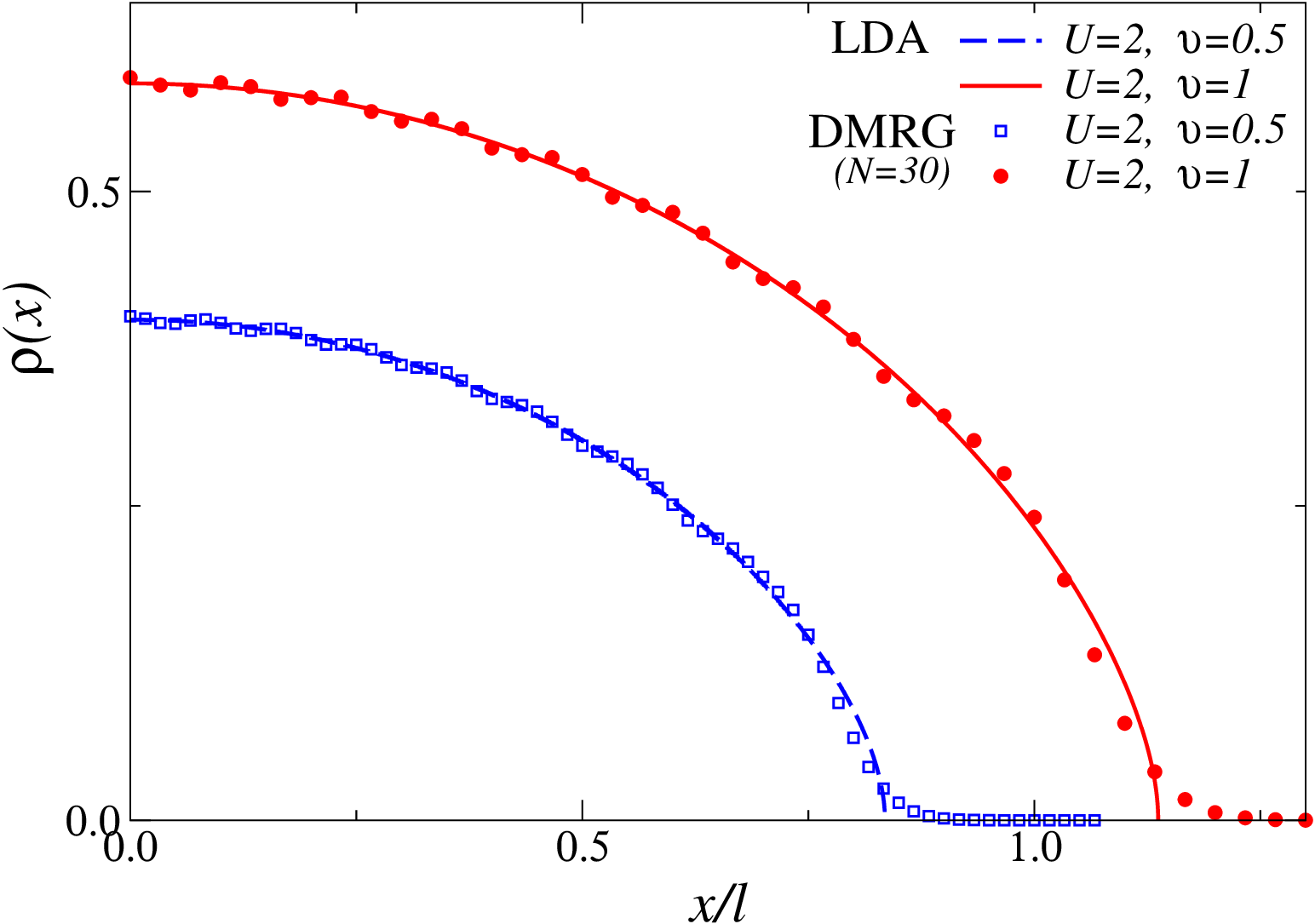}
\caption{(Color online) Comparison of DMRG and LDA results of the
  particle density, for $U=-2$ (top) and $U=2$ (bottom) and
  $\upsilon=0.5,\,1$.  Since the curves are symmetric for $x\to -x$,
  we show only data for $x\ge 0$.  }
\label{rholda}
\end{figure}

\begin{figure}[tbp]
\includegraphics*[scale=\graphicscale]{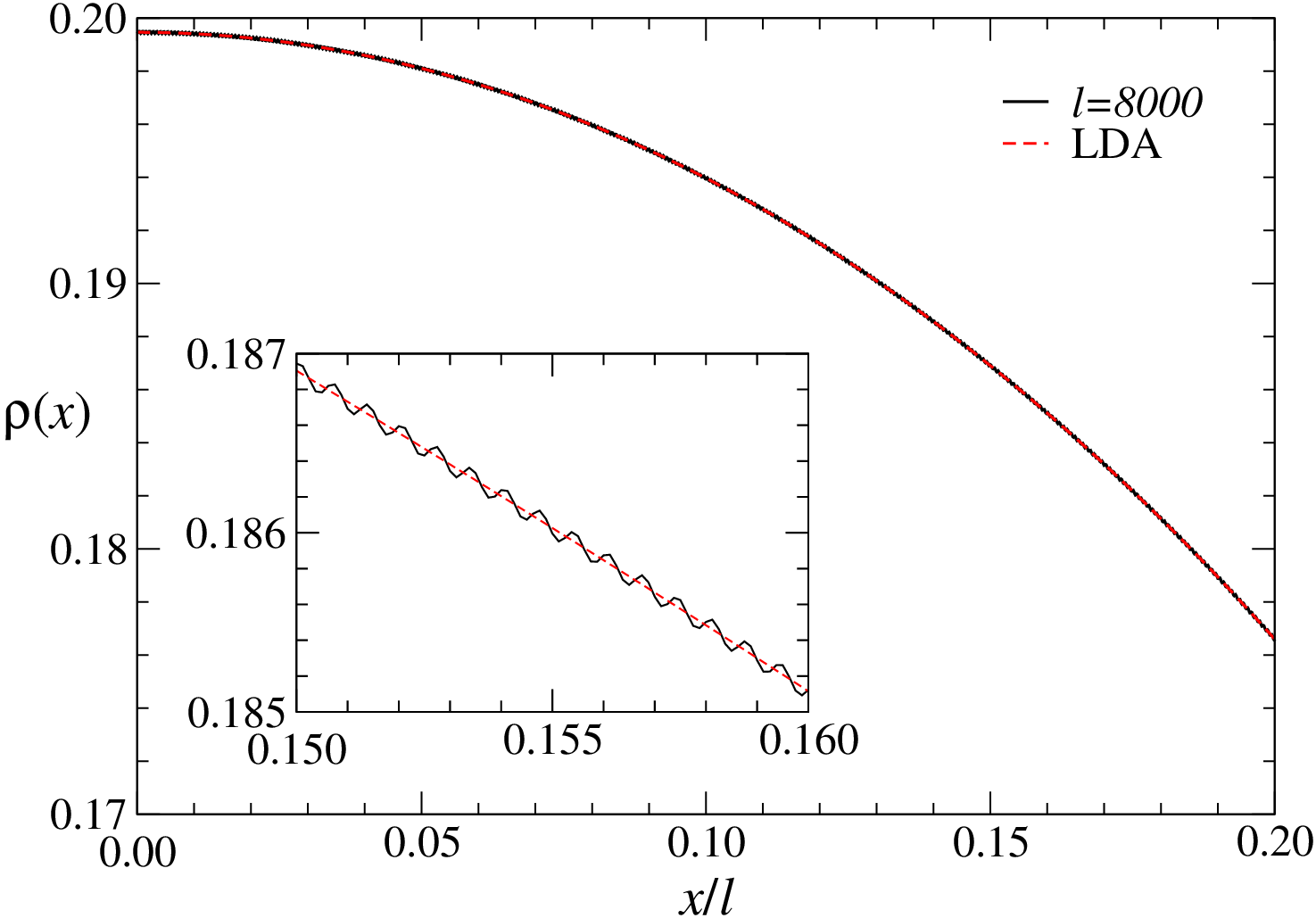}
\includegraphics*[scale=\graphicscale]{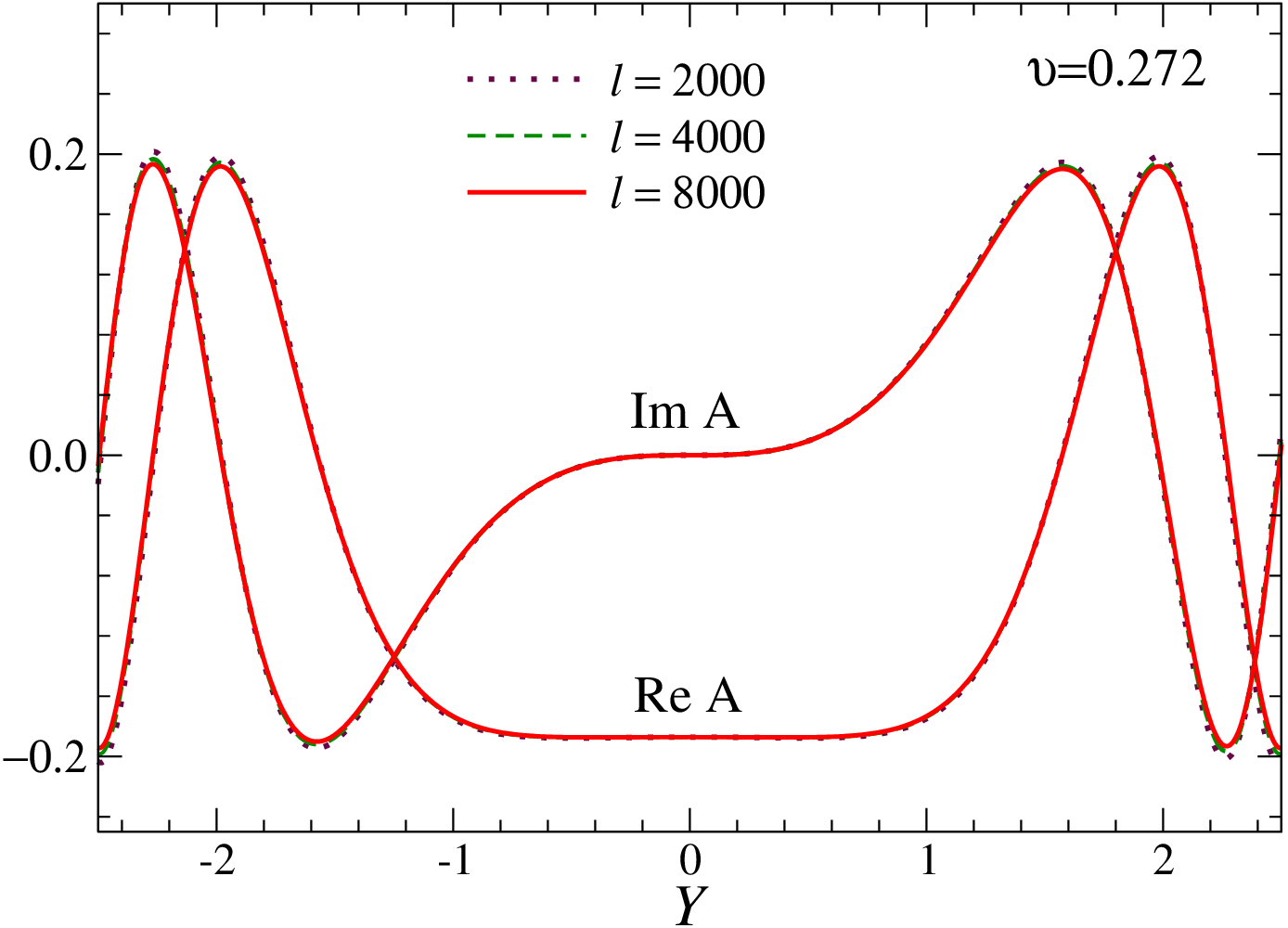}
\includegraphics*[scale=\graphicscale]{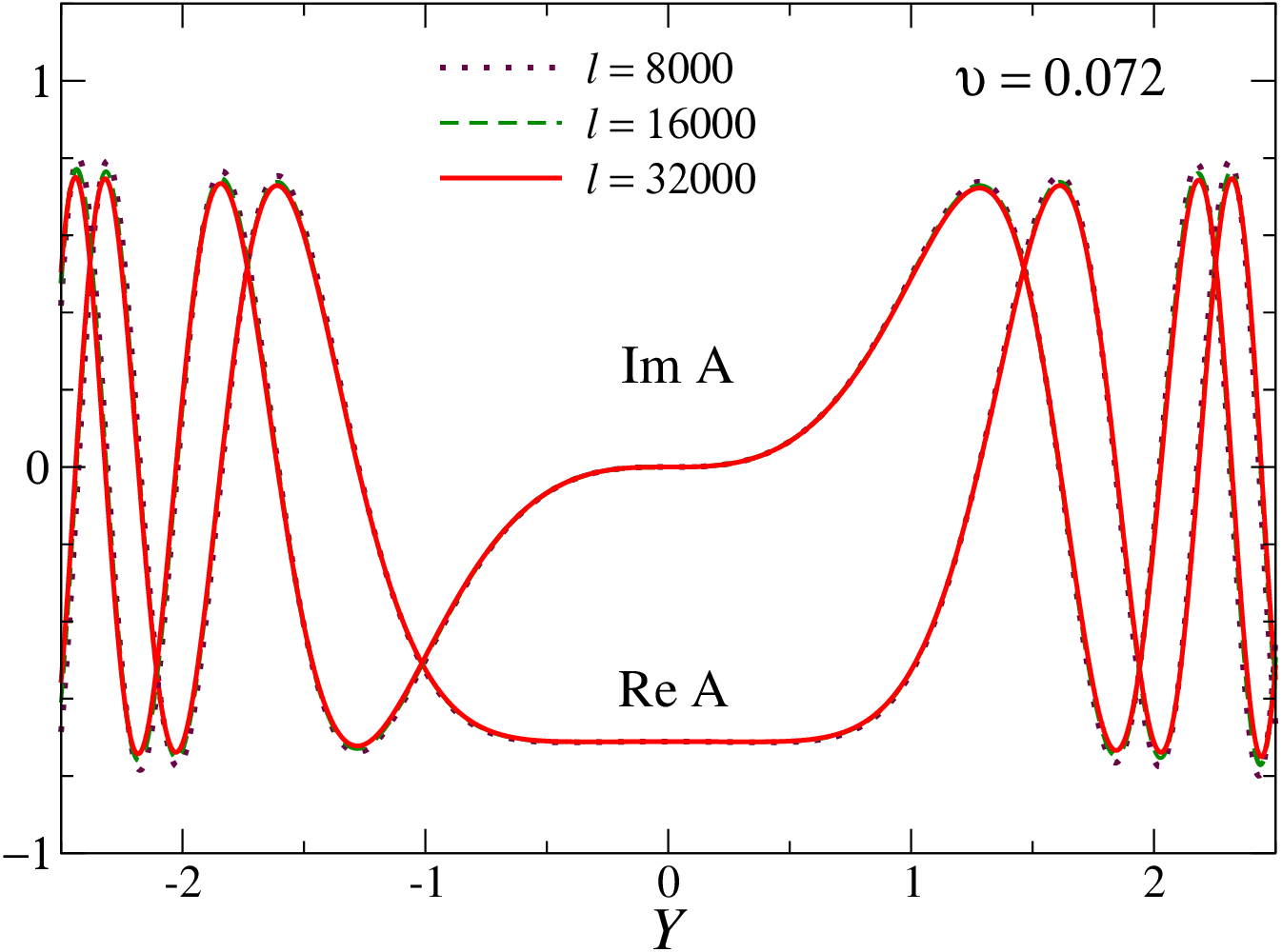}
\caption{(Color online) The top figure shows the particle density for
  $\upsilon=0.272$ and $\ell=8000$ and its LDA approximation (the
  inset is necessary to observe the tiny differences). The other
  figures show the real and imaginary part of the complex function
  $A(Y,\upsilon)$ in Eq.~(\ref{eq:rhox,sf}) for two values of
  $\upsilon$, $\upsilon=0.272$ and $\upsilon=0.072$ corresponding
  respectively to $q=0.6399...$ and $q=1.253...$.  }
\label{reima}
\end{figure}

Assuming that LDA provides the exact large-$\ell$ limit, we may derive
the asymptotic relation between the particle number and the chemical
potential:
\begin{eqnarray}
\upsilon\equiv N/\ell &\approx&   \ell^{-1} \sum_x 
\widetilde{\rho}_{\rm lda}(x/\ell,U,\mu_\upsilon) 
\label{ntlda}\\
&\approx&
\int_{-\infty}^\infty dz\,\widetilde{\rho}_{\rm lda}(z,U,\mu_\upsilon) = 
\Upsilon(U,\mu_\upsilon). 
\nonumber
\end{eqnarray}
In Fig.~\ref{cmuplot} we show the function $\Upsilon(U,\mu)$ for some
values of $U$. We use this relation to derive the chemical potential
corresponding to the value $\upsilon$ we are interested in. In
practice, for a given value of $\upsilon$, we consider the LDA
\begin{equation}
\rho_{{\rm lda}}(x,U,\upsilon) \equiv 
\widetilde{\rho}_{\rm lda}(x,U,\mu_\upsilon) 
\label{ldanu}
\end{equation}
with $\mu_\upsilon$ obtained using the equation
\begin{equation}
\upsilon = \Upsilon(U,\mu_\upsilon).
\label{mufm1}
\end{equation}
The eventual consistency of the results will support this work
hypothesis.

Before showing results for the Hubbard model, we note that for $U=0$
we must recover the known results for free fermion lattice systems.
As shown in Ref.~\cite{CV-10-XX}, the LDA of the particle density of
1D hard-core bosons, equivalent to a free spinless Fermi gas, provides
the asymptotic behavior in the large-$\ell$ limit keeping $x/\ell$
fixed.  This also applies to the unpolarized Hubbard model for $U=0$,
for which we have two identical free spinless Fermi gases with half
particles.  Thus, the LDA of the particle density at $U=0$ reads
\begin{eqnarray}
\rho_{\rm lda}(x,U=0,\upsilon) = 2 \,
\rho_{\rm lda0}(x,\upsilon/2) \label{nxlda} 
\end{eqnarray}
where
\begin{eqnarray}
&& \rho_{\rm lda0}(x,\upsilon) =
\widetilde{\rho}_{\rm lda0}(x,\mu_\upsilon) = \label{nxldaf} \\
&&\quad\left\{
\begin{array}{l@{\ \ }l@{\ \ }l}
0 & {\rm for} & \mu_{\rm eff}(x) < -2, \\
(1/\pi)\arccos\left[-{\mu_{\rm eff}(x)/2}\right] &
    {\rm for} & -2 \le \mu_{\rm eff}(x) \le 2, \\
1 & {\rm for} & \mu_{\rm eff}(x) > 2, \\
\end{array} 
\right. \nonumber
\end{eqnarray}
with $\mu_{\rm eff}=\mu_\upsilon - V(x)$, 
and $\mu_\upsilon$ is related
to the ratio $\upsilon\equiv N/\ell$ by
\begin{eqnarray}
\upsilon = \int_{-\infty}^\infty dx\,\widetilde{\rho}_{\rm lda0}(x,\mu_\upsilon)  .
\label{cmufree}
\end{eqnarray}
The corrections to the relation (\ref{ntlda}) turn out to be
$O(\ell^{-1})$ for free Fermi gases, thus for the Hubbard model at
$U=0$. We expect that this fact extends to nonzero on-site
interactions.

Fig.~\ref{ldafig} shows some results for the LDA of the particle
density, for some values of $\upsilon$ and $U$, derived using the
Bethe Ansatz for the homogenous system, see App.~\ref{appBA}, and
assuming Eq.~(\ref{ntlda}).  Depending on the values of $\upsilon$ and
$U$, the curves show plateaus at integer values corresponding to the
Mott phases of the phase diagram shown in Fig.~\ref{phdiad1}.
Analogous results have been also reported in
Ref.~\cite{RMBS-03,MOYDOM-08}, by quantum Monte Carlo
and DMRG simulations. In Fig.~\ref{rholda} we compare the LDA curves with the
DMRG results for $U=2$ and $\upsilon=1/2,\,1$.  Analogous results are
derived for other values of $U$ and $\upsilon$. We note that the
agreement is satisfactory, but deviations are clearly observed, in
particular at the boundary of the fermion cloud.  We also note that
the oscillations around the asymptotic curve are larger in the case of
attractive interactions.

In order to characterize the deviations from LDA, whether they really
vanish  in the large-$\ell$ limit and how they get suppressed if LDA
becomes asymptotically exact, we consider the difference
\begin{equation}
\Delta\rho \equiv   
\rho(x,\ell,U,\upsilon) -  \rho_{\rm lda}(x/\ell,U,\upsilon).
\label{deltarho}
\end{equation}
We investigate whether and how such a difference gets suppressed in
the large trap-size limit.  In the following we consider the case in
which the whole trapped system is within the metallic phase, but the
analysis can be straightforwardly extended to the other possible
cases.

In the case of free Fermi systems, i.e. $U=0$, the difference
$\Delta\rho$ vanishes as $O(\ell^{-1})$ around the center of the trap,
with a quite intricate scaling within the metallic phase, as shown by
the results of Ref.~\cite{CV-10-XX} for free fermion systems (or
equivalently hard-core Bose-Hubbard systems) at fixed chemical
potential.  In the case of fixed ratio $N/\ell$, an educated guess for
its asymptotic behavior turns out to be
\begin{eqnarray}
&&\ell \, \Delta \rho(x,\ell,U=0,\upsilon)  \approx
{\rm Re}\left[ A(Y,\upsilon) e^{i q x}\right]  + o(1), \label{eq:rhox,sf}\\
&&Y=x \ell^{-2/3},  \quad q = \pi \rho_{\rm lda}(0,0,\upsilon), 
\nonumber
\end{eqnarray}
where $A$ is a nontrivial complex function.  This is supported by
numerical results up to very large trap size, up to $\ell=O(10^4)$ by
exact diagonalization methods, as shown by Fig.~\ref{reima} which
shows results for some values of $\upsilon$.  This is consistent with
the analysis of free fermion systems (or equivalently hard-core
Bose-Hubbard systems) at fixed chemical potential reported in
Ref.~\cite{CV-10-XX}.  Moreover, free Fermi systems show an anomalous
behavior at the boundaries of the trap, where the LDA of the particle
density vanishes, and much larger $O(\ell^{-1/3})$ corrections
arise~\cite{CV-10-XX,CTV-12}.  An interesting issue is whether these
features extend to the interacting $|U|>0$ case.

\begin{figure}[tbp]
\includegraphics*[scale=\graphicscale]{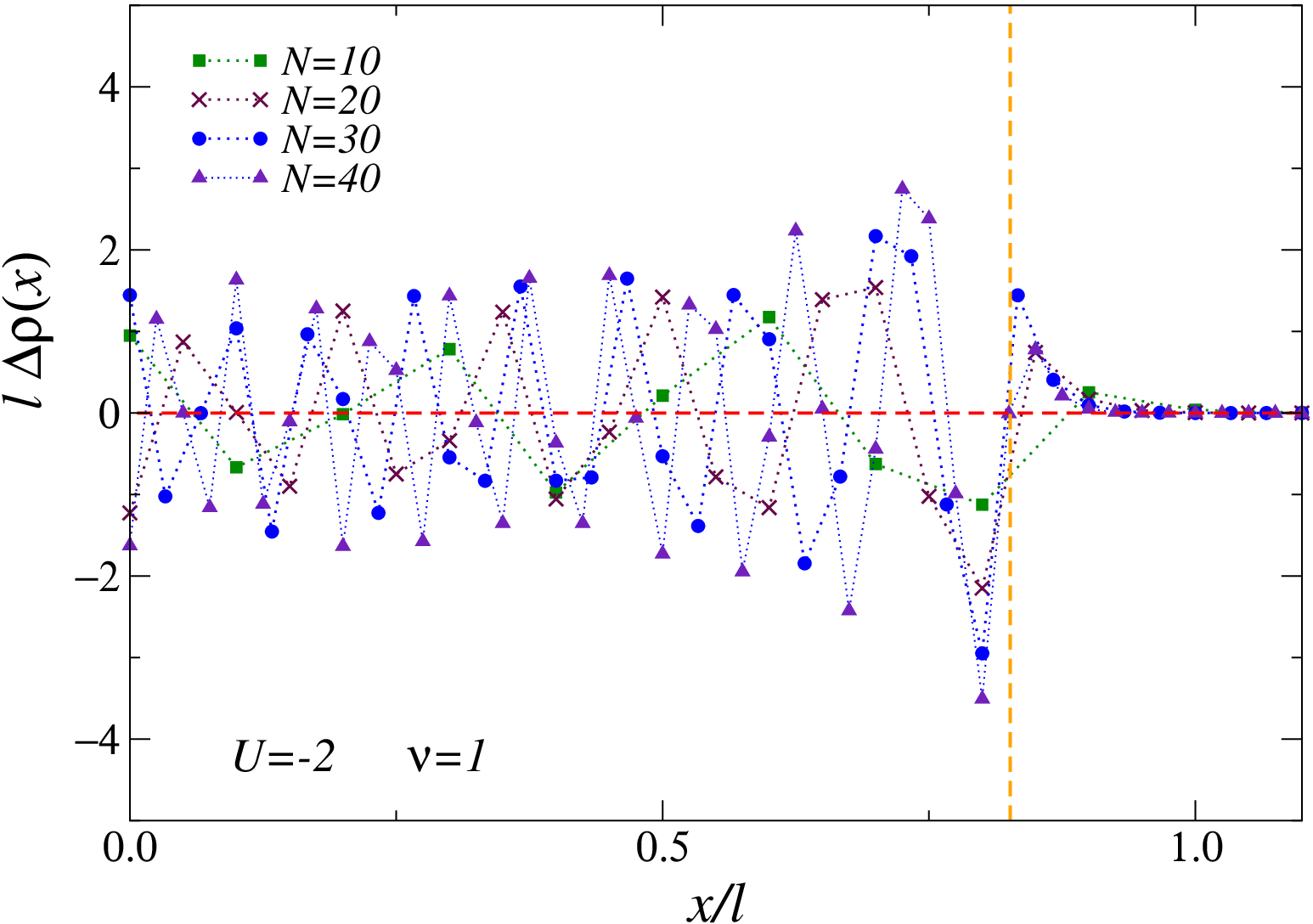}
\includegraphics*[scale=\graphicscale]{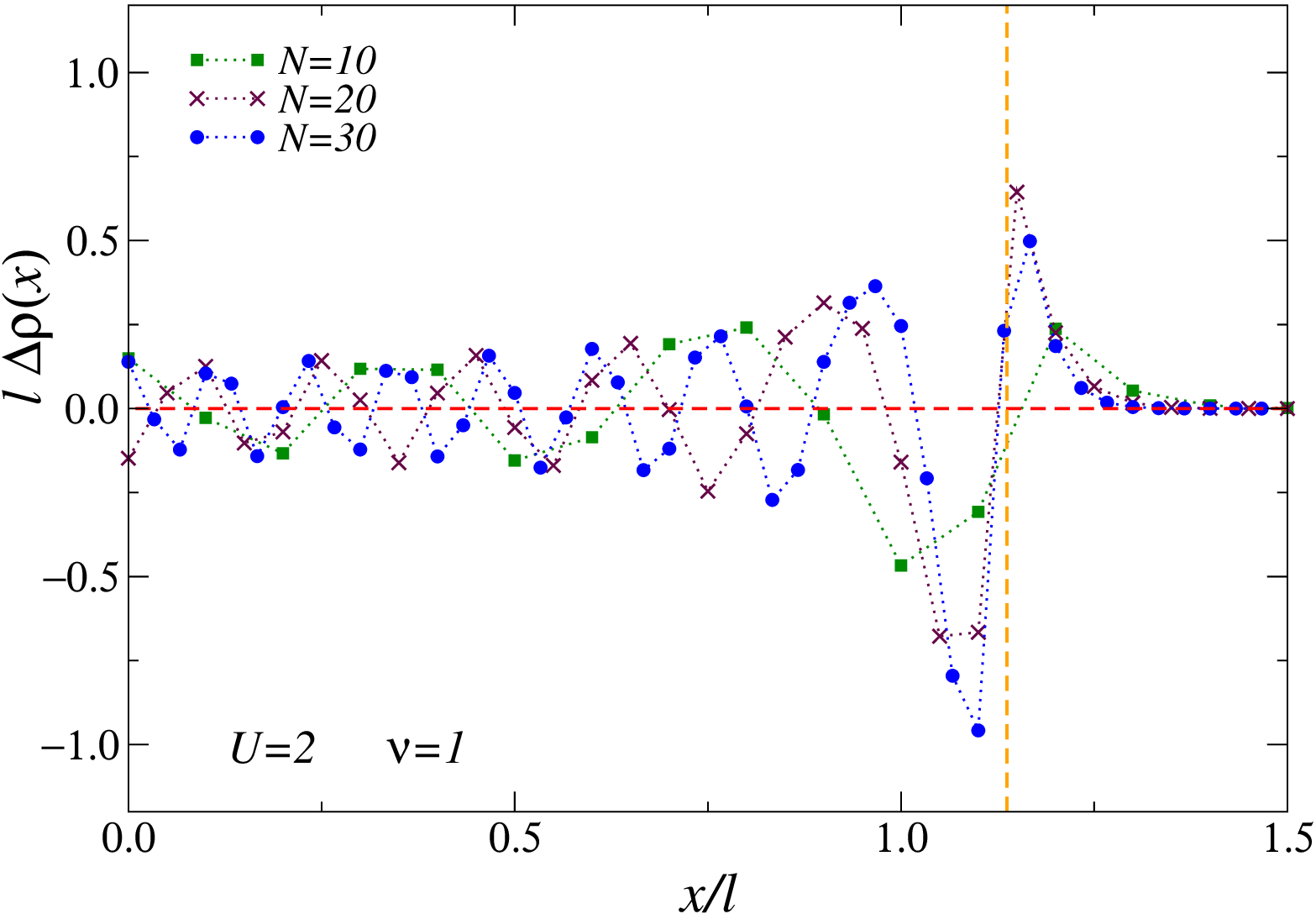}
\caption{(Color online) DMRG data for the product $\ell
  \Delta\rho(x)$, where $\Delta\rho(x)$ is the difference of the
  particle density and its LDA, for $\upsilon=1$, $U=-2$ (top) and
  $U=2$ (bottom).  The vertical dashed line indicates the value
  $X_0=x_0/\ell$ where the LDA of the particle density vanishes.  }
\label{rholdau}
\end{figure}

For this purpose we present DMRG results at fixed $\upsilon\equiv
N/\ell$ and various values of $U$.  Figs.~\ref{rholdau} show some data
of $\ell \Delta\rho$ versus $x/\ell$ for $\upsilon=1$ and $U=\pm 2$.
Analogous results are obtained for other values of $\upsilon$ and/or
$U$.  The difference $\Delta\rho(x)$, defined in Eq.~(\ref{deltarho}),
does not show a simple scaling behavior. However it shows most
features of the free Fermi gas.  Indeed, around the center of the trap
the product $\ell \Delta\rho$ shows an oscillating behavior with an
almost constant amplitude, indicating that $\Delta \rho$ gets
suppressed as $O(\ell^{-1})$ around the center of the trap. Such
oscillations were already reported and discussed in the literature,
see e.g. Refs.~\cite{RPCBTF-08,SBE-11}.  Moreover, the number $n_{\rm
  peak}$ of peaks turns out to increase proportionally to $\ell$ as
suggested by the phase term in Eq.~(\ref{eq:rhox,sf}) (more precisely
at $U=-2,\,2$ we count $n_{\rm peak}\approx N/2 = \upsilon\ell/2$).
We also note that the deviations from LDA get further suppressed,
effectively as $O(\ell^{-2})$, by averaging the oscillations around
the center of the trap, in a relatively large space interval
sufficiently far from the boundaries, for $x/\ell\lesssim 0.2$ say.
Our DMRG results for the Hubbard model are not sufficiently asymptotic
to disentangle terms scaling differently as in Eq.~(\ref{eq:rhox,sf}).

\begin{figure}[tbp]
\includegraphics*[scale=\graphicscale]{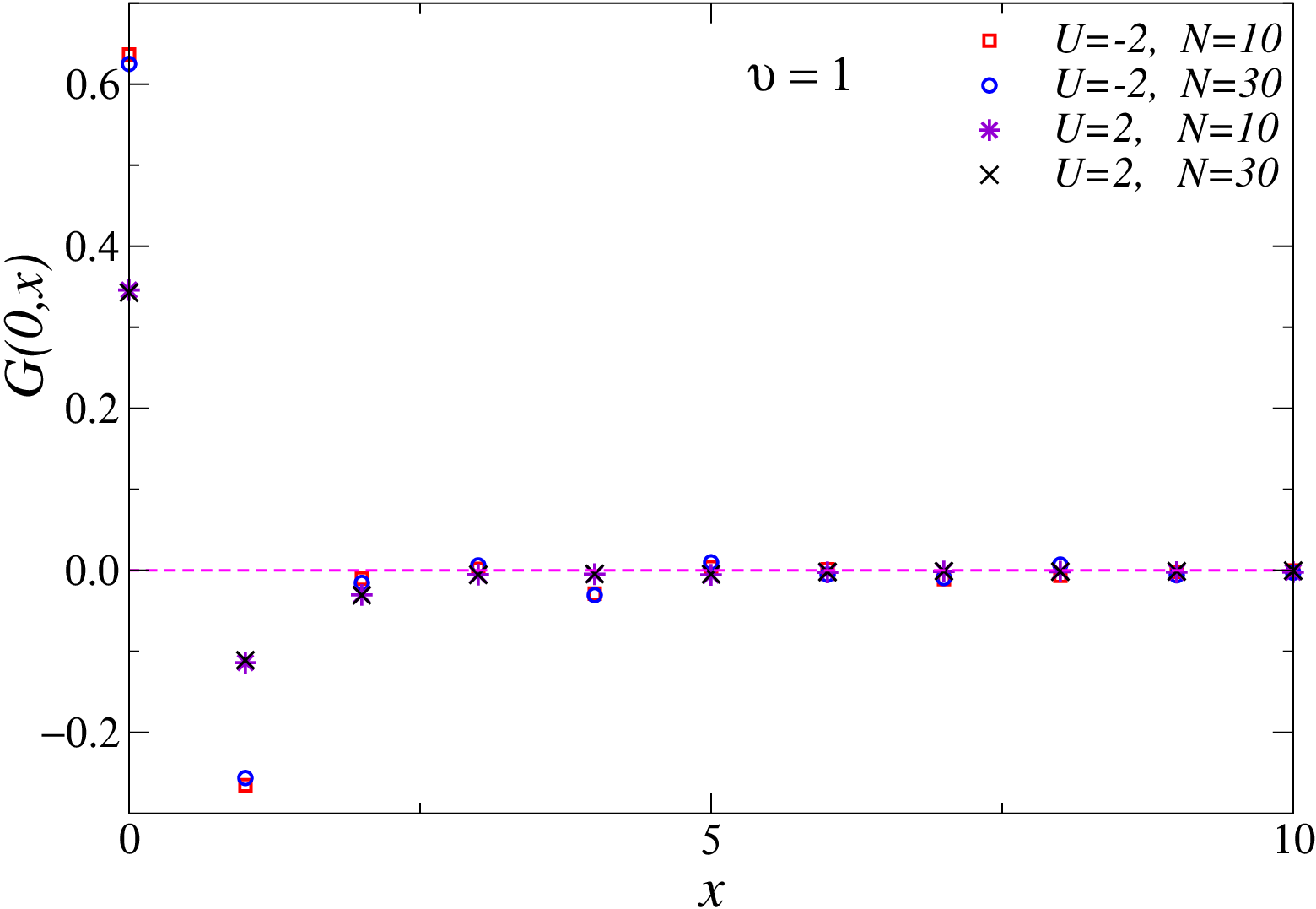}
\caption{(Color online) Results for the connected density-density
  correlation $G(0,x)$ at $U=\pm 2$ and $\upsilon=1$.  For each value
  of $U$ the $N=10$ and $N=30$ data vary very little.  }
\label{gn0u2nu1}
\end{figure}

We finally mention that the connected density-density correlation
$G(0,x)$ turns out to vanish after a few lattice spacings, without
showing any particular dependence on the trap size, see e.g.
Fig.~\ref{gn0u2nu1}.

\subsection{Scaling behavior at the boundary of the trap}
\label{boutra}

\begin{figure}[tbp]
\includegraphics*[scale=\graphicscale]{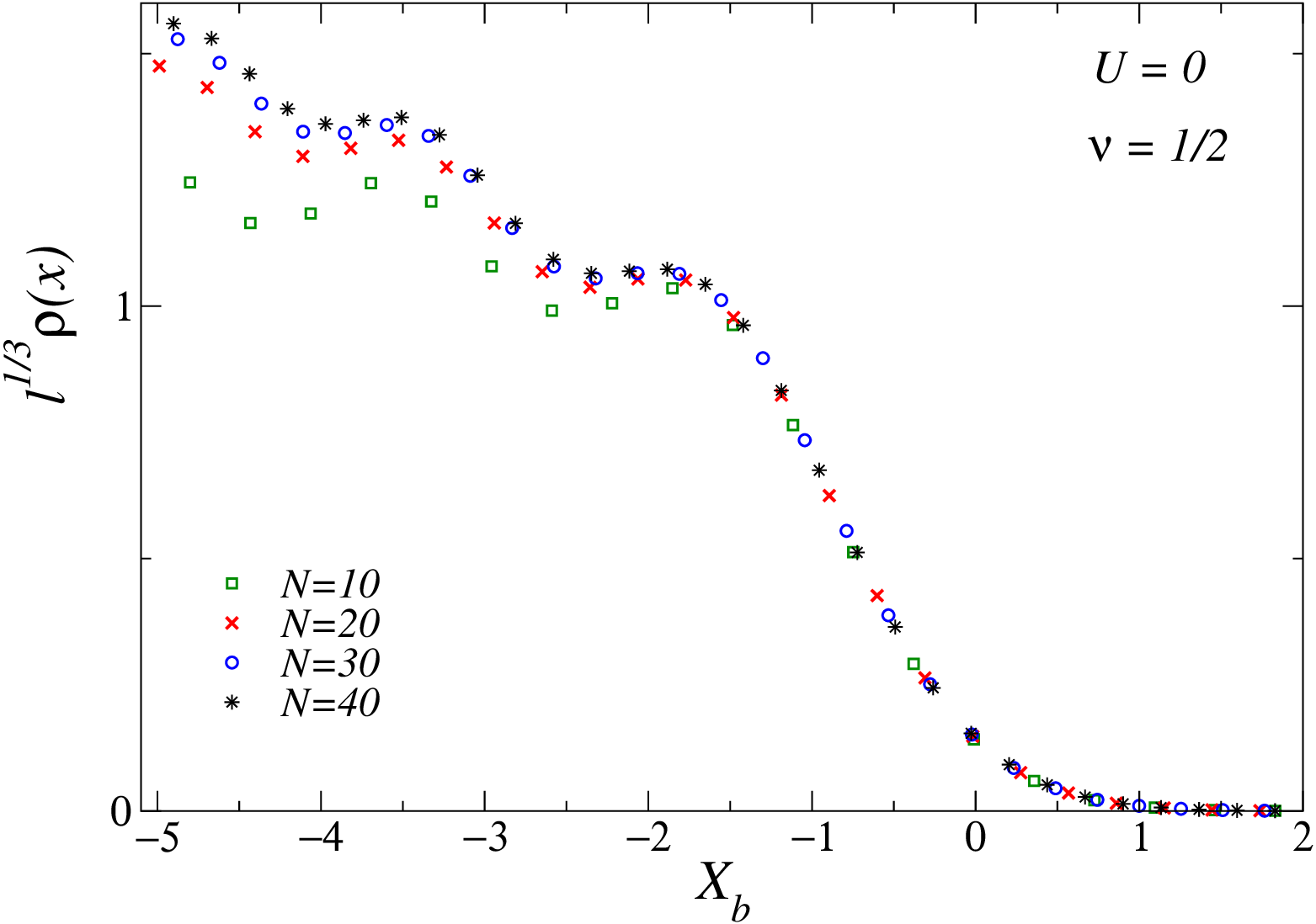}
\includegraphics*[scale=\graphicscale]{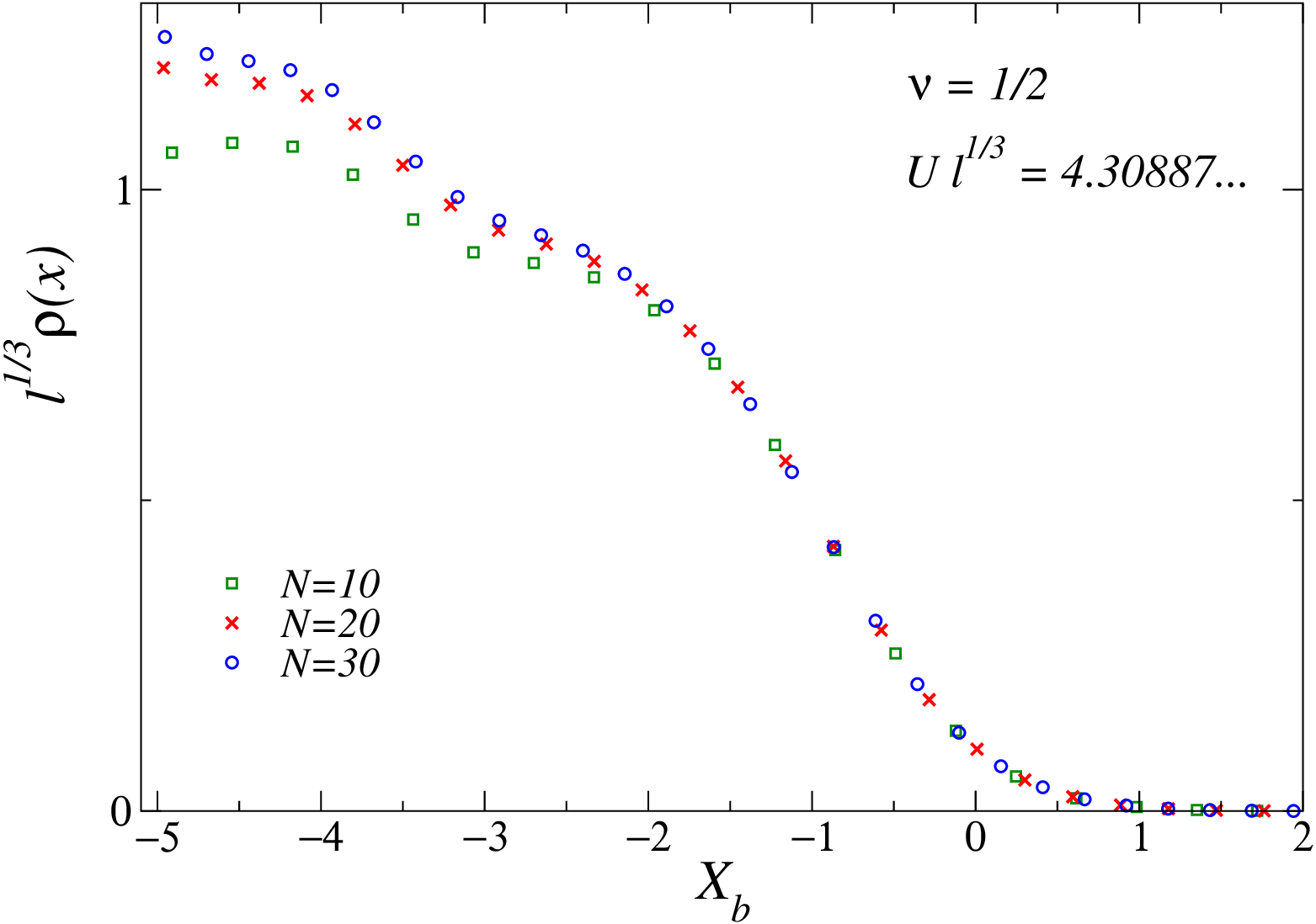}
\caption{(Color online) Scaling at the boundaries of the cloud:
  $\ell^{1/3}\rho(x)$ versus $X_b\equiv (x-x_0)/\ell^{1/3}$ for $U=0$
  (top) and $U\ell^{1/3} = 4.30887...$ (bottom).  }
\label{rhobou}
\end{figure}

As already mentioned, Figs.~\ref{rholdau} show significant deviations
at the boundary of the cloud, where the amplitude of the fluctuations
do not appear constant by clearly increase.  The data indicate that
around the boundaries of the fermion cloud the behavior is
substantially different, and the deviations may have a different
origin.  They can be explained as an effect of the metal-to-vacuum
transition occurring at the boundary of the cloud.  Indeed, for
repulsive interactions, at the spatial points corresponding to
$\mu_{\rm eff}(x)\approx \mu_0=-2$ the system effectively passes from
the vacuum [where $\mu_{\rm eff}(x)\lesssim -2$ for $U>0$] to the
metallic [where $\mu_{\rm eff}(x)\gtrsim -2$] space region.  We thus
expect that, for generic values of $\mu$ and $U>0$, the regions around
$x_0/\ell=\pm X_0$, with
\begin{equation}
X_0 = \sqrt{2+\mu},\qquad \mu_{\rm eff}(X_0)=-2,
\label{xcv}
\end{equation}
develop quantum critical modes.  But we must take into account that
this occurs in the presence of an external space-dependent field.  The
effective chemical potential can be expanded around $x_0$ as
\begin{equation}
\mu_{\rm eff}=\mu-(x/\ell)^2 
= -2 - 2 X_0 {x-x_0\over \ell} + O[(x-x_0)^2].
\label{linV}
\end{equation}
Thus, the behavior around $x_0$ is essentially analogous to that
arising at the vacuum-to-metal transition in the presence of a linear
potential $V_l\sim r/\ell$.  Around $x_0$, critical modes should
appear with length scale $\xi\sim l^\sigma$, where $\sigma$ is the
exponent associated with a linear external potential.  Thus, by
replacing $p=1$ in Eq.~(\ref{thetaexp}), we obtain $\sigma=1/3$.  We
expect that around $x=x_0$ the particle density shows the scaling
behavior
\begin{eqnarray}
&&\rho(x;U,\ell,\upsilon) \approx  \ell^{-1/3} f_\rho(X_b, U_b), \label{xc1}\\
&&X_b =  (x-x_0)/\ell^{1/3},\qquad U_b = U\ell^{1/3}.\label{boures}
\end{eqnarray}
A similar scaling behavior is also expected for the correlation
functions around $x_0$. For example the connected density correlation
is expected to scale as
\begin{eqnarray}
G(x_0,x) =  \ell^{-2/3} f_g(X_b,U_b) \label{xc2}.
\end{eqnarray}
An analogous behavior is found in trapped bosonic systems with
repulsive interaction, see e.g. the results of
Refs.~\cite{CTV-12,CT-12,CV-10-XX}.  The above scaling behaviors are
nicely confirmed by the DMRG data, see e.g. Figs.~\ref{rhobou}, where
we show results for the particle density keeping $\upsilon\equiv N/l$
and $U_b=U\ell^{1/3}$ fixed.

The scaling (\ref{xc1}) implies that $\rho(x)$ gets roughly suppressed
as $\ell^{-1/3}$ at $x_0$ where $\rho_{\rm lda}$ vanishes (assuming
that $f_\rho(0,u)\sim {\rm const}$ for $u\to\pm\infty$).  Therefore,
corrections to LDA are $O(\ell^{-1/3})$ around the boundary, thus much
less suppressed than those around the center of the trap, which are
$O(\ell^{-1})$ and oscillating.  Note that this anomalous scaling
behavior at the boundaries does not necessarily affect the corrections
between the total particle number and its LDA approximation,
cf. Eq.~(\ref{ntlda}), which are expected to be suppressed as
$O(\ell^{-1})$. Indeed, although corrections to the LDA of the
particle density are $O(\ell^{-1/3})$ around the boundary of the trap,
the region where such behavior is observed shrinks as $O(\ell^{-2/3})$
in terms of $x/\ell$ (which means that this boundary critical region
enlarges as $\ell^{1/3}$ around $x_0$).

The above edge scaling behaviors are expected to be universal apart
from a multiplicative constant and normalizations of the arguments of
the scaling functions. They are universal with respect to changes of
the chemical potential $\mu$, thus $\upsilon$, and microscopic
short-ranged interactions, for example adding the nearest-neighbor
interaction (\ref{hddhm}), essentially because they are controlled by
the dilute fixed point of the field theory (\ref{zf}) in the presence
of an external linear field.

We also note that analogous phenomena are expected around the
space region corresponding to a transition between metallic and Mott
phases, at the end of the plateaus of the particle-density curves
shown in Fig.~\ref{ldafig}.  Such a transition is controlled by the
same RG exponents of the dilute fixed points, cf. Eqs.~(\ref{rgdims})
and (\ref{thetaexp}), which can be inferred from the exact solution
for the homogenous system.  Thus scaling equations analogous to
Eqs.~(\ref{xc1}-\ref{xc2}) are expected to apply around the edges of
the density plateaus.

Analogous scaling behaviors are also expected in higher-dimensional
fermionic systems, at the boundaries of their trap. For example, in 3D
systems with a rotational invariant trap, the radial space dependence
of the particle density around the boundary of the trap should behave
\begin{eqnarray}
\rho(r;U,\ell,\upsilon) \approx  \ell^{-d/3} f_\rho(R_b), \label{xc1d3}
\end{eqnarray}
where $R_b = (r-r_0)/\ell^{1/3}$ and $r_0$ is the space distance where
the particle density vanishes asymptotically.  Note the independence
of $U$ of the leading scaling behavior, due to the fact that the
quartic coupling is irrelevant at the three-dimensional dilute fixed
point, see Sec.~\ref{ttsdilp0}.

\section{Summary and conclusions}
\label{conclusions}

We study the effects of an inhomogeneous trap in fermion systems
described by the lattice Hubbard model with an external confining
potential, in the zero-temperature limit.  This issue is of
experimental relevance because these systems are investigated in
experiments with ultracold atoms, see
e.g. Refs.~\cite{BDZ-08,GPS-08,Esslinger-10}. Indeed an important
feature of these experiments is the presence of a confining force,
which traps the atoms within a limited spatial region.

We investigate the scaling behavior of the ground state of unpolarized
systems, i.e. with zero global spin, when varying the size $\ell$ of
the trap.  The trap size $\ell$ is naturally defined by writing the
external harmonic potential as $V({\bf x}) = t |{\bf x}|^2/\ell^2$,
cf. Eq.~(\ref{trapsize}), where $t$ is the kinetic constant (tunneling
rate) of the Hubbard model (\ref{hm}). We mostly consider 1D systems,
with both attractive and repulsive interactions, in the dilute regime
when the trap size $\ell$ gets large keeping the particle number $N$
fixed, and in the {\em trap thermodynamic} limit defined as the
large-$N$ limit keeping the ratio $\upsilon\equiv N/\ell$ fixed.  We
discuss the universal behavior of several observables, such as the
particle density, double occupancy, density-density correlation,
one-point correlation, and pair correlation.

For a lattice system of $N$ fermion particles with short-range
interactions, such as the Hubbard model, the trap-size dependence in
the dilute regime shows universal scaling behaviors, which can be
described in the framework of the TSS theory~\cite{CV-10,CV-09}.  The
universal features of the TSS in the dilute regime of Hubbard models
are derived by a RG analysis of the various relevant perturbations at
the dilute fixed point.  In particular, TSS is controlled by the trap
exponent $\theta$, cf. Eq.~(\ref{thetaexp}), related to the RG
perturbation arising from the trapping potential.  This implies that
spatial coordinates ${\bf x}$ must be rescaled as ${\bf X}={\bf
  x}/\ell^\theta$ to get a nontrivial TSS limit when $\ell\to\infty$.
For harmonic traps, which is the relevant case in most experiments,
$\theta=1/2$ in any spatial dimensions.

In three dimensions the dilute fixed point is stable with respect to
the on-site interaction, thus the TSS behavior approaches that of a
free Fermi gas, cf.  Eqs.~(\ref{rhou0}), independently of $U$.  The
on-site interaction gives only rise to $O(\ell^{-\theta}$) relative
corrections.  On the other hand, in the 1D case the on-site
interaction turns out to be relevant. Thus the TSS in the dilute
regime also requires a nontrivial rescaling of the on-site coupling
$U$, indeed the corresponding scaling variable turns out to be
$U_r=U\ell^\theta$.  This RG analysis leads to the universal TSS
reported in Eqs.~(\ref{rhoxs}-\ref{demfs}).  We argue that the leading
TSS behavior reproduces the particle density and correlations of a 1D
Fermi gas defined in the continuum, given by the so-called Gaudin-Yang
model~\cite{Gaudin-67,Yang-67} in the presence of a trapping
potential, cf. Eq.~(\ref{GYmodels}), with $g\sim U_r$.

In order to check, and further characterize, the TSS of the 1D Hubbard
model at fixed $N$, we present results of DMRG simulations for several
values of $N$, the on-site interactions, and the trap size $\ell$ of
the harmonic trap.  The DMRG results are in fully agreement with the
TSS predictions.  In particular, the data show that the TSS functions
of the particle density and density-density correlation crossover
among different asymptotic regimes where these quantities can be
exactly computed, i.e.  for strongly repulsive interactions where they
approach those of a spinless Fermi gas, for weak interactions those of
a free Fermi gas, and for strongly attractive interactions they match
those of a gas of hard-core bosonic molecules.  In these 1D systems
the formation of a gas of molecules of pair fermions presents some
analogies to the formation of a 3D molecule BEC and the BCS-BEC
crossover for 3D Fermi systems, which has been recently observed in
experiments with ultracold Fermi gases, see e.g.
Refs.~\cite{Esslinger-10,RGJ-04,Zwer-etal-04,Kin-etal-04,Chin-etal-04,
  Bou-etal-04,Par-etal-05,HDL-07}.

The large-$N$ behavior of the TSS functions show a further nontrivial
large-$N$ scaling.  We argue that the space dependence of the particle
density and the correlations show a substantially different scaling
behavior, see Eqs.~(\ref{lnrho1}), (\ref{g0s2}), and
(\ref{gn0s2}). Indeed, while the large-$N$ scaling behavior of the
particle density is realized keeping $X/N^{1/2}$ fixed, the large-$N$
behavior of the TSS functions of the correlations is obtained keeping
$N^{1/2} X$ fixed, thus remaining significant only at short distance.

The 1D Hubbard model in the trap thermodynamic limit is asymptotically
equivalent to introducing a chemical-potential term, as in
Eq.~(\ref{chempot}).  We address the issue of the accuracy of the LDA,
which approximates the space-dependent particle density in a trap by
the particle density of the homogenous system at the corresponding
value of the effective chemical potential.  
LDA is routinely used in the analyses of the numerical and experimental
data of inhomogeneous particle systems.  However, a quantitative
analysis of the deviations from LDA, and therefore of its accuracy, is
required to get a robust confidence of its results, especially when
the actual trap size is not very large and/or the system is close to
criticality.

In the 1D Hubbard model the validity of LDA can be accurately checked
because the particle density of homogenous systems can be exactly
computed by Bethe-Ansatz methods.  We show that LDA becomes exact in
the large trap-size limit keeping $x/\ell$ fixed, with power-law
suppressed corrections.  When the trapped system is in the metallic
phase, the corrections to LDA turn out to be $O(\ell^{-1})$ and
oscillating around the center of the trap. Actually they may be
further suppressed by appropriate averages around the center of the
trap, likely to $O(\ell^{-2})$.  However, they become much larger at
the boundary of the fermion cloud, where they get suppressed as
$O(\ell^{-1/3})$ only.  Such an anomalous behavior at the boundary of
the trapped fermion cloud is explained, and described, by a quantum
critical behavior at the metal-to-vacuum transition occurring at the
boundaries of the trap.

In our study we consider in particular the observables related to the
particle density, double occupancy, and their correlations,
determining the scaling behaviors of their space dependence along the
trap.  These features can be experimentally investigated by in situ
imaging techniques for atomic quantum
gases,~\cite{GZHC-09,Bakr-etal-10,Sherson-etal-10} which allow to
probe the spatial dependence of the particle density, and also density
fluctuations and density-density correlations, see, e.g.,
Refs.~\cite{HZHTGC-11,Endres-etal-11,Cheneau-etal-12}.

The study of the scaling behavior of trapped interacting Fermi systems
may be extended considering also unbalanced systems. In this case
attractive interactions give rise to new states such as the
Fulde-Ferrel-Larkin-Ovchinnikov (FFLO) state~\cite{FFLO}, in which an
unbalance in the populations with different spin leads to the
formation of Cooper pairs with nonzero momentum.  Experimental
evidence of such states has been recently reported, see e.g.
\cite{FFLOexp}.  We expect that Hubbard models describing trapped
imbalanced Fermi gases present a nontrivial scaling behavior as well,
sharing most features with the balanced case. In particular, the
extension of our analysis to imbalanced fermions in the dilute regime
should be straightforward, a universal TSS should be observed,
analogously to that found in the balanced case, but 
with TSS functions also depending on the total spin.

In our paper we mostly focus on 1D systems, but the general features
should also apply to higher-dimensional systems, in particular those
related to the universal TSS in the dilute regime, the accuracy of the
LDA, and the scaling behavior at the boundaries of the trap.

\appendix

\section{Mapping between 1D Hubbard and 
 two-flavor Bose-Hubbard models}
\label{equibh}

We discuss the exact mapping between the 1D Hubbard model and
the Bose-Hubbard model with two bosonic fields.  We consider the
two-flavor Bose-Hubbard open chain of length $L$ with
Hamiltonian
\begin{eqnarray}
H_{\rm BH} &=& 
- \sum_{\langle ij\rangle,\alpha} J_\alpha 
(b_{\alpha i}^\dagger b_{\alpha j} + b_{\alpha j}^\dagger b_{\alpha i})
\\
&&
+ \sum_{\alpha i} U_\alpha n_{\alpha i}(n_{\alpha i}-1)
+ U_{12} \sum_i n_{1i}n_{2i}, \label{bhtwo}
\nonumber
\end{eqnarray}
where $b_{i\alpha}$ is a bosonic creation operator and $\alpha=1,2$
and $n_{i\alpha} \equiv b_{i\alpha}^\dagger b_{i\alpha}$ is the
particle density.  We are interested in the hard-core limit
$U_\alpha\to\infty$, where $n_{\alpha i}$ is limited to the values 0
and 1.  

Following Ref.\ \onlinecite{HR-11},
we first map the Bose-Hubbard model into a spin-1/2 system through the
Holstein-Primakoff transformation
\begin{eqnarray}
\sigma^+_{\alpha i} &=& b^\dagger_{\alpha i} 
\sqrt{1 - b^\dagger_{\alpha i} b_{\alpha i}}, \nonumber\\
\sigma^-_{\alpha i} &=& \sqrt{1 - b^\dagger_{\alpha i} b_{\alpha i}}
\; b_{\alpha i},
\nonumber \\
\sigma^z_{\alpha i} &=& b^\dagger_{\alpha i} b_{\alpha i} - {1\over2}.
\label{Eq:HP}
\end{eqnarray}
Then the spin-1/2 system can be mapped into a Hubbard model through
the Jordan-Wigner transformation~\cite{LSM}
\begin{equation}
\sigma^+_{\alpha i} = c^\dagger_{\alpha i} R_{\alpha i}, \quad
\sigma^-_{\alpha i} = R_{\alpha i} c_{\alpha i}, \quad
\sigma^z_{\alpha i} = c^\dagger_{\alpha i} c_{\alpha i} - {1\over2},
\label{Eq:JW}
\end{equation}
where
\begin{eqnarray}
R_{1i} &=& (-1)^{\sum_{j<i} n_{1j}}, \quad
R_{2i} = (-1)^{N_1 + \sum_{j<i} n_{2j}},
\nonumber \\
N_\alpha &=& \sum_i n_{\alpha i}.
\end{eqnarray}
The resulting Hamiltonian is exactly the Hubbard model Hamiltonian (1).
The above mapping also applies if additional interactions only
involving $n_{\alpha i}$ are present, e.g., a trapping potential as in
Eq.\ (4) or a nearest-neighbor $nn$ coupling (extended Hubbard model).

Concerning the boundary conditions, while the case of OBC is quite
trivial, we note that periodic boundary conditions in $b$ are not
equivalent to periodic boundary conditions in $c$.

\section{The Bethe \emph{Ansatz}}
\label{appBA}

The Bethe-Ansatz method allows us to solve exactly several 1D models,
like the Hubbard model, the XXZ model, the Gaudin--Yang model,
etc. This method reduces the complex many-particle scattering process
between fermions to the composition of many two-particle processes,
where particle momenta are not changed, but merely reassigned to
different particles.  This physical picture allows us to introduce a
wavefunction ansatz where the role of momenta permutations is
explicit: the ansatz keeps in account all possible outcoming
scattering states, weighted with appropriate
amplitudes. Model-specific conditions are used to give conditions
about them: in the Hubbard Model~\cite{1DHM} this is achieved using
boundary conditions, the first quantization Schr\"odinger equation,
and symmetry conditions, obtaining a number of relations called
Lieb-Wu equations.

While these equations are in general difficult to solve, in the
thermodynamic limit their solutions are approximated by particular
solution patterns in the complex plane: assuming that all solutions
can be approximated by these patterns allows us to rewrite the Lieb-Wu
equations as integral equations which allow us to compute the
thermodynamic quantities of the model. In the absence of external
magnetic fields and at $T = 0$, the particle density $\rho(\mu)$ for
$U>0$ can be determined solving the integral-equation system
\begin{eqnarray}
\kappa_Q(k) &=& -2 \cos k - \mu + \\
&&+\int_{-Q}^{+Q} dk'\, R \left( \sin k' - \sin k \right) 
\kappa_Q(k') \cos k', \nonumber\\
 \Pi_Q(k) &=& \frac{1}{2 \pi} + \cos k \int_{-Q}^{+Q} dk'\,R \left( \sin k' - \sin k 
\right) \Pi_Q(k'), \nonumber \\
\rho(Q) &=& \int_{-Q}^{+Q} \Pi_Q(k) \, dk, \nonumber \\
R(x) &\equiv& \int_{- \infty}^{+ \infty} \frac{e^{i \omega x}}{1 + e^{U|\omega|/2}} \, 
\frac{d \omega}{2 \pi},\nonumber
\end{eqnarray}
where $|k|<Q$, $\kappa_Q(k)$ is the dressed energy of the relevant
elementary excitations of the model over the ground state and
$\Pi_Q(k)$ is the density of particles in the momentum space.  The
equation $\kappa_Q(k = \pm Q) = 0$ can be decomposed in further
integral equations, which allow us to link the free parameter $Q \in [0,
  \pi]$ to the chemical potential $\mu$, obtaining $\rho(\mu) =
\rho[Q(\mu)]$.

The attractive case ($U < 0$) can be reduced to the repulsive
one\cite{1DHM} applying appropriate transformations to the Bethe
Ansatz Equations and states: in this case, the integral equations
defining the density $\rho(\mu)$ of the model at $T = 0$ and in
absence of magnetic field become
\begin{eqnarray}    
&& \epsilon_{\Lambda}(\lambda) + \int_{-\Lambda}^{+\Lambda} \frac{|U|}
{2 \pi [ U^2/4 + ( \lambda - \nu)^2 ]} 
\epsilon_{\Lambda}(\nu) \; d \nu = \nonumber \\
&&= 4 \, \mathrm{Re}[1 - \left( \lambda + iU/4 \right)^2]^{1/2} + 2 \mu, 
\nonumber \\
&& \epsilon_{\Lambda}(\lambda = \pm \Lambda) = 0,
\nonumber \\
&& \sigma_{\Lambda}(\lambda) + \int_{-\Lambda}^{+ \Lambda} \frac{|U|}
{2 \pi [U^2/4 + ( \lambda - \nu )^2]} \sigma_{\Lambda}(\nu) \; d \nu =\nonumber \\
&&= \frac{1}{\pi} \, \mathrm{Re}[1 - \left( \lambda + i U/4 \right)^2]^{-1/2}, \nonumber \\
&& \rho(\Lambda) = 2 \int_{-\Lambda}^{+\Lambda} \sigma_{\Lambda}(\lambda) \; d\lambda,
\label{Um0eqs}
\end{eqnarray}
where $|\lambda| < \Lambda$.
In this case $\epsilon_{\Lambda}(\lambda)$ is the dressed energy of
the model excitations, and is function of the spin rapidity $\lambda$
(fermionic spin rapidities are, as fermionic momenta, quantum numbers
parametrizing the Bethe Ansatz solutions). $\sigma_{\Lambda}(\lambda)$
is the particle density in the spin rapidity space. As previously, the
relation $\epsilon_{\Lambda}(\lambda = \pm \Lambda) = 0$ can be
decomposed in further integral equations, which allow us to find a
relation between the positive parameter $\Lambda$ and the chemical
potential: this yields the relation $\rho(\mu) = \rho[\Lambda(\mu)]$.

\end{document}